\documentclass[cernpreprint,texmf,texlive=2014,UKenglish,txfonts,siunitx=false]{atlasdoc} 
\usepackage[centering,scale=0.75]{geometry}

\hypersetup{colorlinks,breaklinks,pdftitle={ATLAS Draft Cover},pdfauthor={ATLAS Collaboration}}
\hypersetup{linkcolor=blue,citecolor=blue,filecolor=black,urlcolor=blue}

\usepackage{graphicx}
\usepackage{amsbsy}
\usepackage{amssymb}
\usepackage{amstext}
\usepackage{multirow}
\usepackage{lscape}
\usepackage{lineno}
\usepackage{rotating}
\usepackage{slashed}
\usepackage{cite}
\usepackage{xspace}

\newcommand{\dyjets}{\ensuremath{Z/\gamma^{*}+\text{jets}}}
\newcommand{\zjets}{\ensuremath{Z+\text{jets}}}
\newcommand{\sherpa}{{\sc Sherpa}}
\newcommand{\mll}{\ensuremath{m_{\ell\ell}}}
\newcommand{\madgraph}{{\sc MadGraph}}

\newcommand{\SF}{same-fla\-vour\xspace}
\newcommand{\OS}{op\-po\-site-sign\xspace}
\newcommand{\SFOS}{\SF\OS\xspace}

\AtlasTitle{Search for supersymmetry in events containing a same-flavour opposite-sign dilepton pair, jets, and large missing transverse momentum in $\sqrt{s}=8$~\TeV\ \
$pp$ collisions with the ATLAS detector}

\PreprintIdNumber{CERN-PH-EP-2015-038}

\AtlasJournal{Eur. Phys. J. C}

\usepackage{atlasphysics}

\AtlasAbstract{ %
Two searches for supersymmetric particles in final states containing a \SF\ \OS\ lepton pair, jets and large missing transverse momentum are presented. 
The proton--proton collision data used in these searches were collected at a centre-of-mass energy $\sqrt{s}=8$ TeV by the ATLAS detector 
at the Large Hadron Collider and corresponds to an integrated luminosity of 20.3~fb$^{-1}$.
Two leptonic production mechanisms are considered:
decays of squarks and gluinos with $Z$ bosons in the final state, resulting in a peak in the dilepton invariant mass distribution around the $Z$-boson mass; 
and decays of neutralinos (e.g. $\tilde{\chi}^{0}_{2} \rightarrow \ell^{+}\ell^{-}\tilde{\chi}^{0}_{1}$), 
resulting in a kinematic endpoint in the dilepton invariant mass distribution.
For the former, an excess of events above the expected Standard Model background is observed,
with a significance of 3 standard deviations.
In the latter case, the data are well-described by the expected Standard Model background. 
The results from each channel are interpreted in the context of several supersymmetric models involving the production of squarks and gluinos.
}

\begin{document}

\section{Introduction}
\label{sec:intro}
Supersymmetry (SUSY)
\cite{Miyazawa:1966,Ramond:1971gb,Golfand:1971iw,Neveu:1971rx,Neveu:1971iv,Gervais:1971ji,Volkov:1973ix,Wess:1973kz,Wess:1974tw}
is an extension to the Standard Model (SM) that introduces supersymmetric particles (sparticles), which
differ by half a unit of spin from their SM partners.
The squarks ($\tilde{q}$) and sleptons ($\tilde{\ell}$) are the scalar partners of the quarks and leptons, and
the gluinos ($\tilde{g}$) are the fermionic partners of the gluons.
The charginos (${\tilde{\chi}}_{i}^{\pm}$ with $i=1,2$) and 
neutralinos (${\tilde{\chi}}_{i}^{0}$ with $i=1,2,3,4$) are the mass eigenstates (ordered from the lightest to the heaviest) formed from 
the linear superpositions of the SUSY partners of the
Higgs and electroweak gauge bosons. SUSY models in which the gluino, higgsino and top squark 
masses are not much higher than the \TeV~scale can provide a solution to the SM hierarchy problem 
\cite{Witten:1981nf,Dine:1981za,Dimopoulos:1981au,Sakai:1981gr,Kaul:1981hi,Dimopoulos:1981zb}.
 
If strongly interacting sparticles have masses not higher than the \TeV~scale, they should be produced with
observable rates at the Large Hadron Collider (LHC). In the minimal supersymmetric extension of the SM,   
such particles decay into jets, possibly leptons, and
the lightest sparticle (LSP). If the LSP is stable due to R-parity conservation \cite{Fayet:1976et,Fayet:1977yc,Farrar:1978xj,Fayet:1979sa,Dimopoulos:1981zb} and only weakly
interacting, it escapes detection, leading to missing transverse momentum
($\mathbf{p}_{\mathrm{T}}^\mathrm{miss}$ and its magnitude \met) in the final state.
In this scenario, the LSP is a dark-matter candidate \cite{Goldberg:1983nd,Ellis:1983ew}.

Leptons may be produced in the cascade decays of squarks and gluinos via several mechanisms.
Here two scenarios that always produce leptons (electrons or muons) in \SF\, \OS\ (SFOS) pairs are considered:  
the leptonic decay of a $Z$ boson, $Z\to\ell^+\ell^-$,
and the decay $\tilde{\chi}_{2}^{0}\to\ell^+\ell^-\tilde{\chi}_{1}^{0}$,
which includes contributions from 
$\tilde{\chi}_{2}^{0} \to \tilde{\ell}^{\pm(*)}\ell^\mp \to  \ell^+\ell^-\tilde{\chi}_{1}^{0}$ and 
$\tilde{\chi}_{2}^{0} \to Z^*\tilde{\chi}_{1}^{0} \to  \ell^+\ell^-\tilde{\chi}_{1}^{0}$.
In models with generalised gauge-mediated (GGM) supersymmetry breaking with a gravitino LSP ($\tilde{G}$),
$Z$ bosons may be produced via the decay $\tilde{\chi}_{1}^{0} \to Z \tilde{G}$. 
$Z$ bosons may also result from the decay $\tilde{\chi}_{2}^{0} \to Z \tilde{\chi}_{1}^{0}$, 
although the GGM interpretation with the decay $\tilde{\chi}_{1}^{0} \to Z \tilde{G}$ 
is the focus of the $Z$ boson final-state channels studied here. 
The $\tilde{\chi}_2^0$ particle may itself be produced in the decays of the squarks or gluinos,
e.g.\ $\tilde{q} \to q \tilde{\chi}_2^0$ and $\tilde{g} \to q \bar{q} \tilde{\chi}_2^0$.

These two SFOS lepton production modes are distinguished by their distributions of dilepton invariant mass ($m_{\ell\ell}$).
The decay $Z\to\ell^+\ell^-$ leads to a peak in the $m_{\ell\ell}$ distribution around the $Z$ boson mass, while the decay
$\tilde{\chi}_{2}^{0}\to\ell^+\ell^-\tilde{\chi}_{1}^{0}$ leads to a rising distribution in 
$m_{\ell\ell}$ that terminates at a kinematic endpoint (``edge'')~\cite{Hinchliffe:1996iu},
because events with larger $\mll$ values would violate energy conservation in the decay of the $\tilde{\chi}_{2}^{0}$ particle.
In this paper, two searches are performed that separately target these two signatures.
A search for events with a SFOS lepton pair consistent with
originating from the decay of a $Z$ boson (on-$Z$ search) targets SUSY
models with $Z$ boson production. 
A search for events with a SFOS lepton pair inconsistent
with $Z$ boson decay (off-$Z$ search) targets the decay $\tilde{\chi}_{2}^{0}\to\ell^+\ell^-\tilde{\chi}_{1}^{0}$.

Previous searches for physics beyond the Standard Model (BSM) in the $Z+\mathrm{jets}+\met$ final state have been performed by the
CMS Collaboration~\cite{Chatrchyan:2012qka,CMS-edge}. 
Searches for a dilepton mass edge have also been performed
by the CMS Collaboration~\cite{Chatrchyan:2012te,CMS-edge}.
In the CMS analysis performed with $\sqrt{s}=8$~\TeV\ data reported in Ref.~\cite{CMS-edge}, an excess of events above the SM background
with a significance of 2.6 standard deviations was observed.

In this paper, the analysis is performed on the full 2012 ATLAS~\cite{Aad:2008zzm} dataset at a 
centre-of-mass energy of 8 \TeV, corresponding to an integrated luminosity of 20.3~\ifb.

\section{The ATLAS detector}
\label{sec:atlas}
ATLAS is a multi-purpose detector consisting of a tracking system, electromagnetic and hadronic calorimeters and a muon system. 
The tracking system comprises an inner detector (ID) immersed in a 2~T axial field supplied by the central solenoid magnet surrounding it. 
This sub-detector provides position and momentum measurements of charged particles over the pseudorapidity\footnote{ATLAS uses a right-handed coordinate system with its origin at the nominal interaction point (IP) in the centre of the detector and the $z$-axis along the beam pipe. The $x$-axis points from the IP to the centre of the LHC ring, and the $y$-axis points upward. Cylindrical coordinates $(r,\phi)$ are used in the transverse plane, $\phi$ being the azimuthal angle around the beam pipe. The pseudorapidity is defined in terms of the polar angle $\theta$ as $\eta=-\ln\tan(\theta/2)$. The opening angle $\Delta R$ in $\eta$--$\phi$ space is defined as $\Delta R=\sqrt{(\Delta\eta)^2+(\Delta\phi)^2}$.} range $|\eta|<2.5$. 
The electromagnetic calorimetry is provided by liquid argon (LAr) sampling calorimeters using lead absorbers, covering the central region ($|\eta|<3.2$). 
Hadronic calorimeters in the barrel region ($|\eta|<1.7$) use scintillator tiles with steel absorbers, 
while the pseudorapidity range $1.5<|\eta|<4.9$ is covered using LAr technology with copper or tungsten absorbers. 
The muon spectrometer (MS) has coverage up to $|\eta|<2.7$ and is built around the three superconducting toroid magnet systems. 
The MS uses various technologies to provide muon tracking and identification as well as dedicated muon triggering
for the range $|\eta|<2.4$. 

The trigger system~\cite{atlastrigger} comprises three levels. 
The first of these (L1) is a hardware-based trigger that uses only a subset of calorimeter and muon system information. 
Following this, both the second level (L2) and event filter (EF) triggers, constituting the software-based high-level trigger, 
include fully reconstructed event information to identify objects. 
At L2, only the regions of interest in $\eta$--$\phi$ identified at L1 are scrutinised, whereas complete event information from all detector sub-systems is available at the EF.

\section{Data and Monte Carlo samples}
\label{sec:data-selection}
The data used in this analysis were collected by ATLAS during 2012.  
Following requirements based on beam and detector conditions and data quality, the complete dataset corresponds to an integrated luminosity of 20.3~fb$^{-1}$,  
with an associated uncertainty of 2.8~\%. 
The uncertainty is derived following the same methodology as that detailed in Ref.~\cite{2011lumi}.

Dedicated high-transverse-momentum (\pt) single-lepton triggers are used in conjunction with the lower-\pt\ dilepton triggers to increase the trigger efficiency at high lepton \pt. 
The required leading-lepton \pt\ threshold is 25~\GeV, whereas the sub-leading lepton threshold can be as low as 10~\GeV, 
depending on the lepton \pt\ threshold of the trigger responsible for accepting the event. 
To provide an estimate of the efficiency for the lepton selections used in these analyses, trigger efficiencies are calculated using \ttbar\ Monte Carlo (MC) simulated event samples for leptons with $\pt>14$~\GeV. 
For events where both leptons are in the barrel (endcaps), the total efficiency of the trigger configuration for a two-lepton selection is approximately 96~\%, 88~\% and 80~\% (91~\%, 92~\% and 82~\%) for $ee$, $e\mu$ and $\mu\mu$ events, respectively. Although the searches in this paper probe only
same-flavour final states for evidence of SUSY, the $e\mu$ channel is used to select control samples in data for background estimation purposes.

Simulated event samples are used to validate the analysis techniques and aid in the estimation of SM backgrounds, as well as to provide predictions for BSM signal processes.  
The SM background samples~\cite{DYNNLO1,DYNNLO2,Lai:2010vv,ttbarxsec1,ttbarxsec2,Kidonakis:2010a,Kidonakis:2010b,Pumplin:2002vw,Campbell:2012,Lazopoulos:2008,diboson1,diboson2} used are listed in Table~\ref{tab:MC}, as are
the parton distribution function (PDF) set, underlying-event tune and cross-section calculation order in $\alpha_{\text{s}}$ used to normalise the event yields for these samples. 
Samples generated with {\sc MadGraph5} 1.3.28~\cite{Alwall:2011uj} are interfaced with {\sc Pythia} 6.426~\cite{PYTHIA} to simulate the parton shower. 
All samples generated using {\sc Powheg}~\cite{PowhegBOX1,PowhegBOX2,PowhegBOX3} use {\sc Pythia} to simulate the parton shower, 
with the exception of the diboson samples, which use {\sc Pythia8}~\cite{Pythia8}. 
{\sc Sherpa}~\cite{sherpa} simulated samples use {\sc Sherpa}'s own internal parton shower and fragmentation methods, as well as the {\sc Sherpa} default underlying-event tune~\cite{sherpa}. 
The standard ATLAS underlying-event tune, 
{\sc AUET2}~\cite{AUET2}, is used for all other samples with the exception of the {\sc Powheg+\-Pythia} samples, 
which use the {\sc Perugia2011C}~\cite{pythiaperugia} tune.

The signal models considered include simplified models and a generalised gauge-mediated super\-symmetry-breaking model. 
In the simplified models, squarks and gluinos are directly pair-pro\-duced, 
and these subsequently decay to the LSP via two sets of intermediate particles. 
The squarks and gluinos decay with equal probability to the next-to-lightest neutralino or the lightest chargino,
where the neutralino and chargino are mass-de\-gen\-er\-ate and have masses taken to be the average of the squark or gluino mass and the LSP mass. 
The intermediate chargino or neutralino then decays via sleptons (or sneutrinos) to two leptons of the same flavour and the lightest neutralino, 
which is assumed to be the LSP in these models. 
Here, the sleptons and sneutrinos are mass-degenerate and have masses taken to be the average of the chargino or neutralino and LSP masses. 
An example of one such process, 
$pp\to\tilde{g}\tilde{g}\to(q\bar{q}\tilde{\chi}_2^0)(q\bar{q}\tilde{\chi}_1^\pm),\- \tilde{\chi}_2^0\to\ell^+\ell^-\tilde{\chi}_1^0,$ $\tilde{\chi}_1^\pm\to\ell^\pm\nu\tilde{\chi}_1^0$ 
is illustrated on the left in Fig.~\ref{fig:models}, where $\ell=e,\mu,\tau$ with equal branching fractions for each lepton flavour.
The dilepton mass distribution for leptons produced from the $\tilde{\chi}_2^0$ in these models is a rising distribution that 
terminates at a kinematic endpoint, 
whose value is given by $m_{\mathrm{max}}$ $\approx m(\tilde{\chi}_2^0)-m(\tilde{\chi}_1^0) = 1/2(m(\tilde{g}/\tilde{q})-m(\tilde{\chi}_1^0))$.
Therefore, signal models with small values of $\Delta m = m(\tilde{g}/\tilde{q})-m(\tilde{\chi}_1^0)$ produce events with small
dilepton masses; those with large $\Delta m$ produce events with large dilepton mass.

For the model involving squark pair production, the left-handed partners of the $u$, $d$, $c$ and $s$ quarks have the same mass.
The right-handed squarks and the partners of the $b$ and $t$ quarks are decoupled. 
For the gluino-pair model, an effective three-body decay for $\tilde{g}\to q \bar{q} \tilde{\chi}_1^0$
is used, with equal branching fractions for $q=u,d,c,s$.
Exclusion limits on these models are set based on the squark or gluino mass and the LSP mass,
with all sparticles not directly involved in the considered decay chains effectively being decoupled. 

In the general gauge mediation models, the gravitino is the LSP and the next-to-lightest SUSY particle (NLSP) is a higgsino-like neutralino. 
The higgsino mass parameter, $\mu$, and the gluino mass are free parameters. 
The U(1) and SU(2) gaugino mass parameters, $M_1$ and $M_2$, are fixed to be 1~TeV, 
and the masses of all other sparticles are set at $\sim 1.5$~TeV. 
In addition, $\mu$ is set to be positive to make $\tilde{\chi}^0_1 \rightarrow Z \tilde{G}$ the dominant NLSP decay.
The branching fraction for $\tilde{\chi}^0_1 \rightarrow Z \tilde{G}$ varies with $\tan\beta$, the ratio of the vacuum expectation value for the two Higgs doublets, and so two different values of $\tan\beta$ are used. 
At $\tan\beta=1.5$, the branching fraction for $\tilde{\chi}^0_1 \rightarrow Z \tilde{G}$ is large 
(about 97~\%)~\cite{Meade:2009qv}, 
whereas setting $\tan\beta=30$ results in a considerable contribution (up to 40~\%) from $\tilde{\chi}^0_1 \rightarrow h \tilde{G}$.
In these models, $h$ is the lightest CP-even SUSY Higgs boson, with $m_h=126$~GeV and SM-like branching fractions.
The dominant SUSY-particle production mode in these scenarios is the strong production of gluino pairs, which subsequently decay to the LSP via several intermediate particles. 
An example decay mode is shown in the diagram on the right in Fig.~\ref{fig:models}. 
The gravitino mass is set to be sufficiently small such that the NLSP decays are prompt. 
The decay length $c\tau_{\text{NLSP}}$ (where $\tau_{\text{NLSP}}$ is the lifetime of the NLSP) can vary depending on $\mu$, and is longest at $\mu=120$~GeV, where it is $2$~mm, decreasing to $c\tau_{\text{NLSP}}<0.1$~mm
for $\mu \geq 150$~\GeV. The finite NLSP lifetime is taken into account in the MC signal acceptance and efficiency determination.

All simplified models are produced using {\sc MadGraph5} 1.3.33 with the \texttt{CTEQ6L1} PDF set, interfaced with {\sc Pythia} 6.426.
The scale parameter for MLM matching~\cite{Mangano:2006rw} is set at a quarter of the mass of the lightest strongly produced sparticle in the matrix element. 
The SUSY mass spectra, gluino branching fractions and the gluino decay width for the GGM scenarios are calculated using 
{\sc Suspect~2.41}~\cite{Djouadi:2002ze} and {\sc Sdecay~1.3}~\cite{Muhlleitner:2003vg}.
The GGM signal samples are generated using {\sc Pythia} 6.423 with the \texttt{MRST2007 LO$^{\ast}$}~\cite{Sherstnev:2007nd} PDF set.
The underlying event is modelled using the {\sc AUET2} tune for all signal samples.
Signals are normalised to cross sections calculated at next-to-leading order (NLO) in $\alpha_s$, 
including the resummation of soft gluon emission at next-to-leading-logarithmic accuracy (NLO+NLL)~\cite{Beenakker:1996ch,Kulesza:2008jb,Kulesza:2009kq,Beenakker:2009ha,Beenakker:2011fu}. 

A full ATLAS detector simulation~\cite{:2010wqa} using {\sc GEANT4}~\cite{Agostinelli:2002hh} is performed for most of the SM background MC samples. 
The signal and remaining SM MC samples use a fast simulation~\cite{atlfast}, which employs
a combination of a parameterisation of the response of the ATLAS electromagnetic and hadronic calorimeters and {\sc GEANT4}. 
To simulate the effect of multiple $pp$ interactions occurring during the same (in-time) or a nearby (out-of-time) bunch-crossing, called pile-up, 
minimum-bias interactions are generated and overlaid on top of the hard-scattering process. 
These are produced using {\sc Pythia8} with the A2 tune~\cite{a2tune}.
MC-to-data corrections are made to simulated samples to account for small differences in lepton identification and reconstruction efficiencies, and the efficiency and misidentification rate associated with the algorithm used to distinguish jets containing $b$-hadrons.
\begin{figure*}[htb]
\centering
\includegraphics[width=0.3\textwidth]{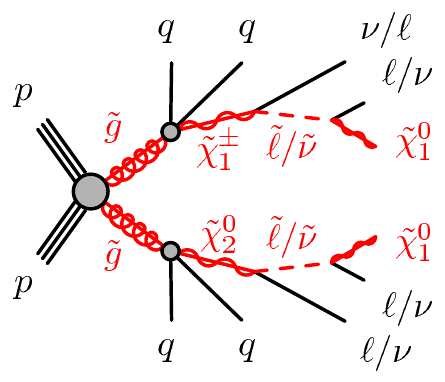}
\qquad
\includegraphics[width=0.3\textwidth]{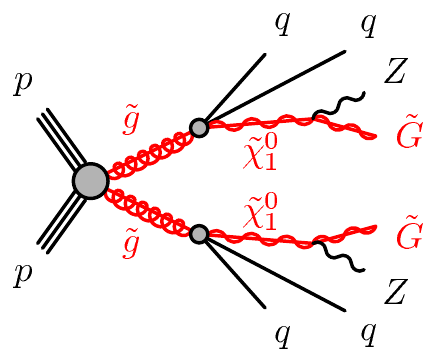}
\caption{Decay topologies for example signal processes. 
A simplified model involving gluino pair production, with the gluinos following two-step decays via sleptons to neutralino LSPs is shown on the left. 
The diagram on the right shows a GGM decay mode, where gluinos decay via neutralinos to gravitino LSPs.}
\label{fig:models}
\end{figure*}

\begin{table*}[ht]
\begin{center}
\scriptsize
\begin{tabular}{l c c c c c }
\hline
Physics process &  Generator  & Parton & Cross section & Tune & PDF set\\
                &             & Shower &              &      & \\
\noalign{\smallskip}\hline\noalign{\smallskip}
$Z/\gamma^{*}(\rightarrow \ell \ell)$ + jets   & \sherpa\ 1.4.1           & \sherpa\ 1.4.1  &NNLO \cite{DYNNLO1,DYNNLO2}       & \sherpa\ default     &NLO CT10~\cite{Lai:2010vv}\\
$t\bar{t}$                                     & {\sc Powheg-Box} r2129   & {\sc Pythia} 6.426 & NNLO+NNLL \cite{ttbarxsec1,ttbarxsec2}          &\sc{Perugia2011C}     &NLO CT10\\
Single-top ($Wt$)                              & {\sc Powheg-Box} r1556   & {\sc Pythia} 6.426 & Approx. NNLO \cite{Kidonakis:2010a,Kidonakis:2010b}& \sc{Perugia2011C}    &NLO CT10\\ 
$t+Z$                                          & \madgraph5 1.3.28        & {\sc Pythia} 6.426 & LO  & AUET2 &CTEQ6L1~\cite{Pumplin:2002vw}\\
$t\bar{t}+W$ and $t\bar{t}+Z$                  & \madgraph5 1.3.28        & {\sc Pythia} 6.426 & NLO \cite{Campbell:2012,Lazopoulos:2008} & AUET2 & CTEQ6L1\\
$t\bar{t}+WW$                                  & \madgraph5 1.3.28        & {\sc Pythia} 8.165 & LO  & AUET2 &CTEQ6L1\\
$WW$,                      & \multirow{2}{*}{{\sc powheg-box} r1508} & \multirow{2}{*}{{\sc Pythia} 8.163} & \multirow{2}{*}{NLO \cite{diboson1,diboson2}} & \multirow{2}{*}{AUET2} & \multirow{2}{*}{NLO CT10} \\
$WZ$ and $ZZ$ &&& \\ 
\noalign{\smallskip}\hline\noalign{\smallskip}
\end{tabular}
\caption{Simulated background event samples used in this analysis with the corresponding generator, cross-section order in $\alpha_{\text{s}}$ used to
normalise the event yield, underlying-event tune and PDF set. 
}
\label{tab:MC}
\end{center}
\end{table*}

\section{Physics object identification and selection}
\label{sec:objects}
Electron candidates are reconstructed using energy clusters in the electromagnetic calorimeter matched to ID tracks.
Electrons used in this analysis are assigned either ``baseline'' or ``signal'' status. 
Baseline electrons are required to have transverse energy $E_{\text{T}}>10$~GeV, satisfy the ``medium'' criteria described in Ref.~\cite{electronref} and reside within $|\eta|<2.47$ and not in the range $1.37<|\eta|<1.52$. 
Signal electrons are further required to be consistent with the primary vertex and isolated with respect to other objects in the event, 
with a \pt-dependent isolation requirement. 
The primary vertex is defined as the reconstructed vertex with the highest $\sum p_{\text{T}}^2$, 
where the summation includes all particle tracks with $\pt>400$~\MeV\ associated with a given reconstructed vertex.
Signal electrons with $E_{\text{T}}<25$~GeV must additionally satisfy the more stringent shower shape, 
track quality and matching requirements of the ``tight'' selection criteria in Ref.~\cite{electronref}.
For electrons with $E_{\text{T}}<25$~GeV ($\geq 25$~GeV), the sum of the transverse momenta of all charged-particle tracks with $\pt>400$~\MeV\ associated with the primary vertex, 
excluding the electron track, 
within $\Delta R = 0.3$ ($0.2$) surrounding the electron must be less than 16~\% (10~\%) of the electron \pt.
Electrons with $E_{\text{T}}<25$~GeV must reside within a distance $|z_0\sin\theta| < 0.4$~mm of the primary vertex along the direction of the beamline\footnote{The distance of closest approach between a particle object and the primary vertex in the longitudinal (transverse) plane is denoted by $z_0$ ($d_0$).}.
The significance of the transverse-plane distance of closest approach of the electron to the primary vertex must be $|d_0/\sigma_{d_0}|<5$.
For electrons with $E_{\text{T}} \geq 25$~\GeV, $|z_0|$ is required to be $<2$~mm and $|d_0|<1$~mm.

Baseline muons are reconstructed from either ID tracks matched to a muon segment in the muon spectrometer or combined tracks formed both from the ID and muon spectrometer~\cite{muonref}. 
They are required to be of good quality, as described in Ref.~\cite{ATLAS:2011ad}, and to satisfy $p_{\text{T}}>10$~GeV and $|\eta|<2.4$. 
Signal muons are further required to be isolated, with the scalar sum of the $\pt$ of charged particle tracks associated with the primary vertex, excluding the muon track, 
within a cone of size $\Delta R<0.3$ surrounding the muon being less than 12~\% of the muon \pt\ for muons with $p_{\text{T}}<25$~GeV. 
For muons with $p_{\text{T}} \geq 25$~GeV, the scalar sum of the $\pt$ of charged-particle tracks associated with the primary vertex, excluding the muon track, 
within $\Delta R<0.2$ surrounding the muon must be less than 1.8~GeV. 
Signal muons with $p_{\text{T}}<25$~GeV must also have $|z_0\sin\theta| \leq 1$~mm and $|d_0/\sigma_{d_0}|<3$. 
For the leptons selected by this analysis, the $d_0$ requirement is typically several times less restrictive than the $|d_0/\sigma_{d_0}|$ requirement.

Jets are reconstructed from topological clusters in the calorimeter using the anti-$k_{t}$ algorithm~\cite{Cacciari:2008gp} with a distance parameter of 0.4.
Each cluster is categorised as being electromagnetic or ha\-dron\-ic in origin according to its shape~\cite{CSCbook}, so as to account for the differing calorimeter response for electrons/photons and hadrons. 
A cluster-level correction is then applied to electromagnetic and had\-ron\-ic energy deposits using correction factors derived from both MC simulation and data. 
Jets are corrected for expected pile-up contributions~\cite{jetPU} and further calibrated to account for the calorimeter response with respect to the true jet energy~\cite{JES,JES2}. 
A small residual correction is applied to the jets in data to account for differences between response in data and MC simulation.
Baseline jets are selected with $p_{\text{T}}>20$~GeV. 
Events in which these jets do not pass specific jet quality requirements are rejected so as to remove events affected by detector noise and non-collision backgrounds~\cite{Aad:2013zwa,ATLAS-CONF-2012-020}. 
Signal jets are required to satisfy $p_{\text{T}}>35$~GeV and $|\eta|<2.5$. 
To reduce the impact of jets from pileup to a negligible level, 
jets with $p_{\text{T}}<50$~GeV within $|\eta|<2.4$ are further required to have a jet vertex fraction $|\mathrm{JVF}|> 0.25$. 
Here the JVF is the \pt-weighted fraction of tracks matched to the jet that are associated with the primary vertex~\cite{ATLAS:2010069}, with jets without any associated tracks being assigned $\text{JVF}=-1$. 

The MV1 neural network algorithm~\cite{MV1} identifies jets containing $b$-hadrons using the impact parameters of associated tracks and any reconstructed secondary vertices.
For this analysis, the working point corresponding to a 60~\% efficiency for tagging $b$-jets in simulated \ttbar\ events is used, resulting in a charm quark rejection factor of approximately 8 and a light quark/gluon jet rejection factor of about $600$. 
To ensure that each physics object is counted only once, an overlap removal procedure is applied.
If any two baseline electrons reside within $\Delta R=0.05$ of one another, the electron with lower $E_{\text{T}}$ is discarded. 
Following this, any baseline jets within $\Delta R=0.2$ of a baseline electron are removed. 
After this, any baseline electron or muon residing within $\Delta R=0.4$ of a remaining baseline jet is discarded. 
Finally, to remove electrons originating from muon bremsstrahlung, any baseline electron within $\Delta R=0.01$ of any remaining baseline muon is removed from the event. 

The \met\ is defined as the magnitude of the vector sum of the transverse momenta of all photons, electrons, muons, baseline jets and an additional ``soft term''~\cite{met_conf}. 
The soft term includes clusters of energy in the calorimeter not associated with any calibrated object, 
which are corrected for material effects and the non-compensating nature of the calorimeter.
Reconstructed photons used in the \met\ calculation are required to satisfy the ``tight'' requirements of Ref.~\cite{ATLAS:2012ana}.

\section{Event selection}
\label{sec-selection}
Events selected for this analysis must have at least five tracks with $p_{\text{T}}>400$~MeV associated with the primary vertex.
Any event containing a baseline muon with $|z_0 \sin\theta|>0.2$~mm or $|d_0|>1$~mm is rejected, to remove cosmic-ray events. 
To reject events with fake \met, those containing poorly measured muon candidates, characterised by large uncertainties on the measured momentum,
are also removed. 
If the invariant mass of the two leading leptons in the event is less than $15$~GeV the event is vetoed to suppress low-mass particle decays and Drell--Yan production. 

Events are required to contain at least two signal leptons (electrons or muons). If more than two signal leptons are
present, the two with the largest values of \pt\ are selected.
These leptons must pass one of the leptonic triggers, with the two leading leptons being matched, within $\Delta R<0.15$, 
to the online trigger objects that triggered the event in the case of the dilepton triggers. 
For events selected by a single-lepton trigger, one of the two leading leptons must be matched to the online trigger object in the same way.
The leading lepton in the event must have $p_{\text{T}}>25$~GeV and the sub-leading lepton is required to have $p_{\text{T}}>10$--14~\GeV, 
depending on the \pt\ theshold of the trigger selecting the event. 
For the off-$Z$ analysis, the sub-leading lepton $p_{\text{T}}$ threshold is increased to $20$~GeV. 
This is done to improve the accuracy of the method for estimating flavour-symmetric backgrounds, discussed in Sect.~\ref{sec:fl-sym}, in events with small dilepton invariant mass. 
For the same reason, the $m_{\ell\ell}$ threshold is also raised to $20$~GeV in this search channel. 
The two leading leptons must be oppositely charged, with the signal selection requiring that these be same-flavour (SF) lepton pairs. 
The different-flavour (DF) channel is also exploited to estimate certain back\-grounds, such as that due to \ttbar\ production.
All events are further required to contain at least two signal jets, since this is the minimum expected jet multiplicity for the
signal models considered in this analysis. 

Three types of region are used in the analysis. 
Control regions (CRs) are used to constrain the SM backgrounds. 
These backgrounds, estimated in the CRs, are first extrapolated to the validation regions (VRs) as a cross check and then to the signal regions (SRs), 
where an excess over the expected background is searched for.

GGM scenarios are the target of the on-$Z$ search, where the $\tilde{G}$ from $\tilde{\chi}^0_1 \rightarrow (Z/h)+ \tilde{G}$ decays is expected to result in \met. 
The $Z$ boson mass window used for this search is $81<m_{\ell\ell}<101$~GeV. 
To isolate GGM signals with high gluino mass and high jet activity the on-$Z$ SR, SR-Z, is defined using requirements 
on \met\ and $H_{\text{T}} = \sum_i p_{\text{T}}^{\text{jet},i} + p_{\text{T}}^{\text{lepton},1} + p_{\text{T}}^{\text{lepton},2}$, 
where \HT\ includes all signal jets and the two leading leptons. 
Since $b$-jets are often, but not always, expected in GGM decay chains, no requirement is placed on $b$-tagged jet multiplicity. 
Dedicated CRs are defined in order to estimate the contribution of various SM backgrounds to the SR. 
These regions are constructed with selection criteria similar to those of the SR, differing either in mll or MET ranges, or in lepton flavour requirements.
A comprehensive discussion of the various methods used to perform these estimates follows in Sect.~\ref{sec:bgEst}. 
For the SR and CRs, detailed in Table~\ref{tab:regions-z}, 
a further requirement on the azimuthal opening angle between each of the leading two jets and the \met\ ($\Delta\phi(\text{jet}_{1,2},\met)$)  
is introduced to reject events with jet mismeasurements contributing to large fake \met. 
This requirement is applied in the SR and two CRs used in the on-$Z$ search, 
all of which have high \met\ and $H_{\text{T}}$ thresholds, at 225~GeV and 600~GeV, respectively. 
Additional VRs are defined at lower \met\ and $H_{\text{T}}$ to cross-check the SM background estimation methods. 
These are also sumarised in Table~\ref{tab:regions-z}. 
The SR selection results in an acceptance times efficiency of 2--4~\%, including leptonic $Z$ branching fractions, for GGM signal models with $\mu>400$~\GeV.

In the off-$Z$ analysis, a search is performed in the $Z$ boson sidebands.
The $Z$ boson mass window vetoed here is larger than that selected in the on-$Z$ analysis ($m_{\ell\ell} \notin [80,110]$~GeV) to maximise $Z$ boson rejection. 
An asymmetric window is chosen to improve the suppression of boosted $Z\to\mu\mu$ events with muons whose momenta are overestimated,
leading to large \met. 
In this search, four SRs are defined by requirements on jet multiplicity, $b$-tagged jet multiplicity, and \met.
The SR requirements are optimised for the simplified models
of pair production of squarks (requiring at least two jets) and gluinos (requiring at least four jets) discussed in Sect.~\ref{sec:data-selection}.
Two SRs with a $b$-veto provide the best sensitivity in the simplified models considered here, since the signal $b$-jet content
is lower than that of the dominant \ttbar\ background.
Orthogonal SRs with a requirement of at least one $b$-tagged jet target other signal models not explicitly considered here,
such as those with bottom squarks that are lighter than the other squark flavours. 
For these four SRs, the requirement $\met>200$~\GeV\ is imposed. 
In addition, one signal region with requirements similar to those used in the CMS search~\cite{CMS-edge} is defined (SR-loose). 
These SRs and their respective CRs, which have the same jet and \met\ requirements, but select different $m_{\ell\ell}$ ranges or lepton flavour combinations, 
are defined in Table~\ref{tab:regions-offz}.

The most sensitive off-$Z$ SR for the squark-pair (gluino-pair) model is SR-2j-bveto (SR-4j-b\-veto).
Because the value of the $\mll$ kinematic endpoint depends on unknown model parameters,
the analysis is performed over multiple $\mll$ ranges for these two SRs.
The dilepton mass windows considered for the SR-2j-bveto and SR-4j-bveto regions are
presented in Sect.~\ref{sec:interpretation}.
For the combined $ee+\mu\mu$ channels, 
the typical signal acceptance times efficiency values for the squark-pair (gluino-pair) model in the SR-2j-bveto (SR-4j-bveto) region are 
0.1--10~\% (0.1--8~\%) over the full dilepton mass range.

The on-$Z$ and off-$Z$ searches are optimised for different signal models and as such are defined with orthogonal SRs. 
Given the different signatures probed, there are cases where the CR of one search may overlap with the SR of the other.
Data events that fall in the off-$Z$ SRs can comprise up to 60~\% of the top CR for the on-$Z$ analysis (CRT, defined in Table~\ref{tab:regions-z}). 
Data events in SR-Z comprise up to 36~\% of the events in the CRs with $80<\mll<110$~\GeV\ that are used to normalise the \zjets\ background in the
off-$Z$ analysis, but the potential impact on the background prediction is small because the \zjets\ contribution is a small fraction of
the total background.
For the following analysis, each search assumes only signal contamination from the specific signal model they are probing. 

\begin{table*}[htbp]
\begin{center}
\footnotesize
 \begin{tabular*}{\textwidth}{@{\extracolsep{\fill}}lccccccccc}
   \noalign{\smallskip}\hline\noalign{\smallskip}
   {\bf On-$Z$} &  {\bf \met}   &  {\bf $\HT$}  &  {\bf $n_{\text{jets}}$}  & {\bf $m_{\ell\ell} $}  &  {\bf SF/DF}  &  {\bf \met~sig.}        &  {\bf $f_{\text{ST}}$} & {\bf $\Delta\phi(\text{jet}_{12},\met)$ } \\
   {\bf Region}               &  {\bf [\GeV]} &  {\bf [\GeV]} &        &          {\bf [\GeV]}           &         &  {\bf $[\sqrt{\GeV}]$}  &   &   \\
   \noalign{\smallskip}\hline\noalign{\smallskip}
   \multicolumn{2}{l}{Signal regions} &&&&&&& \\
   \noalign{\smallskip}\hline\noalign{\smallskip}
   SR-Z  &  $> 225$  &  $> 600$  &  $\geq 2$  & $81 < m_{\ell\ell} < 101$  &  SF  &  -  &  -  & $>0.4$ \\
   \noalign{\smallskip}\hline\noalign{\smallskip}
   \multicolumn{2}{l}{Control regions} &&&&&&& \\
   \noalign{\smallskip}\hline\noalign{\smallskip}
   Seed region  &        -  &  $> 600$  &  $\geq 2$   &  $81 < m_{\ell\ell} < 101$  &  SF  &  $< 0.9$  &  $< 0.6$ & - \\
   CRe$\mu$     &  $> 225$  &  $> 600$  &  $\geq 2$   &  $81 < m_{\ell\ell} < 101$  &  DF  &  -  &  -  & $>0.4$ \\
   CRT          &  $> 225$  &  $> 600$  &  $\geq 2$   &  $m_{\ell\ell} \notin [81,101]$  &  SF  &  -  &  - & $>0.4$  \\
   \noalign{\smallskip}\hline\noalign{\smallskip}
   \multicolumn{2}{l}{Validation regions} &&&&&&& \\
   \noalign{\smallskip}\hline\noalign{\smallskip}
   VRZ  &   $< 150$  &  $> 600$   &  $\geq 2$  &    $81 < m_{\ell\ell} < 101$  &  SF  &  -  &  - & - \\
   VRT  &  150--225  &  $> 500 $  &  $\geq 2$  &    $m_{\ell\ell} \notin [81,101]$  &  SF  &  -  &  -  & $>0.4$\\
   VRTZ &  150--225  &  $> 500 $  &  $\geq 2$  &    $81 < m_{\ell\ell} < 101$  &  SF  &  -  &  -  & $>0.4$ \\
   \noalign{\smallskip}\hline\noalign{\smallskip}
\end{tabular*}   
 \caption{Overview of all signal, control and validation regions used in the on-$Z$ search.
 More details are given in the text. 
 The \met~significance and the soft-term fraction $f_{\text{ST}}$ needed in the seed regions for the jet smearing method
 are defined in Sect.~\ref{sec:zjets}. The flavour combination of the dilepton pair is denoted as either ``SF'' for same-flavour or ``DF'' for different flavour.}
 \label{tab:regions-z}
\end{center}
\end{table*}

\begin{table*}[htbp]
\begin{center}
\footnotesize
 \begin{tabular*}{\textwidth}{@{\extracolsep{\fill}}lcccccc}    
   \noalign{\smallskip}\hline\noalign{\smallskip}
   {\bf Off-$Z$}   &  {\bf \met}   &  {\bf $n_{\text{jets}}$}  & {\bf $n_{\text{b-jets}}$} & {\bf $m_{\ell\ell} $}  &  {\bf SF/DF}  \\
   {\bf Region}        &  {\bf [\GeV]} &         &             &   {\bf [\GeV]}      &          \\
   \noalign{\smallskip}\hline\noalign{\smallskip}
   \multicolumn{2}{l}{Signal regions} &&&&& \\
   \noalign{\smallskip}\hline\noalign{\smallskip}
   SR-2j-bveto  &  $> 200$     &  $\geq 2$  & = 0     & $m_{\ell\ell} \notin [80,110]$  &  SF      \\
   SR-2j-btag   &  $> 200$     &  $\geq 2$  & $\geq1$ & $m_{\ell\ell} \notin [80,110]$  &  SF      \\
   SR-4j-bveto  &  $> 200$     &  $\geq 4$  & = 0     & $m_{\ell\ell} \notin [80,110]$  &  SF      \\
   SR-4j-btag   &  $> 200$     &  $\geq 4$  & $\geq1$ & $m_{\ell\ell} \notin [80,110]$  &  SF      \\
   SR-loose      &  $> (150,100)$     &  $(2, \geq 3)$  &  -      & $m_{\ell\ell} \notin [80,110]$  &  SF      \\
   \noalign{\smallskip}\hline\noalign{\smallskip}
   \multicolumn{2}{l}{Control regions} &&&&& \\
   \noalign{\smallskip}\hline\noalign{\smallskip}
   CRZ-2j-bveto  &  $> 200$     &  $\geq 2$  & = 0     & $80 < m_{\ell\ell} < 110$  &  SF     \\
   CRZ-2j-btag   &  $> 200$     &  $\geq 2$  & $\geq1$ & $80 < m_{\ell\ell} < 110$  &  SF     \\
   CRZ-4j-bveto  &  $> 200$     &  $\geq 4$  & = 0     & $80 < m_{\ell\ell} < 110$  &  SF     \\
   CRZ-4j-btag   &  $> 200$     &  $\geq 4$  & $\geq1$ & $80 < m_{\ell\ell} < 110$  &  SF     \\
   CRZ-loose     &  $> (150,100)$     &  $(2,\geq 3)$  &  -      & $80 < m_{\ell\ell} < 110$  &  SF     \\
   \noalign{\smallskip}\hline\noalign{\smallskip}
   CRT-2j-bveto  &  $> 200$     &  $\geq 2$  & = 0     & $m_{\ell\ell} \notin [80,110]$  &  DF     \\
   CRT-2j-btag   &  $> 200$     &  $\geq 2$  & $\geq1$ & $m_{\ell\ell} \notin [80,110]$  &  DF     \\
   CRT-4j-bveto  &  $> 200$     &  $\geq 4$  & = 0     & $m_{\ell\ell} \notin [80,110]$  &  DF     \\
   CRT-4j-btag   &  $> 200$     &  $\geq 4$  & $\geq1$ & $m_{\ell\ell} \notin [80,110]$  &  DF     \\
   CRT-loose     &  $> (150,100)$     &  $(2,\geq 3)$  &  -      & $m_{\ell\ell} \notin [80,110]$  &  DF     \\
   \noalign{\smallskip}\hline\noalign{\smallskip}
   \multicolumn{2}{l}{Validation regions} &&&&& \\
   \noalign{\smallskip}\hline\noalign{\smallskip}
   VR-offZ   &  100-150  & = 2  & - & $m_{\ell\ell} \notin [80,110]$  &  SF  \\
   \noalign{\smallskip}\hline\noalign{\smallskip}
 \end{tabular*}
 \caption{Overview of all signal, control and validation regions used in the off-$Z$
 analysis. For SR-loose, events with two jets (at least three jets) are required to satisfy \met\ $>$ 150 (100) \GeV.
Further details are the same as in Table~\ref{tab:regions-z}.}
 \label{tab:regions-offz}
\end{center}
\end{table*}

\section{Background estimation}
\label{sec:bgEst}
The dominant background processes in the signal regions, and those that are expected to be most difficult to model using MC simulation, 
are estimated using data-driven techniques.
With SRs defined at large \met, any contribution from \dyjets\ will be a consequence of artificially high \met\ in the event due to, for example, jet mis\-meas\-ure\-ments. 
This background must be carefully estimated, particularly in the on-$Z$ search, since the peaking \dyjets\ background
can mimic the signal. 
This background is expected to constitute, in general, less than 10~\% of the total background in the off-$Z$ SRs and have a negligible contribution to SR-Z. 

In both the off-$Z$ and on-$Z$ signal regions, the dominant backgrounds come from 
so-called ``fla\-vour-sym\-met\-ric'' processes, 
where the dileptonic br\-anching fractions to $ee$, 
$\mu\mu$ and $e\mu$ have a 1:1:2 ratio such that the same-flavour contributions can be estimated using information from the different-flavour contribution. 
This group of backgrounds is dominated by \ttbar\ and also includes $WW$, single top ($Wt$) and 
$Z\rightarrow\tau\tau$ production, 
and makes up $\sim 60$~\% ($\sim 90$~\%) of the predicted background in the on-$Z$ (off-$Z$) SRs. 

Diboson backgrounds with real $Z$ boson production, while small in the off-$Z$ regions, 
contribute up to $25$~\% of the total background in the on-$Z$ regions. 
These backgrounds are estimated using MC simulation, 
as are ``rare top'' backgrounds, including $t\bar{t}+W(W)/Z$ (i.e. $t\bar{t}+W$, $t\bar{t}+Z$ and $t\bar{t}+WW$) and $t+Z$ processes.
All backgrounds that are estimated from MC simulation are subject to carefully 
assessed theoretical and experimental uncertainties. 

Other processes, including those that might be present due to mis-reconstructed jets entering as leptons, can contribute up to $10$~\% ($6$~\%) in the on-$Z$ (off-$Z$) SRs. 
The background estimation techniques followed in the on-$Z$ and off-$Z$ searches are similar,
with a few well-motivated exceptions.

\subsection{Estimation of the $Z/\gamma^{*}+$~jets background}
\label{sec:zjets}
\subsubsection{\dyjets\ background in the off-$Z$ search}

In the off-$Z$ signal regions, the background from \dyjets\ is due to off-shell $Z$ bosons and photons, or to on-shell $Z$ bosons
with lepton momenta that are mismeasured. 
The region with dilepton mass in the range $80<m_{\ell\ell}<110$ GeV is not considered as a search region. 
To estimate the contribution from \dyjets\ outside of this range, dilepton mass shape templates are derived from \dyjets\ MC events.
These shape templates are normalised to data in control regions with the same selection as the corresponding signal regions,
but with the requirement on $m_{\ell\ell}$ inverted to $80<m_{\ell\ell}<110$~GeV, to select a sample enriched 
in \dyjets\ events. 
These CRs are defined in Table~\ref{tab:regions-offz}.

\subsubsection{\dyjets\ background in the on-$Z$ search}
The assessment of the peaking background due to \dyjets\ in the on-$Z$ signal regions requires careful consideration. 
The events that populate the signal regions result from mismeasurements of physics objects where, 
for example, one of the final-state jets has its energy underestimated, resulting in an overestimate of the total \met\ in the event. 
Due to the difficulties of modelling instrumental \met\ in simulation, MC events are not relied upon alone for the estimation of the \dyjets\ background. 
A data-driven technique is used as the nominal method for estimating this background. 
This technique confirms the expectation from MC simulation that the $Z+\mathrm{jets}$ background is negligible in the SR. 

The primary method used to model the \dyjets\ background in SR-Z is the so-called ``jet smearing'' method, which is described in detail in Ref.~\cite{Aad:2012fqa}. 
This involves defining a region with \dyjets\ events containing well-measured jets (at low \met), known as the ``seed'' region. 
The jets in these events are then smeared using functions that describe the detector's jet \pt\ response and $\phi$ resolution as a function of jet \pt, 
creating a set of pseudo-data events. 
The jet-smearing method provides an estimate for the contribution from events containing both fake \met, from object mismeasurements, and real \met, 
from neutrinos in heavy-flavour quark decays, by using different response functions for light-flavour and $b$-tagged jets. 
The response function is measured by comparing gen\-er\-a\-tor-level jet \pt\ to reconstructed jet \pt\ in {\sc Py\-thia8} dijet MC events, generated using the CT10 NLO PDF set. 
This function is then tuned to data, based on a dijet balance analysis in which the \pt\ asymmetry is used to constrain the width of the Gaussian core. 
The non-Gaussian tails of the response function are corrected based on $\geq 3$-jet events in data, 
selected such that the \met\ in each event points either towards, or in the opposite direction to one of the jets. 
This ensures that one of the jets is clearly associated with the \met, and the jet response can then be described in terms of the \met\ and reconstructed jet \pt. 
This procedure results in a good estimate of the overall jet response. 

In order to calculate the \met\ distribution of the pseudo-data, the \met\ is recalculated using the new (smeared) jet \pt\ and $\phi$. 
The distribution of pseudo-data events is then normalised to data in the low-\met\ region (10 $<$ \met\ $<$ 50 GeV) of a validation region,
denoted VRZ, after the requirement of $\Delta\phi(\text{jet}_{1,2},\met)>0.4$. 
This is defined in Table~\ref{tab:regions-z} and is designed to be representative of the signal region but at lower \met, where the contamination for relevant GGM signal models is expected to be less than 1~\%. 

The seed region must contain events with to\-pol\-o\-gies similar to those expected in the signal region. 
To ensure that this is the case, the \HT\ and jet multiplicity requirements applied to the seed region remain the same as in the signal region, 
while the \met\ threshold of $225$ GeV is removed, as shown in Table~\ref{tab:regions-z}. Although the seed events should have little to no \met, 
enforcing a direct upper limit on \met\ can introduce a bias in the jet \pt\ distribution in the seed region compared with the signal region. 
To avoid this, a requirement on the \met\ significance, defined as:

\begin{equation}
\met\ \text{sig.} = \frac{\met}{\sqrt{\sum E_{\text{T}}^{\text{jet}} + \sum E_{\text{T}}^{\text{soft}}}} ,
\end{equation}

\noindent is used in the seed region. 
Here $\sum E_{\text{T}}^{\text{jet}}$ and $\sum E_{\text{T}}^{\text{soft}}$ are the summed $E_{\text{T}}$ from the baseline jets and the low-energy calorimeter deposits not associated with final-state physics objects, respectively.
Placing a requirement on this variable does not produce a shape difference between jet \pt\ distributions in the seed and signal regions, 
while effectively selecting well-balanced \dyjets\ events in the seed region. 
This requirement is also found to result in no event overlap between the seed region and SR-Z.

In the seed region an additional requirement is placed on the soft-term fraction, $f_\text{ST}$, 
defined as the fraction of the total \met\ in an event originating from calorimeter energy deposits not associated with a calibrated lepton or jet ($f_\text{ST} = \sum E_{\text{T}}^{\text{miss, Soft}} / \met$), 
to select events with small $f_\text{ST}$. 
This is useful because events with large values of fake \met\ tend to have low soft-term fractions ($f_\text{ST}<0.6$). 

The requirements on the \met\ significance and $f_\text{ST}$ are initially optimised by applying the jet smearing method to \dyjets\ MC events and testing the agreement in the \met\ spectrum between direct and smeared MC events in the VRZ.
This closure test is performed using the response function derived from MC simulation. 

The \dyjets\ background predominantly comes from events where a single jet is grossly mismeas\-ured, since the mismeasurement of additional jets is unlikely, 
and can lead to smearing that reduces the total \met.
The requirement on the opening angle in $\phi$ between either of the leading two jets and the \met, $\Delta\phi(\text{jet}_{1,2},\met)>0.4$, strongly suppresses this background.
The estimate of the \dyjets\ background is performed both with and without this requirement, in order to aid in the interpretation of the results in the SR, as described in Sect.~\ref{sec:results}.
The optimisation of the \met\ significance and $f_\text{ST}$ requirements are performed separately with and without the requirement, 
although the optimal values are not found to differ significantly.

The jet smearing method using the data-cor\-rec\-ted jet response function is validated in VRZ, 
comparing smeared pseudo-data to data. 
The resulting \met\ distributions show agreement within uncertainties assessed based on varying the response function and the \met\ 
significance requirement in the seed region.  
The \met\ distribution in VRZ, with the additional requirement $\Delta\phi(\text{jet}_{1,2},$ $\met)>0.4$, is shown in Fig.~\ref{fig:VRZ}. 
Here the \met\ range extends only up to 100~\GeV, since \ttbar\ events begin to dominate at higher \met\ values.
The pseudo-data to data agreement in VRZ motivates the final determination of the \met\ significance requirement used for the seed region ($\met\ \text{sig.}<0.9$). 
Backgrounds containing real \met, including \ttbar\ and diboson production, are taken from MC simulation for this check. 
The chosen values are detailed in Table~\ref{tab:regions-z} with a summary of the kinematic requirements imposed on the seed and $Z$ validation region.
Extrapolating the jet smearing estimate to the signal regions yields the results detailed in Table~\ref{tab:results-z}. 
The data-driven estimate is compatible with the MC expectation that the $Z+\mathrm{jets}$ background contributes significantly less than one event in SR-Z. 

\begin{figure*}
\centering
\includegraphics[width=0.49\textwidth]{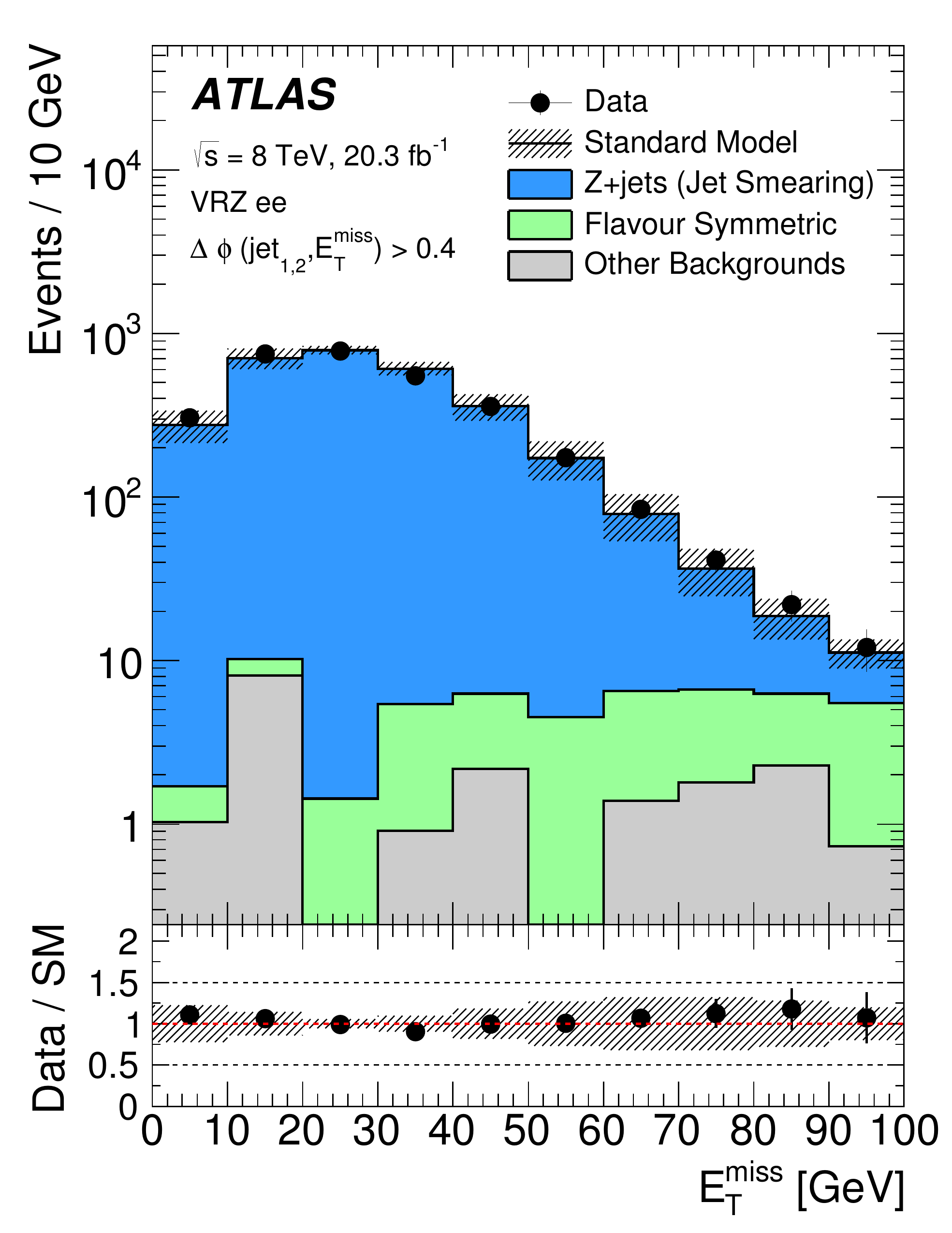}
\includegraphics[width=0.49\textwidth]{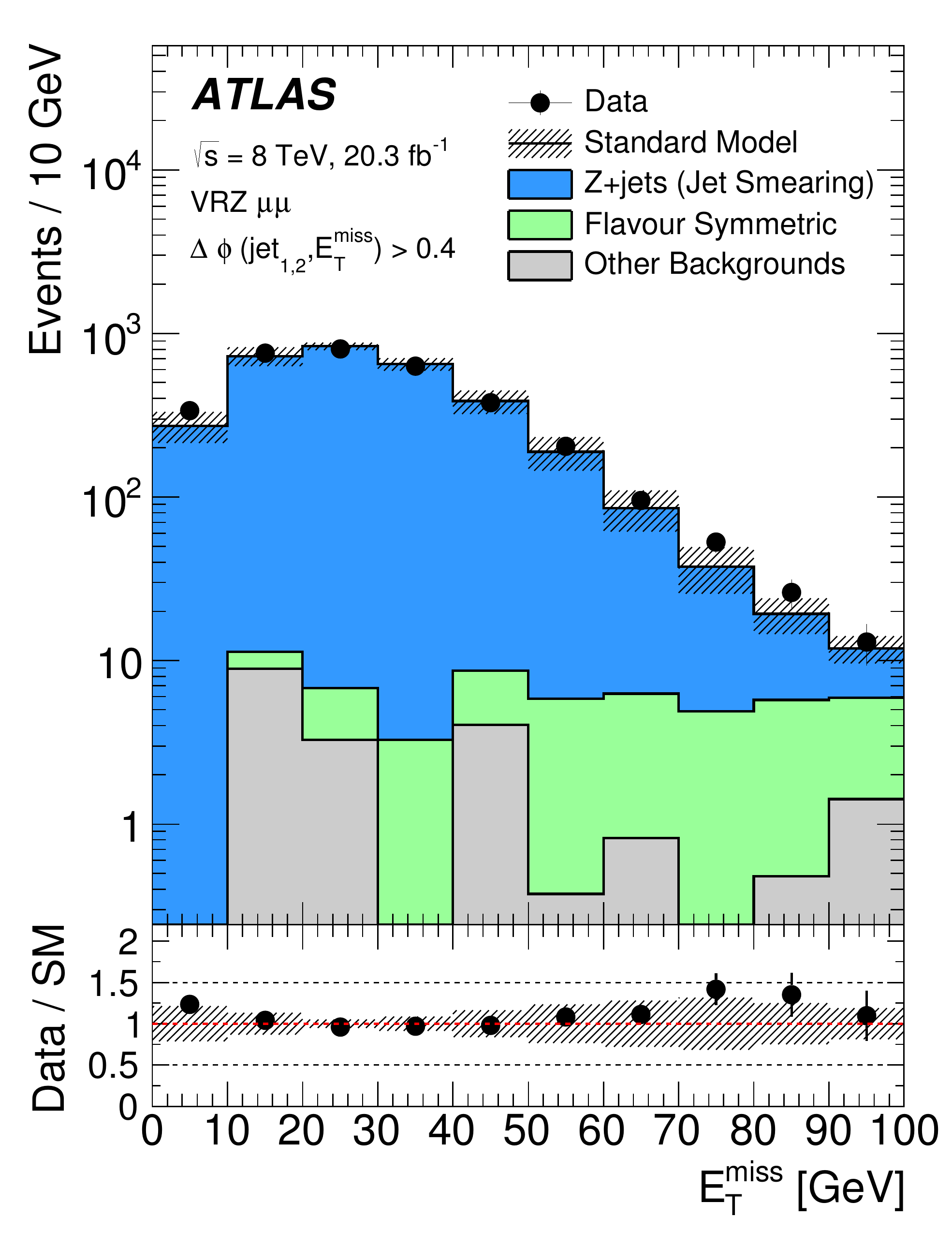}
\caption{Distribution of \met\ in the electron (left) and muon (right) channel in VRZ of the on-$Z$ analysis following the requirement of $\Delta\phi(\text{jet}_{1,2},\met)>0.4$. 
Here the \dyjets\ background (solid blue) is modelled using \pt- and $\phi$-smeared pseudo-data events. 
The hatched uncertainty band includes the statistical uncertainty on the simulated event samples and the systematic uncertainty on the jet-smearing estimate due to the jet response function and the seed selection. 
The backgrounds due to $WZ$, $ZZ$ or rare top processes, as well as from lepton fakes, are included under ``Other Backgrounds''.
}
\label{fig:VRZ}
\end{figure*}

\begin{table}[h]
\centering
\begin{tabular}{ccc}
\noalign{\smallskip}\hline\noalign{\smallskip}
Signal region & Jet-smearing & $Z$+jets MC \\
\noalign{\smallskip}\hline\noalign{\smallskip}
SR-Z $ee$ & $0.05 \pm 0.04$ &  $ 0.05 \pm 0.03 $  \\ [+0.05cm]
SR-Z $\mu\mu$  & $0.02^{+0.03}_{-0.02}$ & $ 0.09 \pm 0.05 $ \\
\noalign{\smallskip}\hline
\end{tabular}
\caption{{\small Number of \dyjets~background events estimated in the on-$Z$ signal region (SR-Z) using the jet smearing method. This is compared with the prediction from the {\sc Sherpa} MC simulation. 
The quoted uncertainties include those due to statistical and systematic effects (see Sect.~\ref{sec:syst}). }} 
\label{tab:results-z}
\end{table}

\subsection{Estimation of the flavour-symmetric backgrounds}
\label{sec:fl-sym}
The dominant background in the signal regions is \ttbar\ production, resulting in two leptons in the final state, 
with lesser contributors including the production of dibosons ($WW$), single top quarks ($Wt$) and $Z$ bosons that decay to $\tau$ leptons. 
For these the so-called ``flavour-symmetry'' method can be used to estimate, 
in a data-driven way, the contribution from these processes in 
the same-flavour channels using their measured contribution to the different-flavour channels. 

\subsubsection{Flavour-symmetric background in the on-$Z$ search}
The flavour-symmetry method uses a control region, CR$e\mu$ in the case of the on-$Z$ search, 
which is defined to be identical to the signal region, but in the different-flavour $e\mu$ channel. 
In CR$e\mu$, the expected contamination due to GGM signal processes of interest is $<3$~\%.

The number of data events observed ($N^{\text{data}}_{e\mu}$) in this control region is corrected by subtracting the expected contribution from backgrounds that are not flavour symmetric. 
The background with the largest impact on this correction is that due to fake leptons, with the estimate provided by the matrix method, described in Sect.~\ref{sec:fakes}, being used in the subtraction.  
All other contributions, which include $WZ$, $ZZ$, $tZ$ and \ttbar$+W(W)/Z$ processes, are taken directly from MC simulation. 
This corrected number, $N^{\text{data,corr}}_{e\mu}$, is related to the expected number in the same-flavour channels, $N^{\text{est}}_{ee/\mu\mu}$, by the following relations: 

\begin{eqnarray}\label{eq:NestData}
\nonumber N_{ee}^{\mathrm{est}} &=& \frac{1}{2}  N_{e\mu}^{\mathrm{data,corr}} k_{ee} \alpha, \\ 
N_{\mu\mu}^{\mathrm{est}} &=& \frac{1}{2} N_{e\mu}^{\mathrm{data,corr}} k_{\mu\mu} \alpha,
\end{eqnarray}

\noindent where $k_{ee}$ and $k_{\mu\mu}$ are electron and muon selection efficiency factors and $\alpha$ accounts for the different trigger efficiencies for same-flavour and different-flavour dilepton combinations. 
The selection efficiency factors are calculated using the ratio of dielectron and dimuon events in 
VRZ according to:

\begin{eqnarray}\label{eq:alpha}
\nonumber k_{ee} = \sqrt{\frac{N_{ee}^{\mathrm{data}}{\rm (VRZ)}}{N_{\mu\mu}^{\mathrm{data}}{\rm (VRZ)}}}, \\ 
\nonumber k_{\mu\mu} = \sqrt{\frac{N_{\mu\mu}^{\mathrm{data}}{\rm (VRZ)}}{N_{ee}^{\mathrm{data}}{\rm (VRZ)}}}, \\ 
\alpha=\frac{\sqrt{\epsilon_{\mathrm{trig}}^{ee} \epsilon_{\mathrm{trig}}^{\mu\mu}}}{\epsilon_{\mathrm{trig}}^{e\mu}},
\end{eqnarray}

\noindent where $\epsilon_{\mathrm{trig}}^{ee}$, $\epsilon_{\mathrm{trig}}^{\mu\mu}$ and $\epsilon_{\mathrm{trig}}^{e\mu}$ are the efficiencies of the dielectron, dimuon and electron--muon trigger configurations, 
respectively, and $N_{ee(\mu\mu)}^{\mathrm{data}}{\rm (VRZ)}$ is the number of $ee$ ($\mu\mu$) data events in VRZ. 
These selection efficiency factors are calculated separately for the cases where both leptons fall within the barrel, both fall within the endcap regions, and for barrel--endcap combinations. 
This is motivated by the fact that the trigger 
efficiencies differ in the central and more forward regions of the detector. 
This estimate is found to be consistent with that resulting from the use of single global $k$ factors,
which provides a simpler but less precise estimate.
In each case the $k$ factors are close to $1.0$, 
and the $N_{ee}^{\mathrm{est}}$ or $N_{\mu\mu}^{\mathrm{est}}$ estimates obtained using $k$ factors from each configuration are consistent with one another to within $0.2\sigma$.

The flavour-symmetric background estimate was chosen as the nominal method prior to examining the data yields in the signal region, 
since it relies less heavily on simulation and provides the most precise estimate. 
This data-driven method is cross-checked using the $Z$ boson mass sidebands ($m_{\ell\ell} \notin [81, 101]$~GeV) 
to fit the \ttbar\ MC events to data in a top control region, CRT.
The results are then extrapolated to the signal region in the $Z$ boson mass window, as illustrated in Fig.~\ref{fig:Z-regions}.  
All other backgrounds estimated using the flavour-symmetry method are taken directly from MC simulation for this cross-check.
Here, \dyjets\ MC events are used to model the small residual \dyjets\ background in the control region, while the jet smearing method provides the estimate in the signal region. 
The normalisation of the \ttbar\ sample obtained from the fit is $0.52 \pm 0.12$ times the nominal MC normalisation, where the uncertainty includes all experimental and theoretical sources of uncertainty as discussed in Sect.~\ref{sec:syst}.
This result is compatible with observations from other ATLAS analyses, which indicate that MC simulation tends to overestimate data 
in regions dominated by $\ttbar$ events accompanied by much jet activity~\cite{Aad:2014kra,Aad:2014wea}. 
MC simulation has also been seen to overestimate contributions from \ttbar\ processes in regions with high \met~\cite{Aad:2015mia}. 
In selections with high \met\, but including lower $H_{\text{T}}$, such as those used in the off-$Z$ analysis, this downwards scaling is less dramatic. 
The results of the cross-check using the $Z$ boson mass sidebands are shown in Table~\ref{tab:results-flsym}, with the sideband fit yielding a prediction slightly higher than, but consistent with, the flavour-symmetry estimate.
This test is repeated varying the MC simulation sample used to model the \ttbar\ background.
The nominal {\sc Powheg+Pythia} \ttbar\ MC sample is replaced with a sample using {\sc Alpgen}, and the fit is performed again.
The same test is performed using a {\sc Powheg} \ttbar\ MC sample that uses {\sc Herwig}, rather than {\sc Pythia}, for the parton shower.
In all cases the estimates are found to be consistent within $1 \sigma$. 
This cross-check using \ttbar\ MC events is further validated in identical regions with intermediate \met\ (150 $<$ \met\ $<$ 225 GeV) and slightly looser \HT\ requirements (\HT$>500$ GeV), as illustrated in Fig.~\ref{fig:Z-regions}. 
Here the extrapolation in $m_{\ell\ell}$ between the sideband region (VRT) and the on-$Z$ region (VRTZ) shows consistent results within approximately $1\sigma$ between 
data and the fitted prediction.

\begin{figure*}
\centering
\includegraphics[width=0.8\textwidth]{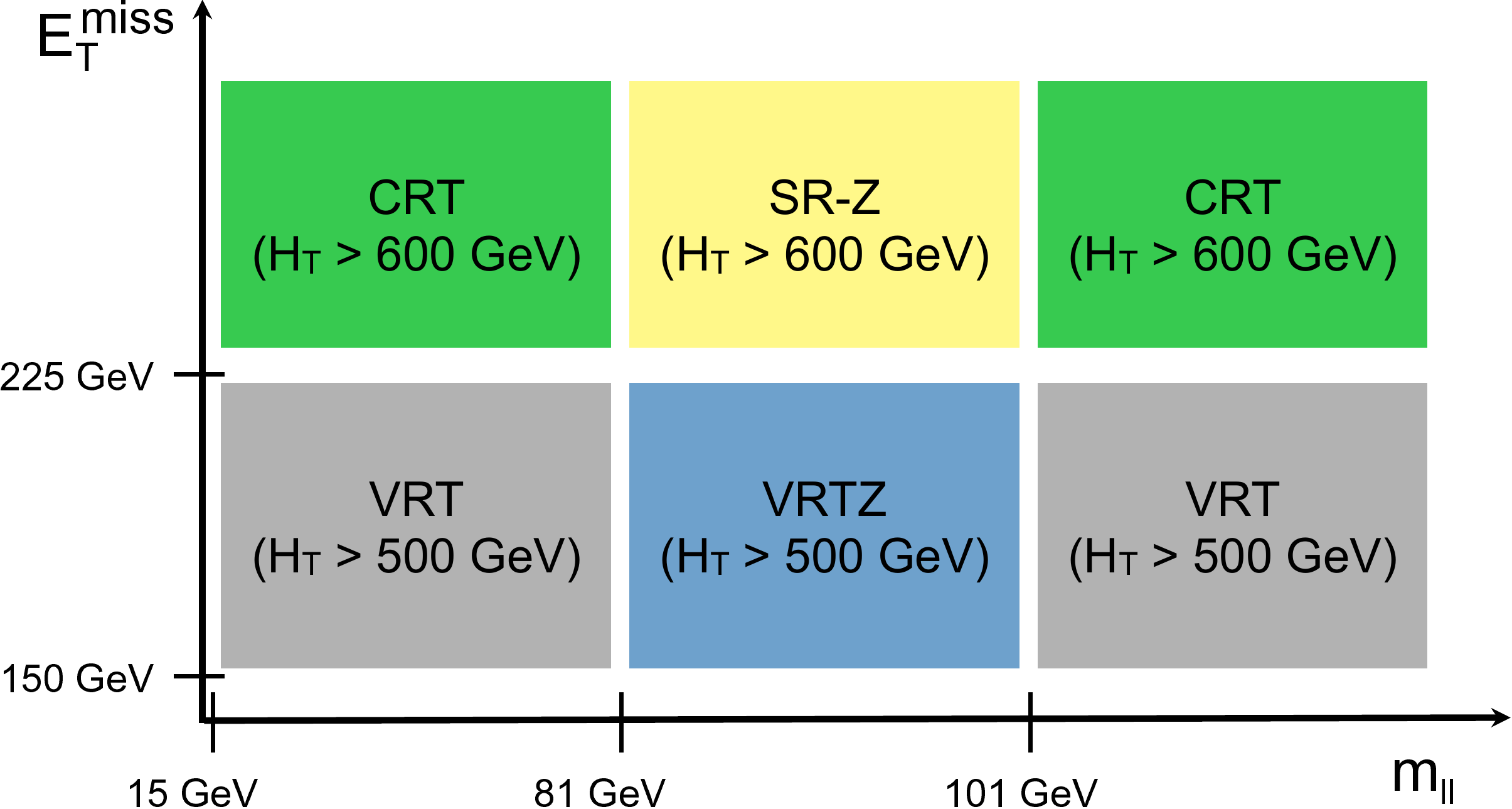}
\caption{Diagram indicating the position in the \met\ versus dilepton invariant mass plane of SR-Z, the control region CRT, and the two validation regions (VRT and VRTZ) used to validate the sideband fit for the on-$Z$ search. VRT and VRTZ have lower \HT\ thresholds than CRT and SR-Z.}
\label{fig:Z-regions}
\end{figure*}

\begin{table}[h]
\centering
\begin{tabular}{ccc}
\noalign{\smallskip}\hline\noalign{\smallskip}
Signal  & Flavour-symmetry  & Sideband fit  \\
region  &                   &   \\
\noalign{\smallskip}\hline\noalign{\smallskip}
SR-Z $ee$ & $2.8 \pm 1.4$ &  $ 4.9 \pm 1.5 $  \\ [+0.05cm]
SR-Z $\mu\mu$  & $3.3 \pm 1.6$ & $ 5.3 \pm 1.9 $ \\
\noalign{\smallskip}\hline
\end{tabular}
\caption{{\small The number of events for the flavour-symmetric background estimate in the on-$Z$ signal region (SR-Z) using the data-driven method based on data in CR$e\mu$. 
This is compared with the prediction for the sum of the flavour-symmetric backgrounds ($WW$, $tW$, \ttbar\ and $Z \rightarrow \tau\tau$) from a sideband fit to data in CRT. 
In each case the combined statistical and systematic uncertainties are indicated. }}
\label{tab:results-flsym}
\end{table}

The flavour-symmetry method is also tested in these VRs. 
An overview of the nominal background predictions, using the flavour-symmetry method, in CRT and these VRs is shown in Fig.~\ref{fig:onZVRs}.
This summary includes CRT, VRT, VRTZ and two variations of VRT and VRTZ. 
The first variation, denoted VRT/VRTZ (high $H_{\text{T}}$), shows VRT/ VRTZ with an increased $H_{\text{T}}$ threshold ($H_{\text{T}}>600$~GeV), 
which provides a sample of events very close to the SR. 
The second variation, denoted VRT/VRTZ (high \met), shows VRT/ VRTZ with the same \met\ cut as SR-Z, but the requirement $400<H_{\text{T}}<600$~GeV is added to provide a sample of events very close to the SR. 
In all cases the data are consistent with the prediction. 
GGM signal processes near the boundary of the expected excluded region are expected to contribute little to the normalisation regions, 
with contamination at the level of up to 4~\% in CRT and 3~\% in VRT. 
The corresponding contamination in VRTZ is expected to be $\sim 10$~\% across most of the relevant parameter space,
increasing to a maximum value of $\sim$50~\% in the region near $m(\tilde{g})=700$~\GeV, $\mu=200$~\GeV.

\begin{figure*}
\centering
\includegraphics[width=1.0\textwidth]{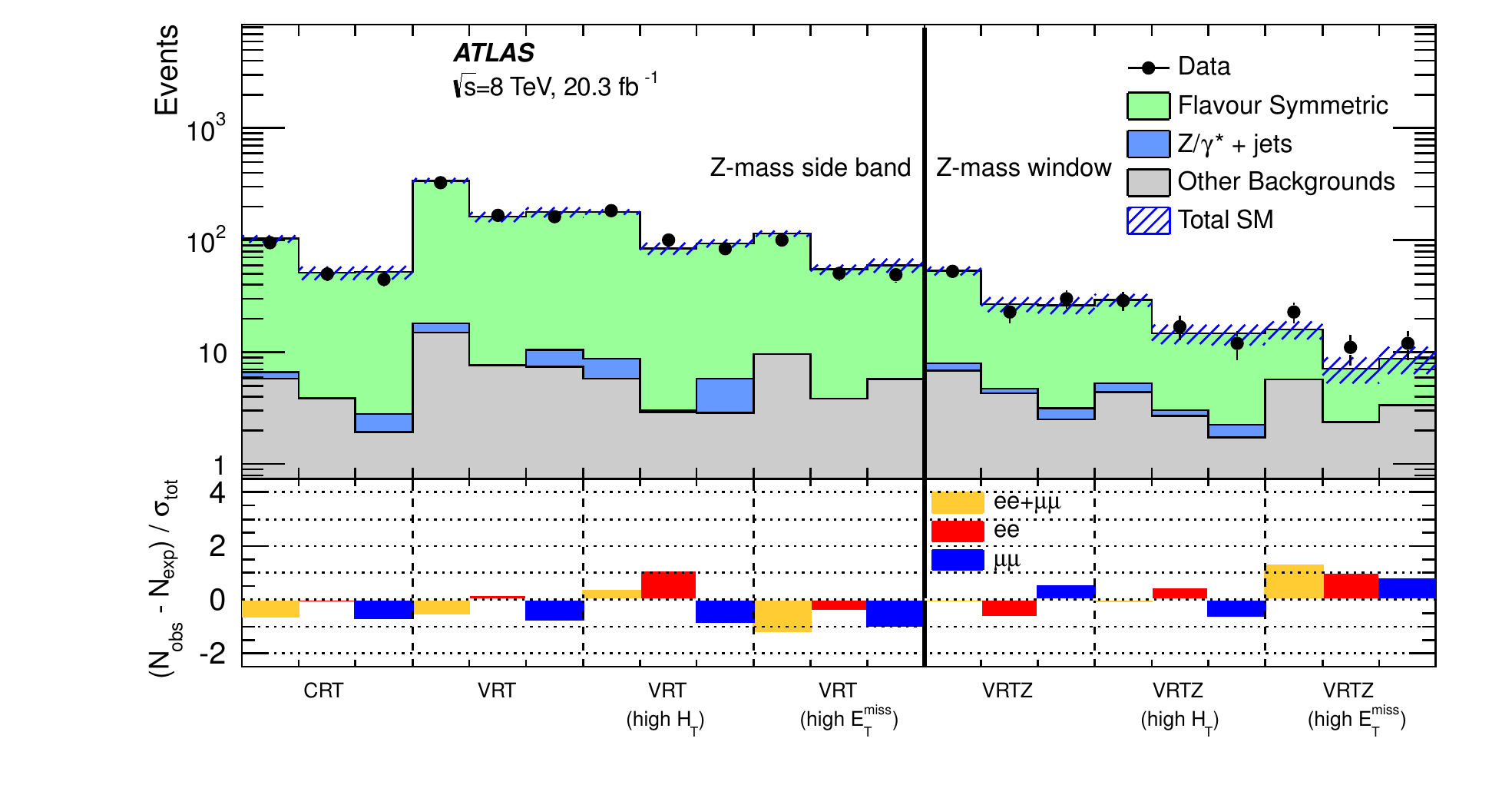}
\caption{The observed and expected yields in CRT and the VRs in the $Z$ boson mass sidebands (left) and the $Z$ boson mass window (right) regions. 
The bottom plot shows the difference in standard deviations between the observed and expected yields. 
The backgrounds due to $WZ$, $ZZ$ or rare top processes, as well as from lepton fakes, are included under ``Other Backgrounds''.
}
\label{fig:onZVRs}
\end{figure*}

\subsubsection{Flavour-symmetric background in the off-$Z$ search}
The background estimation method of Eq.~(\ref{eq:NestData}) is extended to allow a prediction
of the background dilepton mass shape, which is used explicitly to discriminate signal from background in the off-$Z$ search. 
In addition to the $k$ and $\alpha$ correction factors, a third correction
factor $S(i)$ is introduced (where $i$ indicates the dilepton mass bin):

\begin{eqnarray}\label{eq:NestData2}
\nonumber N_{ee}^{\mathrm{est}}(i) &=& \frac{1}{2}  N_{e\mu}^{\mathrm{data,corr}}(i) k_{ee} \alpha S_{ee}(i), \\ 
N_{\mu\mu}^{\mathrm{est}}(i) &=& \frac{1}{2} N_{e\mu}^{\mathrm{data,corr}}(i) k_{\mu\mu} \alpha S_{\mu\mu}(i).
\end{eqnarray}

\noindent These shape correction factors account for different reconstructed dilepton mass shapes in the $ee$, $\mu\mu$, and $e\mu$ 
channels, which result from two effects. First, the offline selection efficiencies for electrons and muons depend differently
on the lepton \pt\ and $\eta$. 
For electrons, the offline selection efficiency increases slowly with \pt, while it has very little dependence on \pt\ for muons.
Second, the combinations of single-lepton and dilepton triggers used for the $ee$, $\mu\mu$, and $e\mu$ channels
have different efficiencies with respect to the offline selection. 
In particular, for $e\mu$ events the trigger efficiency with respect to the offline selection at low \mll\ is 80\%, which is 10--15\% lower
than the trigger efficiencies in the $ee$ and $\mu\mu$ channels.
To correct for these two effects, \ttbar\ MC simulation is used. 
The dilepton mass shape in the $ee$ or $\mu\mu$ channel is compared to that in the $e\mu$ channel,
after scaling the latter by the $\alpha$- and $k$-factor trigger and lepton selection efficiency corrections.
The ratio of the dilepton mass distributions, $N_{ee}(\mll)/N_{e\mu}(\mll)$ or $N_{\mu\mu}(\mll)/N_{e\mu}(\mll)$, is fitted with a second-order polynomial,
which is then applied as a correction factor, along with $\alpha$ and $k$, to the $e\mu$ distribution in data.
These correction factors have an impact on the predicted background yields
of approximately a few percent in the $ee$ channel and up to $\sim$10--15~\% in the $\mu\mu$ channel, depending on the signal region.

The background estimation methodology is validated in a region with exactly two jets and $100<\MET<150$~\GeV (VR-offZ).
The flavour-symmetric category contributes more than 95~\% of the total background in this region.
The dominant systematic uncertainty on the background prediction is the 6~\% uncertainty on the trigger
efficiency $\alpha$-factor.
The observed dilepton mass shapes are compared to the SM expectations in Fig.~\ref{fig:edge-VR}, 
indicating consistency between the data and the expected background yields. 
The observed yields and expected backgrounds in the below-$Z$ and above-$Z$ regions are presented in~\ref{app:edge}.
For signal models near the edge of the sensitivity of this analysis, the contamination from signal events in VR-offZ is less than 3~\%.

\begin{figure*}[!tbh]
\centering
\includegraphics[width=0.48\textwidth]{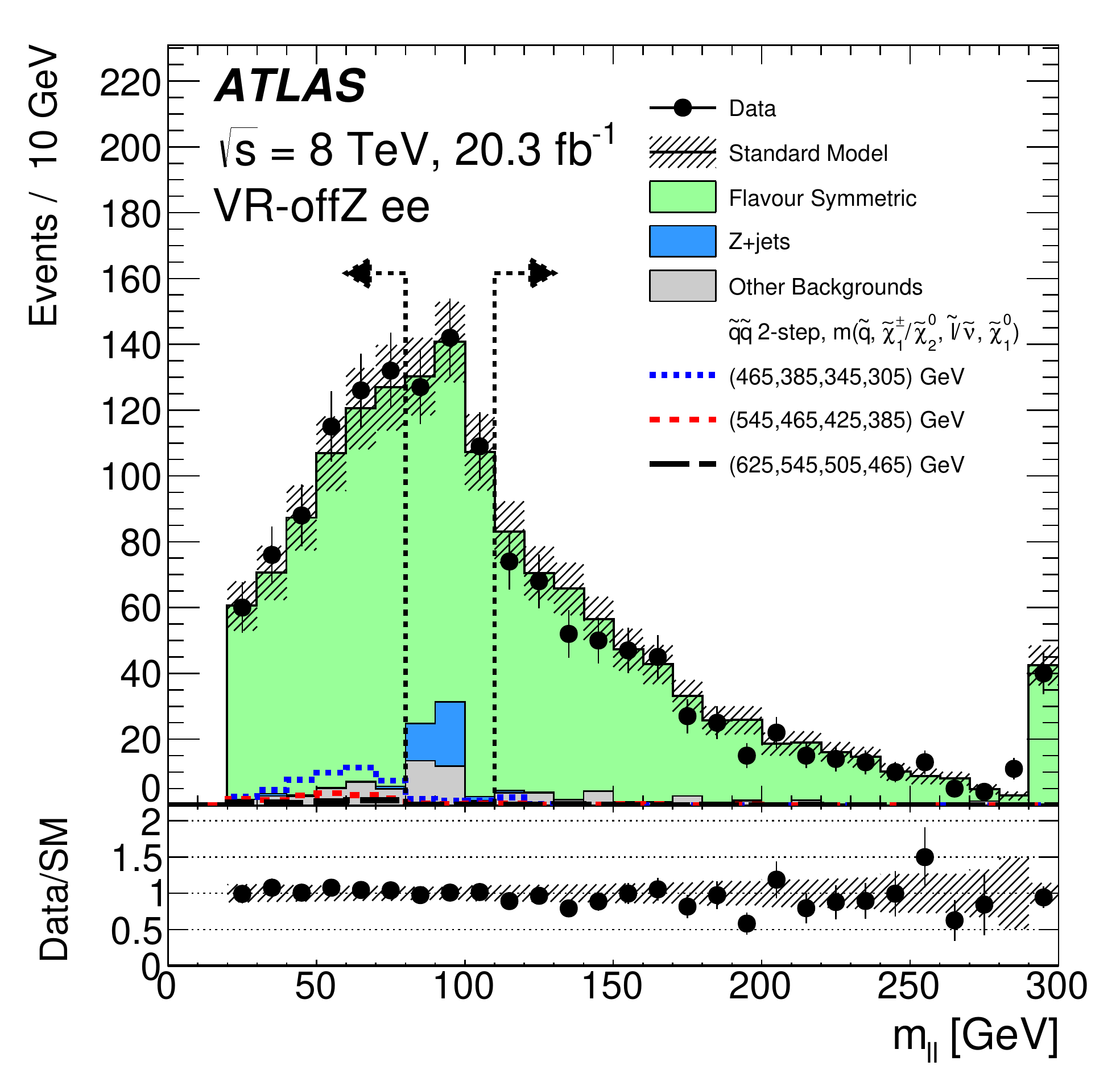}
\includegraphics[width=0.48\textwidth]{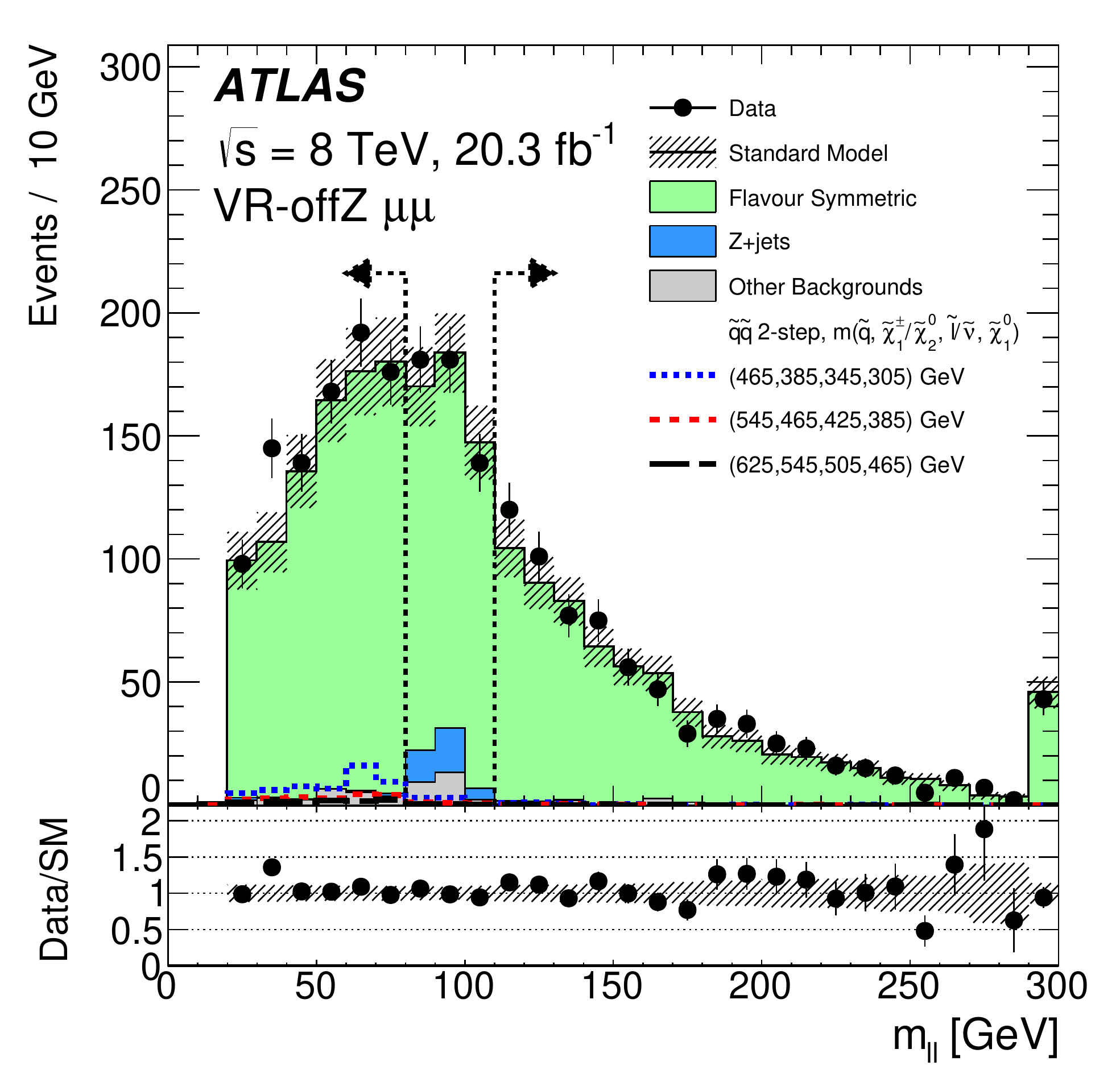}
\caption{ The observed and expected dilepton mass distributions in the electron (left) and muon (right) channel of the
validation region (VR-offZ) of the off-$Z$ search. 
Data (black points) are compared to the sum of expected backgrounds (solid histograms).
The vertical dashed lines indicate the $80<\mll<110$~\GeV\ region, which is used to normalise the \zjets\ background.
Example signal models (dashed lines) are overlaid, with $m(\tilde{q})$, $m(\tilde{\chi}^{0}_{2})/m(\tilde{\chi}^{\pm}_{1})$, $m(\tilde{\ell})/m(\tilde{\nu})$, and $m(\tilde{\chi}^{0}_{1})$ of each benchmark point being indicated in the figure legend. 
The bottom plots show the ratio of the data to expected background. The error bars indicate the statistical uncertainty in data,
while the shaded band indicates the total background uncertainty. The last bin contains the overflow.
\label{fig:edge-VR}}
\end{figure*}

\subsection{Fake-lepton contribution}
\label{sec:fakes}
Events from $W \rightarrow \ell\nu +$jets, semileptonic \ttbar\ and single top ($s$- and $t$-channel) contribute to the background in the dilepton channels due to ``fake'' leptons.
These include leptons from $b$-hadron decays, misidentified hadrons or converted photons, and are estimated from data using a matrix method, 
which is described in detail in Ref.~\cite{matrixmethod}. 
This method involves creating a control sample using baseline leptons, thereby loosening the lepton isolation and identification requirements and increasing the probability of selecting a fake lepton. 
For each control or signal region, the relevant requirements are applied to this control sample, and the number of events with leptons that pass or fail the subsequent signal-lepton requirements are counted. 
Denoting the number of events passing signal lepton requirements by $N_{\text{pass}}$ and the number failing by $N_{\text{fail}}$, 
the number of events containing a fake lepton for a single-lepton selection is given by

\begin{equation}
N_{\text{fake}} = \frac{N_{\text{fail}} - (1/\epsilon^{\text{real}}-1) N_{\text{pass}}}{(1/\epsilon^{\text{fake}} - 1/\epsilon^{\text{real}})},
\end{equation}

\noindent where $\epsilon^{\text{fake}}$ is the efficiency with which fake leptons passing the baseline lepton selection also pass signal lepton requirements and $\epsilon^{\text{real}}$ is the relative identification efficiency (from baseline to signal lepton selection) for real leptons. 
This principle is expanded to a dilepton sample using a four-by-four matrix to account for the various possible real--fake combinations for the two leading leptons in the event.

The efficiency for fake leptons is estimated in control regions enriched with multi-jet events. 
Events are selected if they contain at least one baseline lepton, one signal jet with \pt~$>60$ GeV and low \met\ ($<$30 GeV). 
The background due to processes containing prompt leptons, estimated from MC samples, is subtracted from the total data contribution in this region. 
From the resulting data sample the fraction of events in which the baseline leptons pass signal lepton requirements gives the fake efficiency. 
This calculation is performed separately for events with $b$-tagged jets and those without to take into account the various sources from which fake leptons originate. 
The real-lepton efficiency is estimated using $Z \rightarrow \ell^+ \ell^-$ events in a data sample enriched with leptonically decaying $Z$ bosons.
Both the real-lepton and fake-lepton efficiencies are further binned as a function of \pt\ and $\eta$.

\subsection{Estimation of other backgrounds}
\label{sec:bg}

The remaining background processes, including diboson events with a $Z$ boson decaying to leptons and the $\ttbar+W(W)/Z$ and $t+Z$ backgrounds,
are estimated from MC simulation. 
In these cases the most accurate theoretical cross sections available are used, as summarised in Table~\ref{tab:MC}.
Care is taken to ensure that the flavour-symmetric component of these backgrounds (for events where the two leptons do not originate from the same $Z$
decay) is not double-counted.

\section{Systematic uncertainties}
\label{sec:syst}

Systematic uncertainties have an impact on the predicted signal region yields from the dominant backgrounds, the fake-lepton estimation, 
and the yields from backgrounds predicted using simulation alone.
The expected signal yields are also affected by systematic uncertainties. 
All sources of systematic uncertainty considered are discussed in the following subsections.

\subsection{Experimental uncertainties}
The experimental uncertainties arise from the modelling of both the signal processes and backgrounds estimated using MC simulation.
Uncertainties associated with the jet energy scale (JES) are assessed using both simulation and in-situ measurements~\cite{JES,JES2}. 
The JES uncertainty is influenced by the event topology, flavour composition, jet \pt\ and $\eta$, 
as well as by the pile-up. 
The jet energy resolution (JER) is also affected by pile-up, and is estimated using in-situ measurements~\cite{atlas-jer}.
An uncertainty associated with the JVF requirement for selected jets is also applied by varying the JVF threshold up (0.28) and down (0.21) with respect to the nominal value of 0.25.
This range of variation is chosen based on a comparison of the
efficiency of a JVF requirement in dijet events in data and MC simulation.

To distinguish between heavy-flavour-enriched and light-flavour-enriched event samples, $b$-jet tagging is used. 
The uncertainties associated with the $b$-tag\-ging efficiency and the light/charm quark mis-tag 
rates are measured in \ttbar-enriched samples~\cite{ATLAS-CONF-2014-004,ATLAS-CONF-2012-043} and 
dijet samples~\cite{ATLAS-CONF-2012-040}, respectively.

Small uncertainties on the lepton energy scales and momentum resolutions are measured in 
$Z \rightarrow \ell^+\ell^-$, $J/\psi \rightarrow \ell^+\ell^-$ and $W \rightarrow \ell^{\pm}\nu$ 
event samples~\cite{electronref}. 
These are propagated to the \met\ uncertainty, along with the uncertainties due to the JES and JER. 
An additional uncertainty on the energy scale of topological clusters in the calorimeters not associated with 
reconstructed objects (the \met\ soft term) is also applied to the \met calculation. 

The trigger efficiency is assigned a 5~\% uncertainty following studies comparing the efficiency in simulation to that measured in $Z \rightarrow \ell^+ \ell^-$ events in data.

The data-driven background estimates are subject to uncertainties associated with the methods employed and the limited 
number of events used in their estimation. 
The \dyjets\ background estimate has an uncertainty to account for
differences between pseudo-data and MC events, the choice of seed region definition, 
the statistical precision of the seed region, and the jet response functions used to create the pseudo-data. 
Uncertainties in the flavour-symmetric background estimate include those related to the electron and muon selection efficiency factors $k_{ee}$ and $k_{\mu\mu}$,
the trigger efficiency factor $\alpha$, and, for the off-$Z$ search only, the dilepton mass shape $S(i)$ reweighting factors.
Uncertainties attributed to the subtraction of the non-flavour-symmetric back\-grounds, and those due to limited statistical precision in the $e\mu$ control 
regions, are also included.
Finally, an uncertainty derived from the difference in real-lepton efficiency observed in \ttbar\ and $Z \rightarrow \ell^+ \ell^-$ events is assigned to the fake-background prediction.
An additional uncertainty due to the number of events in the control samples used to derive the real efficiencies and fake rates is assigned to this background,
as well as a 20~\% uncertainty on the MC background subtraction in the control samples.

\subsection{Theoretical uncertainties on background processes}

For all backgrounds estimated from MC simulation, the following theoretical uncertainties are considered. 
The uncertainties due to the choice of factorisation and renormalisation scales are calculated by varying the nominal values by a factor of two. 
Uncertainties on the PDFs are evaluated following the prescription recommended by {\sc PDF4LHC}~\cite{Botje01}.
Total cross-section uncertainties of 22~\%~\cite{Campbell:2012} and 50~\% are applied to \ttbar~$+W$/$Z$ and \ttbar~$+WW$ sub-processes, respectively.
For the \ttbar~$+W$ and \ttbar~$+Z$ sub-processes, an additional uncertainty is evaluated by comparing samples generated with different numbers of partons, 
to account for the impact of the finite number of partons generated in the nominal samples.
For the $WZ$ and $ZZ$ diboson samples, a parton shower uncertainty is estimated by comparing samples showered with {\sc Pythia} and 
{\sc Herwig+Jimmy}~\cite{Corcella:2000bw,Butterworth:1996zw} and cross-section uncertainties of 5~\% and 7~\% are applied, respectively. 
These cross-section uncertainties are estimated from variations of the value of the strong coupling constant, the PDF and the generator scales. 
For the small contribution from $t+Z$, a 50~\% uncertainty is assigned.
Finally, a statistical uncertainty derived from the finite size of the MC samples used in the background estimation process is included.

\subsection{Dominant uncertainties on the background estimates}
The dominant uncertainties in each signal region, along with their values relative to the total background expectation, are summarised in Table~\ref{tab:syst}. 
In all signal regions the largest uncertainty is that associated with the flavour-symmetric background. 
The statistical uncertainty on the flavour-symmetric background due to the finite data yields in the $e\mu$ CRs is 24~\% in the on-$Z$ SR. 
This statistical uncertainty is also the dominant uncertainty for all SRs of the off-$Z$ analysis except for SR-loose, for which
the systematic uncertainty on the flavour-symmetric background prediction dominates.
In SR-Z the combined MC generator and parton shower modelling uncertainty on the $WZ$ background (7~\%), as well as the uncertainty due to the
fake-lepton background (14~\%), are also important.

\begin{table*}[htbp]
\begin{center}
\scriptsize
 \begin{tabular*}{\textwidth}{@{\extracolsep{\fill}}lcccccc}
   \noalign{\smallskip}\hline\noalign{\smallskip}   
   Source &  \multicolumn{6}{c}{Relative systematic uncertainty [\%]} \\
   \noalign{\smallskip}\hline\noalign{\smallskip}   
   & SR-Z & SR-loose & SR-2j-bveto & SR-2j-btag & SR-4j-bveto & SR-4j-btag \\
   \noalign{\smallskip}\hline\noalign{\smallskip}   
   Total systematic uncertainty  & 29 & 7.1 & 13 & 9.3 & 30 & 15 \\
   \noalign{\smallskip}\hline\noalign{\smallskip}   
   Flavour-symmetry statistical           & 24 & 1.7 & 9.3 & 6.2 & 23   & 12   \\
   Flavour-symmetry systematic            &  4 & 5.7 & 6.7 & 5.9 & 11   & 6.6  \\
   \dyjets\                               &  - & 2.1 & 6.3 & 3.5 & 14   & 7.0  \\
   Fake lepton                            & 14 & 3.2 & 1.4 & 1.2 & 1.8  & 2.2  \\
   $WZ$ MC $+$ parton shower              &  7 & - & - & - & - & -  \\
   \noalign{\smallskip}\hline\noalign{\smallskip}   
\end{tabular*}   
 \caption{Overview of the dominant sources of systematic uncertainty on the background estimate in the signal regions. 
 Their relative values with respect to the total background expectation are shown (in~\%). For the off-$Z$ region, 
 the full dilepton mass range is used, and in all cases the $ee+\mu\mu$ contributions are considered together.}
\label{tab:syst}
\end{center}
\end{table*}

\subsection{Theoretical uncertainties on signal processes}

Signal cross sections are calculated to next-to-lead\-ing order in the strong coupling constant, 
adding the resummation of soft gluon emission at NLO+NLL accuracy ~\cite{Beenakker:1996ch,Kulesza:2008jb,Kulesza:2009kq,Beenakker:2009ha,Beenakker:2011fu}. 
The nominal cross section and the uncertainty are taken from an envelope of cross-section predictions using different PDF sets and factorisation and renormalisation scales, as described in Ref.~\cite{Kramer:2012bx}. 
For the simplified models the uncertainty on the initial-state radiation modelling is important in the case of small mass differences during the cascade decays. 
{\sc MadGraph+Pythia} samples are used to assess this uncertainty, with the factorisation and normalisation scale, 
the {\sc MadGraph} parameter used for jet matching, 
the {\sc MadGraph} parameter used to set the QCD radiation scale and the {\sc Pythia} parameter responsible for the value of the QCD scale for final-state radiation, each being varied up and down by a factor of two. 
The resulting uncertainty on the signal acceptance is up to $\sim 25$~\% in regions with small mass differences within the decay chains.

\section{Results}
\label{sec:results}
For the on-$Z$ search, the resulting background estimates in the signal regions, along with the observed event yields, are displayed in Table~\ref{tab:zSR}.
The dominant backgrounds are those due to flavour-symmetric and $WZ$ and $ZZ$ diboson processes. 
In the electron and muon channel combined, $10.6\pm3.2$ events are expected and $29$ are observed.
For each of these regions, a local probability for the background estimate to produce a fluctuation great\-er than or equal to the excess observed in the data 
is calculated using pseudo-experiments.  
When expressed in terms of the number of standard deviations, this value is referred to as the local significance, or simply the significance. 
These significances are quantified in the last column of Table~\ref{tab:upperlimits} and correspond to a $1.7\sigma$ deviation in the 
muon channel and a $3.0\sigma$ deviation in the electron channel, with the combined significance, calculated from the sum of the background predictions and
observed yields in the muon and electron channels, being $3.0\sigma$. 
The uncertainties on the background predictions in the $ee$ and $\mu\mu$ channels are correlated 
as they are dominated by the statistical uncertainty of the $e\mu$ data sample that is used to derive the flavour-symmetric background in both channels. 
Since this sample is common to both channels, 
the relative statistical error on the flavour-symmetric background estimation does not decrease when combining the $ee$ and $\mu\mu$ channels. 
No excess was reported in the CMS analysis of the $Z+\mathrm{jets}+\met$ final state based on $\sqrt{s}=8$~\TeV\ data~\cite{CMS-edge};
however, the kinematic requirements used in that search differ from those used in this paper.

\begin{table*}[!htbp]
\begin{center}
\setlength{\tabcolsep}{0.0pc}
{\small
\begin{tabular*}{\textwidth}{@{\extracolsep{\fill}}lrrr}
\noalign{\smallskip}\hline\noalign{\smallskip}
Channel        & SR-Z $ee$   & SR-Z $\mu\mu$ & SR-Z same-flavour             \\[-0.05cm]
               &        &          & combined             \\[-0.05cm]
\noalign{\smallskip}\hline\noalign{\smallskip}
Observed events         & $16$       & $13$     & $29$        \\
\noalign{\smallskip}\hline\noalign{\smallskip}
Expected background events         & $4.2 \pm 1.6$       & $6.4 \pm 2.2$     & $10.6 \pm 3.2$        \\
\noalign{\smallskip}\hline\noalign{\smallskip}  
       Flavour-symmetric backgrounds   & $2.8 \pm 1.4$        & $3.3 \pm 1.6$             & $6.0 \pm 2.6$        \\
        \dyjets~(jet-smearing)         & $0.05 \pm 0.04$      & $0.02_{-0.02}^{+0.03}$    & $0.07 \pm 0.05$         \\
        Rare top                       & $0.18 \pm 0.06$      & $0.17 \pm 0.06$           & $0.35 \pm 0.12 $ \\
        $WZ$/$ZZ$ diboson              & $1.2 \pm 0.5$        & $1.7 \pm 0.6$       & $2.9 \pm 1.0$       \\
        Fake leptons                   & $0.1_{-0.1}^{+0.7}$     & $1.2_{-1.2}^{+1.3}$      & $1.3_{-1.3}^{+1.7}$  \\ 
 \noalign{\smallskip}\hline\noalign{\smallskip}
\end{tabular*}
}
\end{center}
\caption{{\small Results in the on-$Z$ SRs (SR-Z). 
The flavour symmetric, \dyjets\ and fake-lepton background components are all derived using data-driven estimates described in the text. 
All other backgrounds are taken from MC simulation. 
The displayed uncertainties include the statistical and systematic uncertainty components combined.}}
\label{tab:zSR}
\end{table*}

Dilepton invariant mass and \met\ distributions in the electron and muon on-$Z$ SR are shown in Fig.~\ref{fig:mll}, 
with $H_{\text{T}}$ and jet multiplicity being shown in Fig.~\ref{fig:ht}.  
For the SR selection a requirement is imposed to reject events with $\Delta\phi(\text{jet}_{1,2},\met)<0.4$ 
to further suppress the background from \dyjets\ processes with mismeasured jets.

\begin{figure*}[!htbp]
\centering
\includegraphics[width=0.47\textwidth]{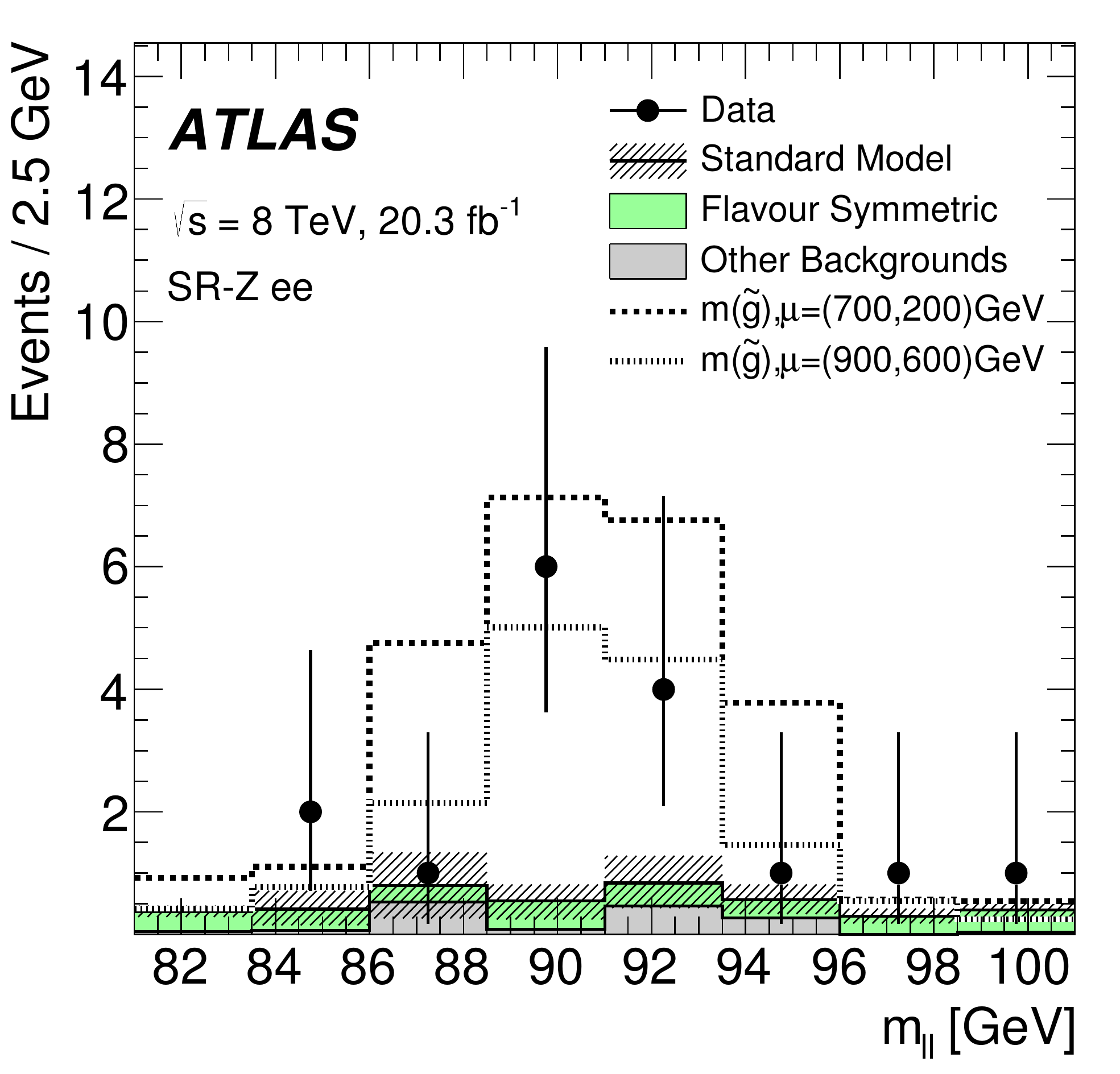}
\includegraphics[width=0.47\textwidth]{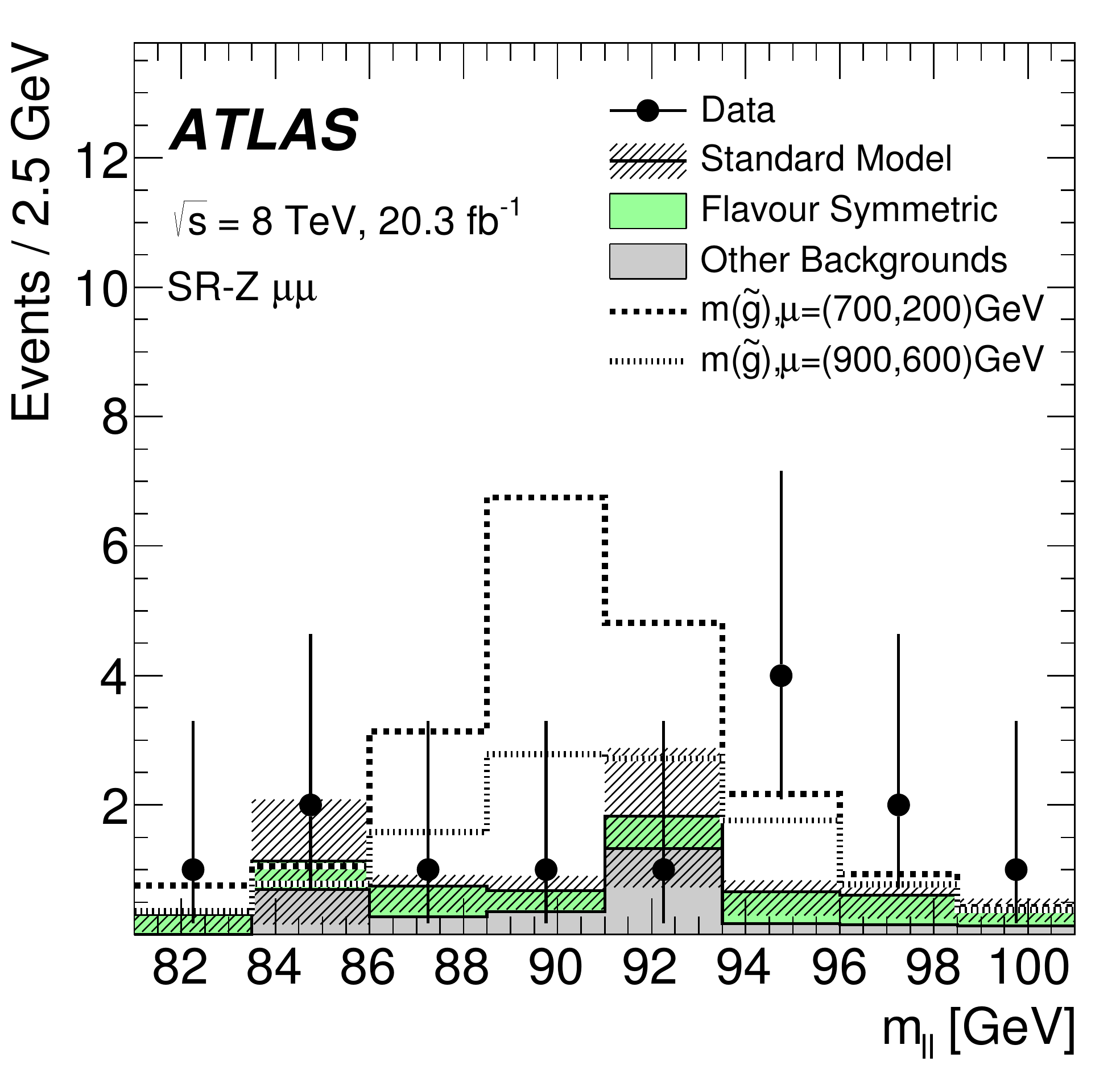}
\includegraphics[width=0.47\textwidth]{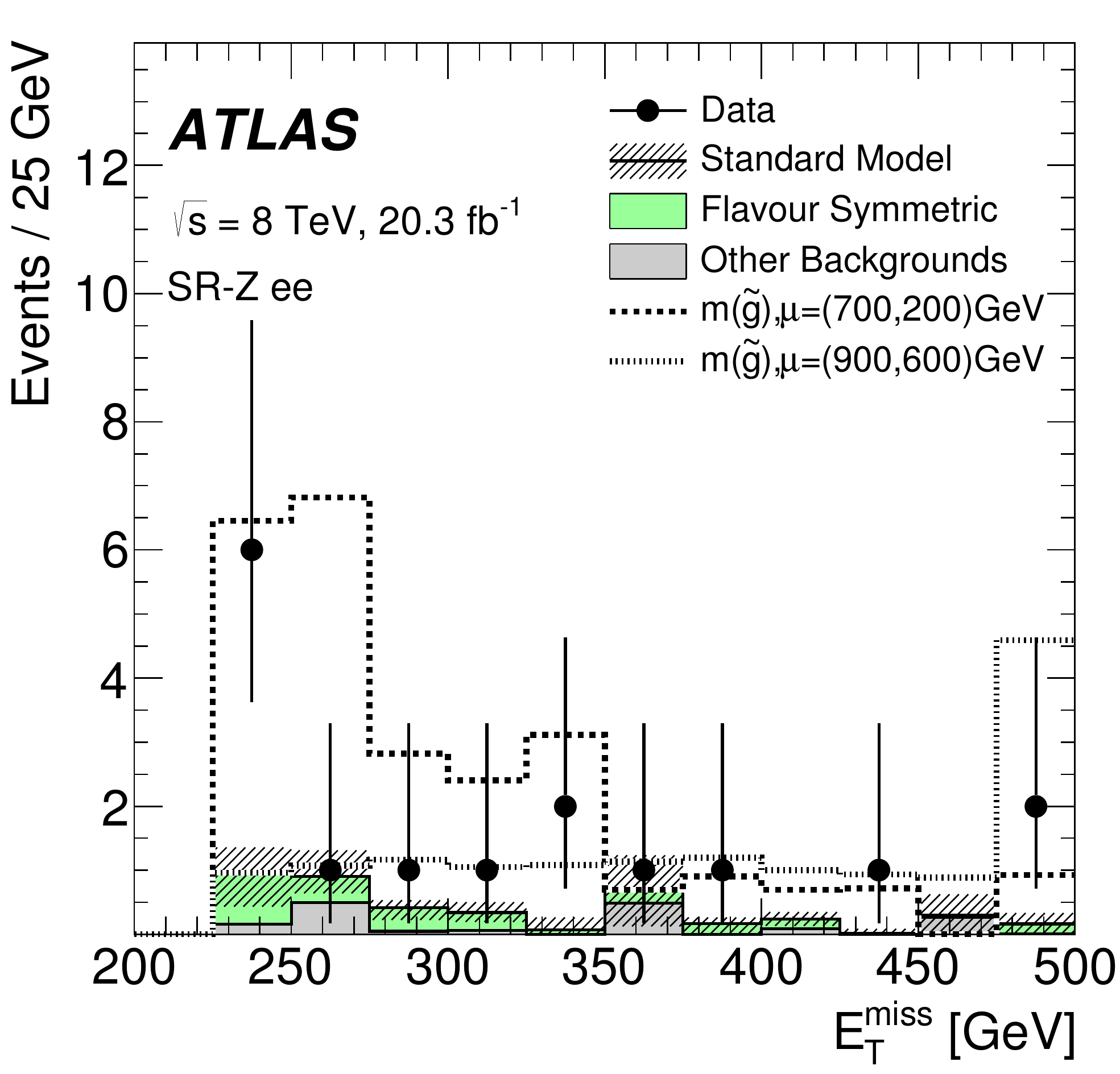}
\includegraphics[width=0.47\textwidth]{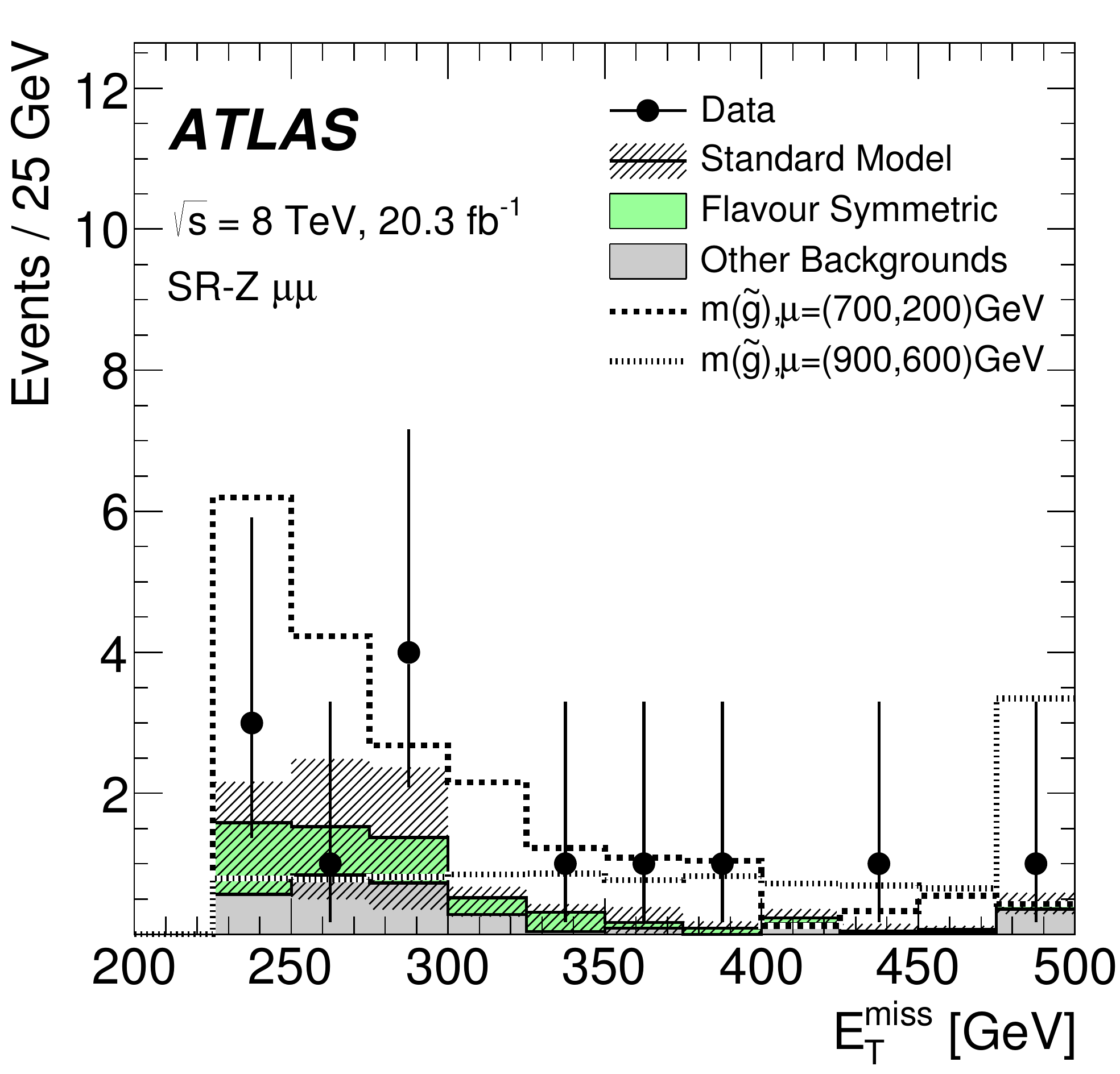}
\caption{The dilepton mass (top) and \met\ (bottom) distributions for the electron (left) and muon (right) channel in the on-$Z$ SRs after having applied the requirement
$\Delta\phi (\text{jet}_{1,2},\MET)>0.4$. All uncertainties are included in the hatched uncertainty band.
Two example GGM ($\tan\beta=1.5$) signal models are overlaid.
For the \met\ distributions, the last bin contains the overflow. 
The backgrounds due to $WZ$, $ZZ$ or rare top processes, as well as from fake leptons, are included under ``Other Backgrounds''.
The negligible contribution from $Z$+jets is omitted from these distributions.
\label{fig:mll}}
\end{figure*}

\begin{figure*}[!htbp]
\centering
\includegraphics[width=0.47\textwidth]{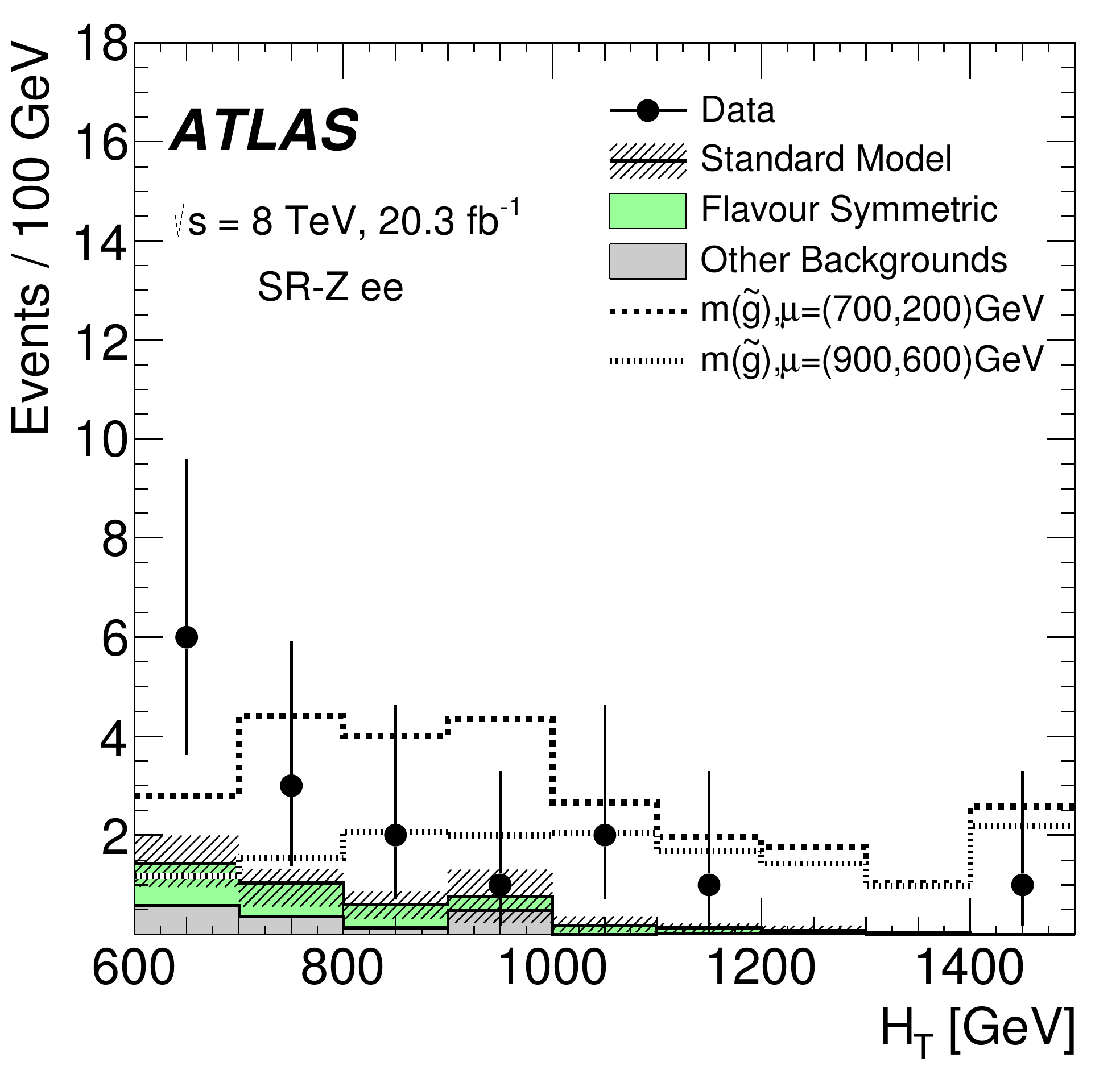}
\includegraphics[width=0.47\textwidth]{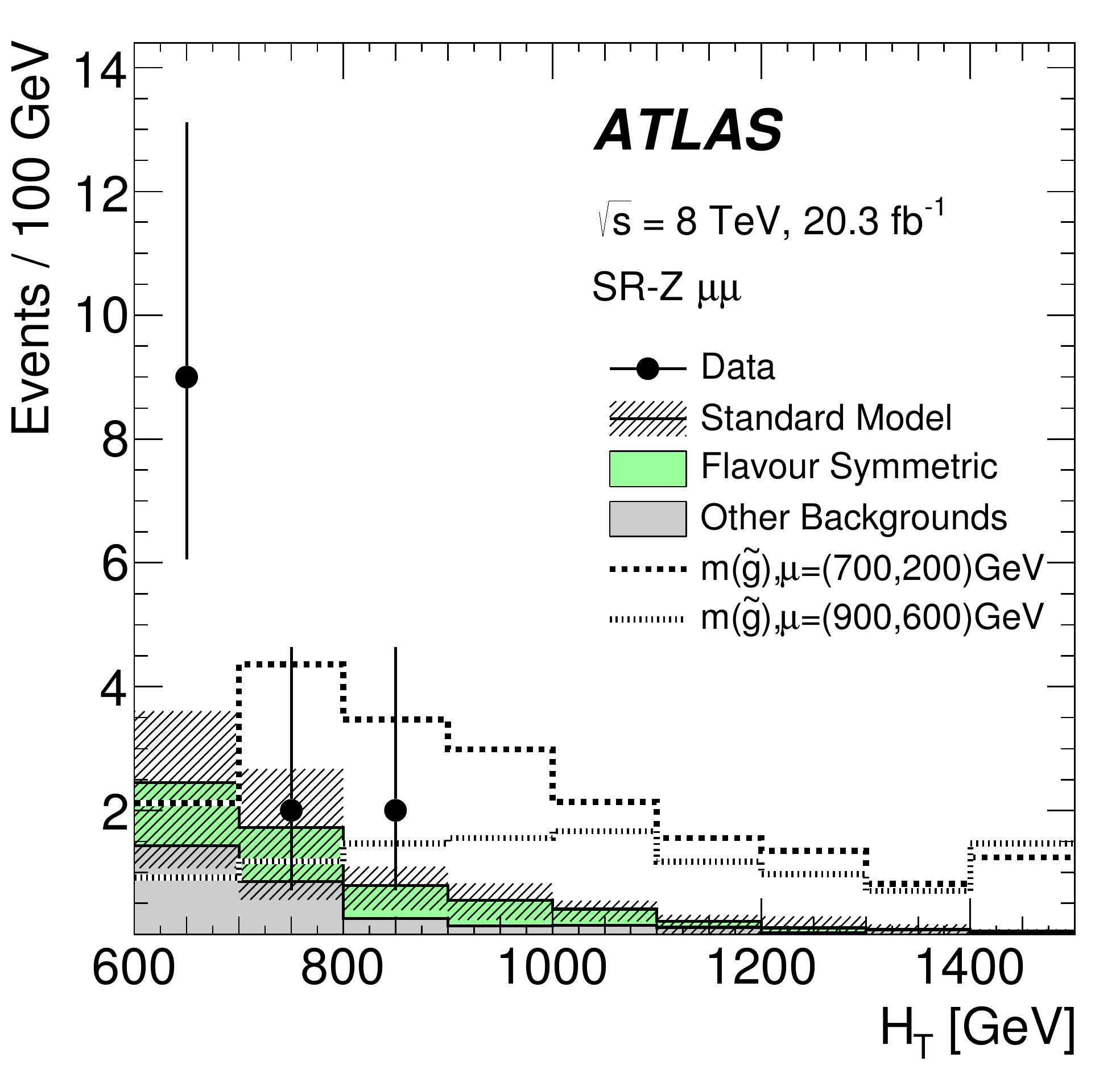}
\includegraphics[width=0.47\textwidth]{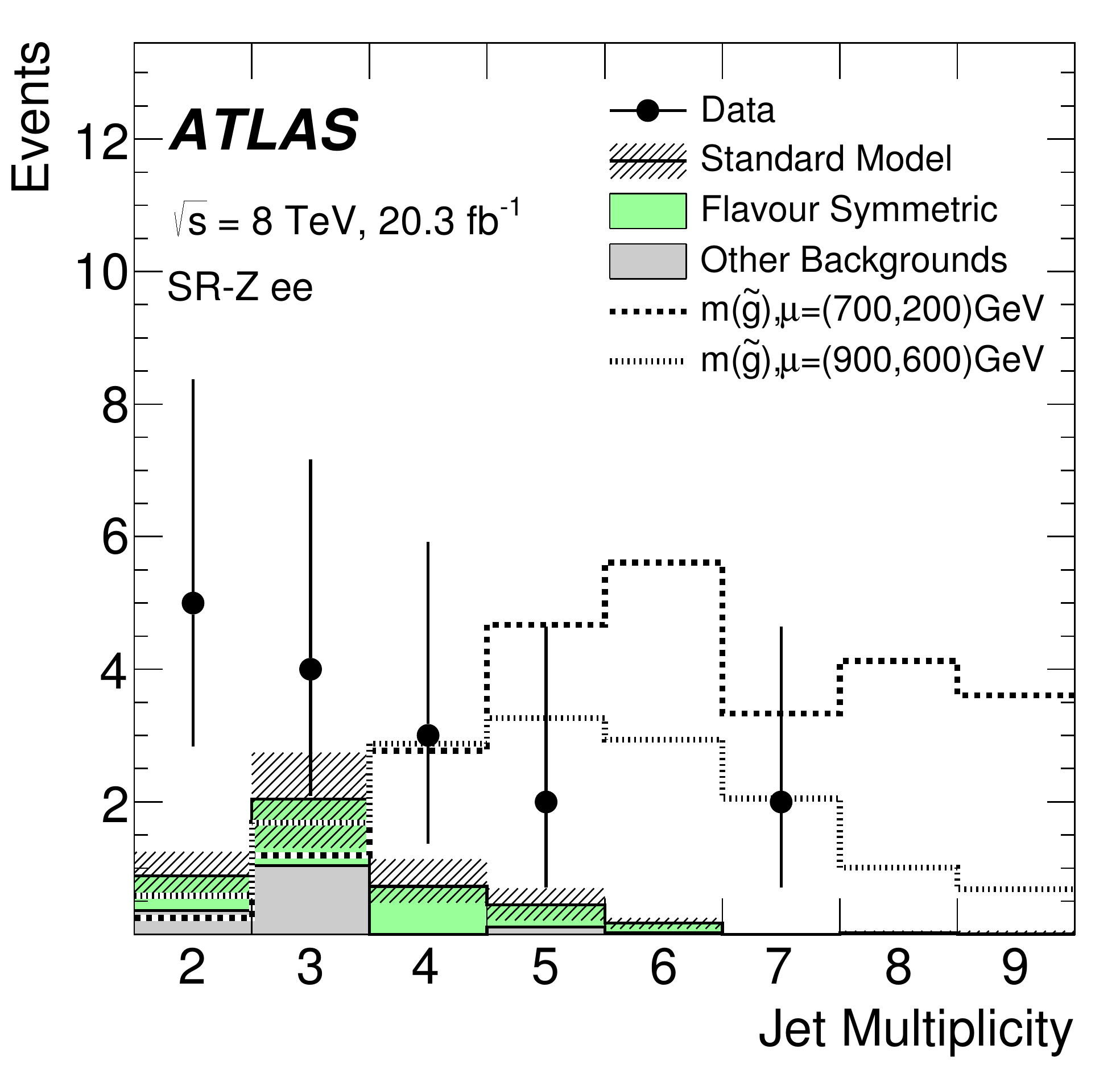}
\includegraphics[width=0.47\textwidth]{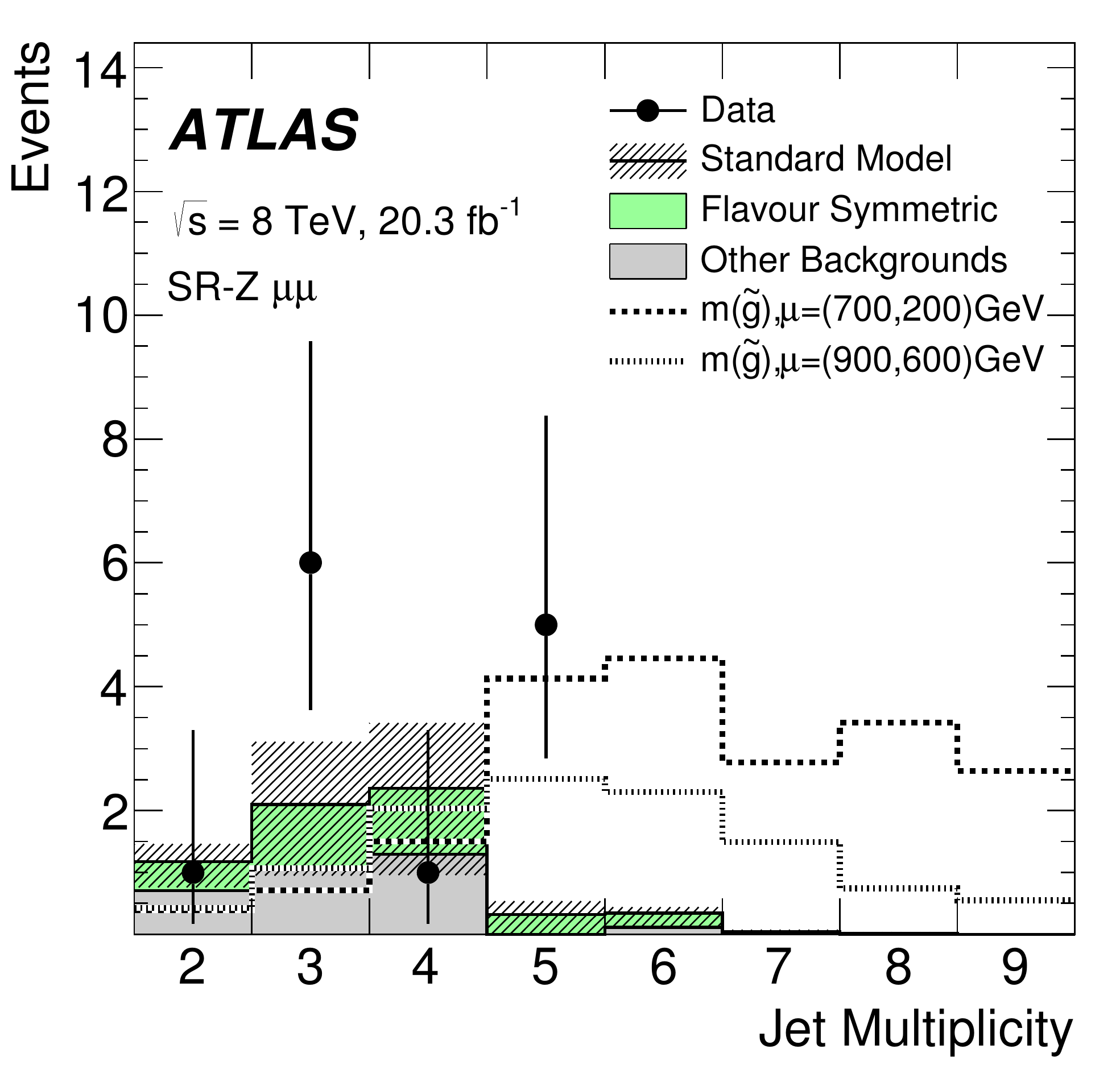}
\caption{The $H_{\text{T}}$ (top) and jet multiplicity (bottom) distributions for the electron (left) and muon (right) channel in the on-$Z$ 
SRs after having applied the requirement $\Delta\phi (\text{jet}_{1,2},\MET)>0.4$. All uncertainties are included in the hatched uncertainty band.
Two example GGM ($\tan\beta=1.5$) signal models are overlaid.
For the $H_{\text{T}}$ distributions, the last bin contains the overflow. 
The backgrounds due to $WZ$, $ZZ$ or rare top processes, as well as from fake leptons, are included under ``Other Backgrounds''.
The negligible contribution from $Z$+jets is omitted from these distributions.
\label{fig:ht}}
\end{figure*}

In Fig.~\ref{fig:jetmet}, 
the distribution of events in the on-$Z$ SR as a function of $\Delta\phi(\text{jet}_{1,2},$ $\met)$ (before this requirement is applied) is shown. 
In these figures the shapes of the flavour-symmetric and \dyjets\ 
backgrounds are derived using MC simulation and the normalisation is taken according to the data driven estimate.

\begin{figure*}[!htbp]
\centering
\includegraphics[width=0.47\textwidth]{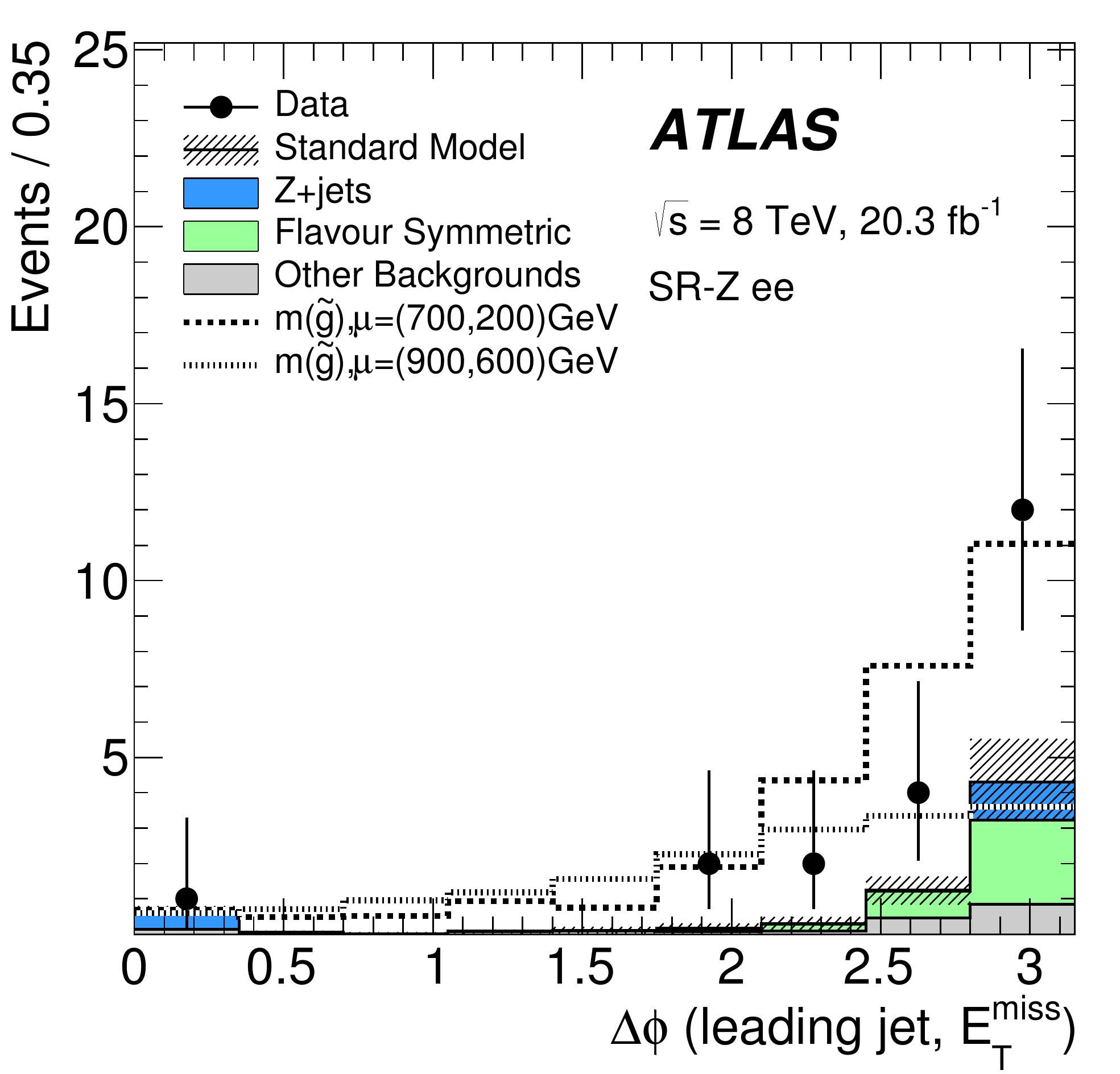}
\includegraphics[width=0.47\textwidth]{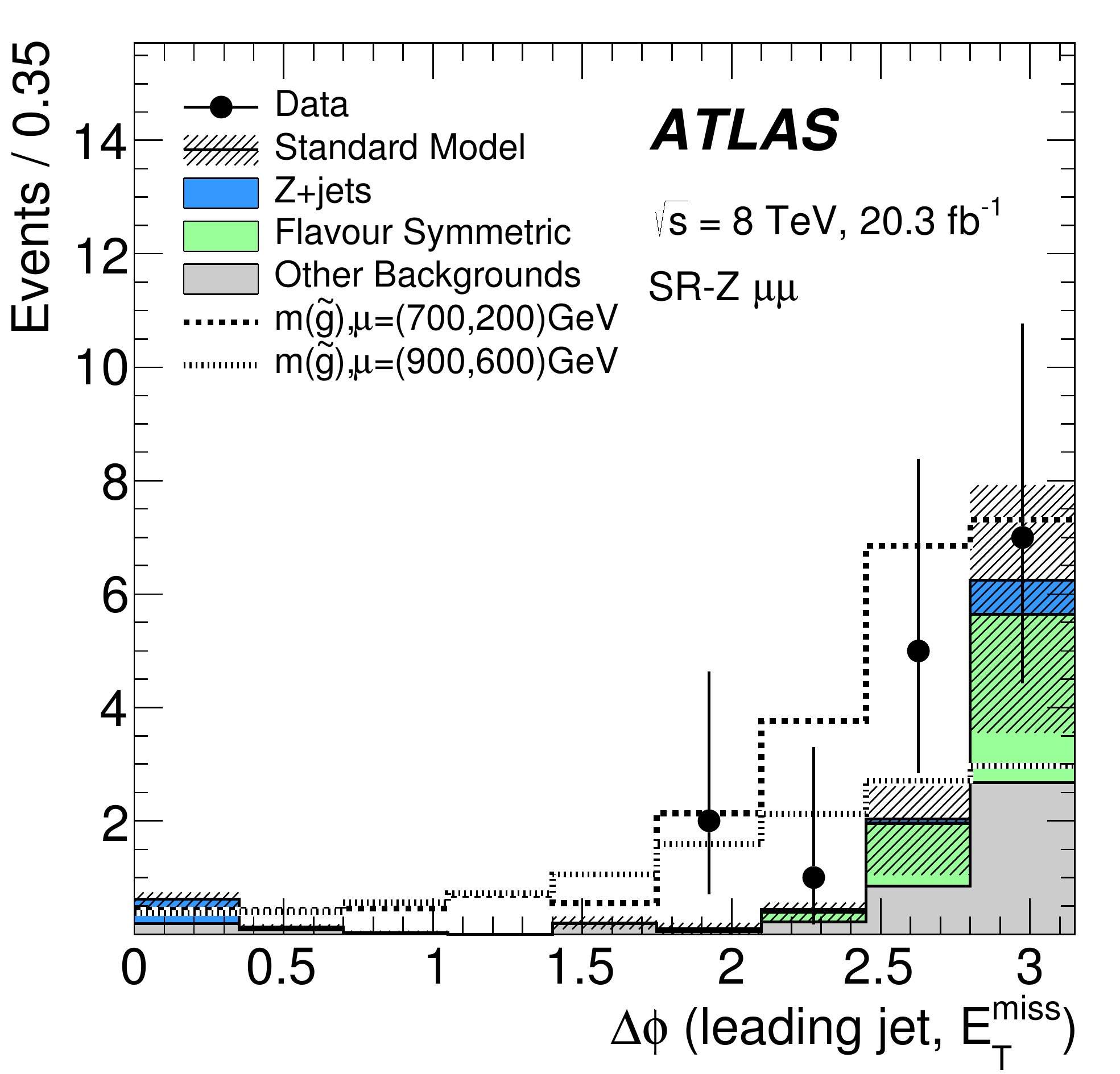}
\includegraphics[width=0.47\textwidth]{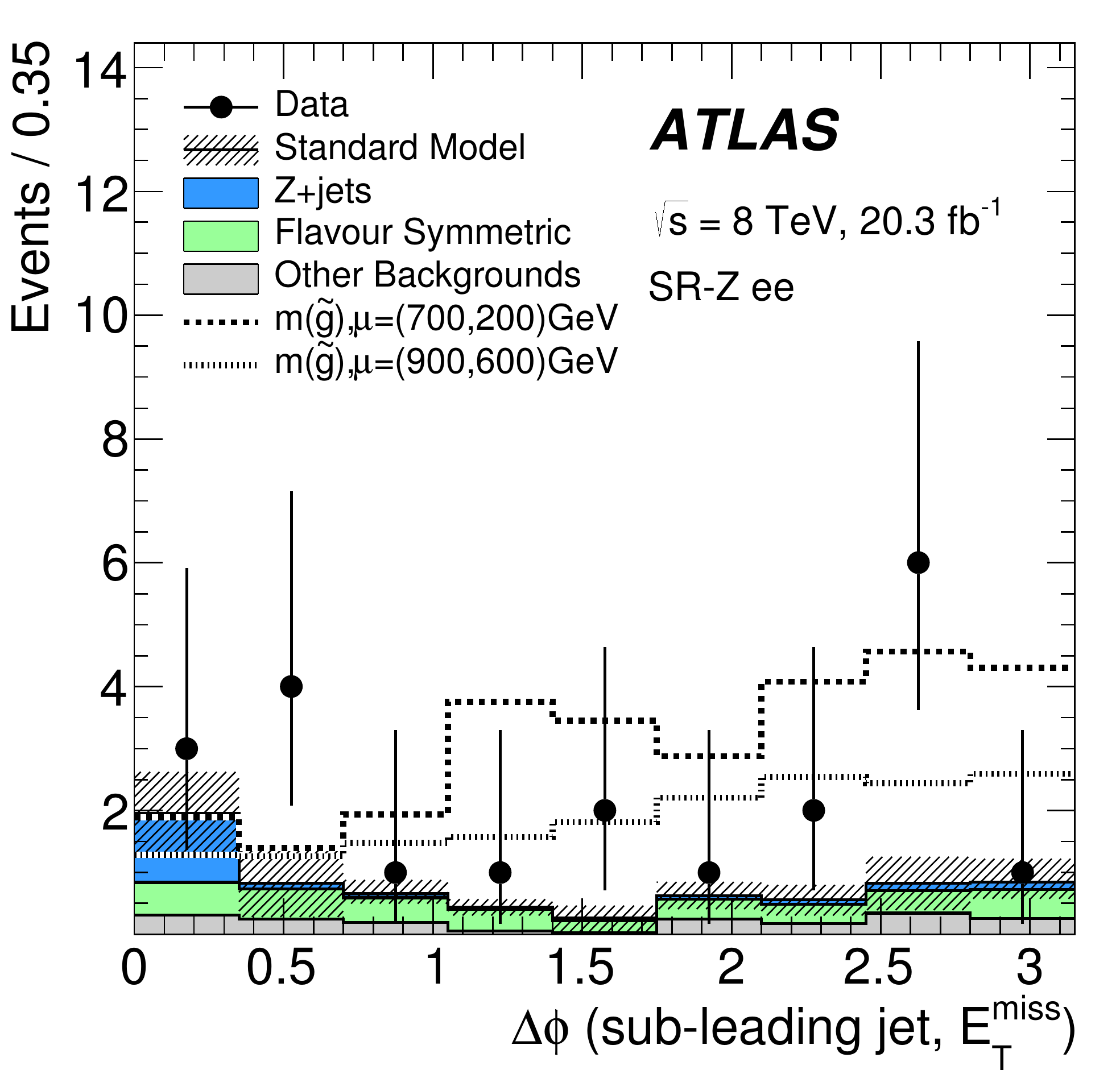}
\includegraphics[width=0.47\textwidth]{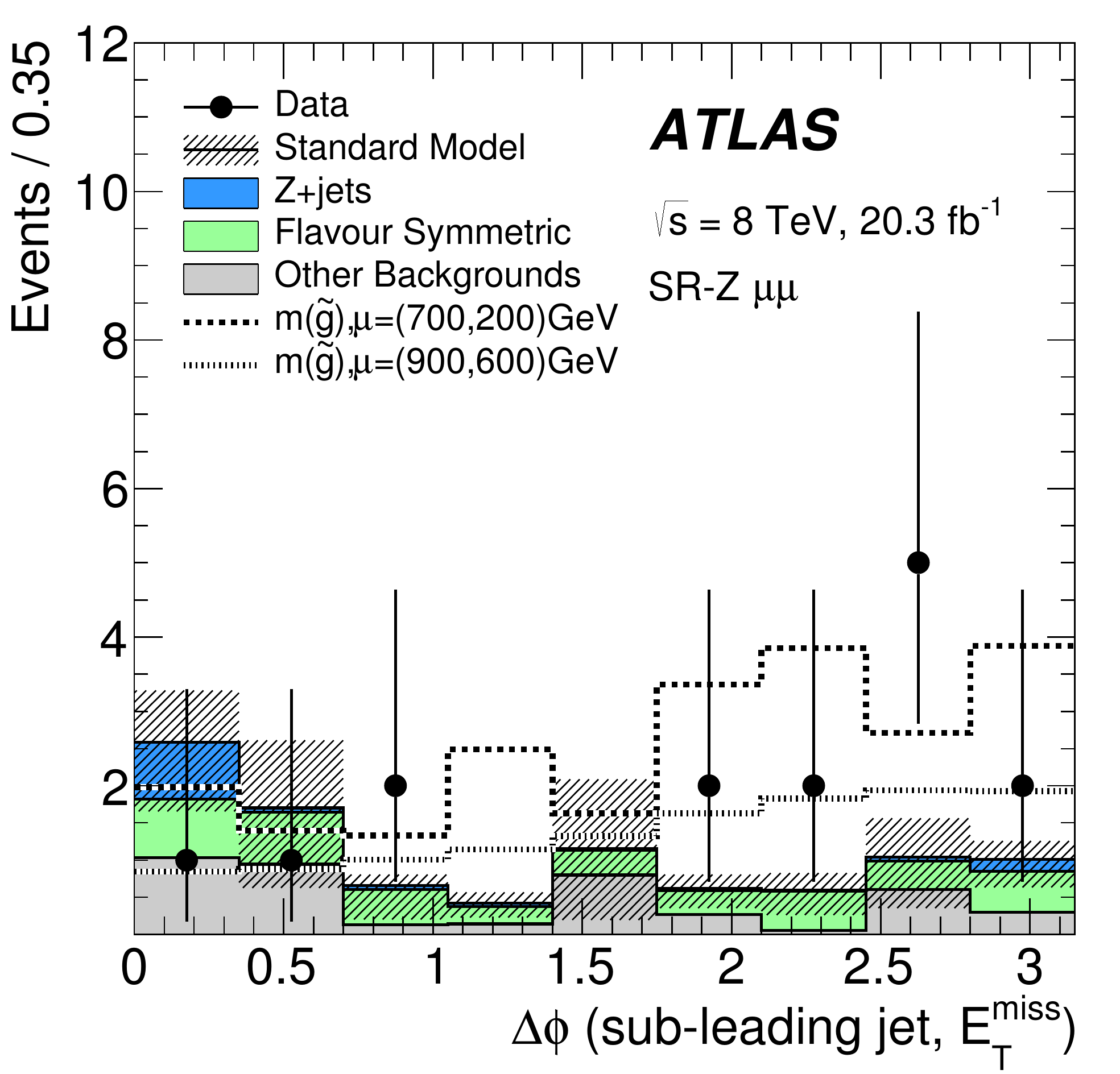}
\caption{The distribution of the $\Delta \phi$ between the leading jet and \met\ (top) and the sub-leading jet and \met\ (bottom) for the electron (left) and muon (right) channel in the on-$Z$ SRs before having applied the requirement $\Delta\phi (\text{jet}_{1,2},\MET)>0.4$. All uncertainties are included in the hatched uncertainty band.
Two example GGM ($\tan\beta=1.5$) signal models are overlaid. 
The backgrounds due to $WZ$, $ZZ$ or rare top processes, as well as from fake leptons, are included under ``Other Backgrounds''.
\label{fig:jetmet}}
\end{figure*}

For the off-$Z$ search, the dilepton mass distributions in the five SRs are presented in Figs.~\ref{fig:edge-SR1} and \ref{fig:edge-SR2}, and summarised in Fig.~\ref{fig:edgeresults}. 
The expected backgrounds and observed yields in the below-$Z$ and above-$Z$ regions for SR-2j-bveto, SR-4j-bveto, and SR-loose
are presented in Tables \ref{tab:SR2jbveto}, \ref{tab:SR4jbveto}, and \ref{tab:SRCMS}, respectively.
Corresponding results for SR-2j-btag and SR-4j-btag are presented in~\ref{app:edge}.
The data are consistent with the expected SM backgrounds in all regions. 
In the SR-loose region with $20<\mll<70$ \GeV, similar to the region in which the 
CMS Collaboration observed a 2.6$\sigma$ excess, 1133 events are observed, compared to an expectation of $1190\pm40\pm70$ events.

\begin{figure*}[!htbp]
\centering
\includegraphics[width=0.44\textwidth]{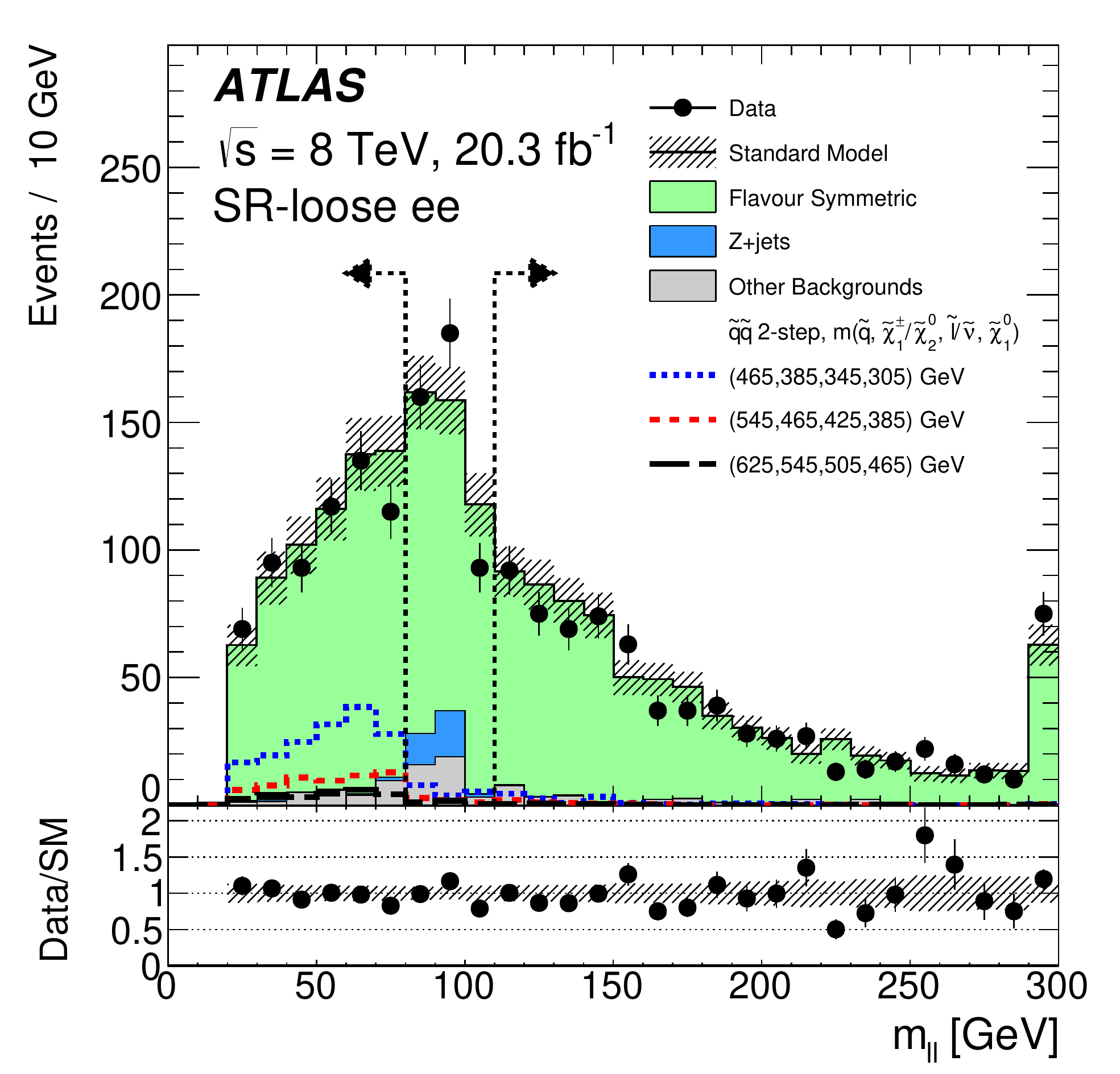}
\includegraphics[width=0.44\textwidth]{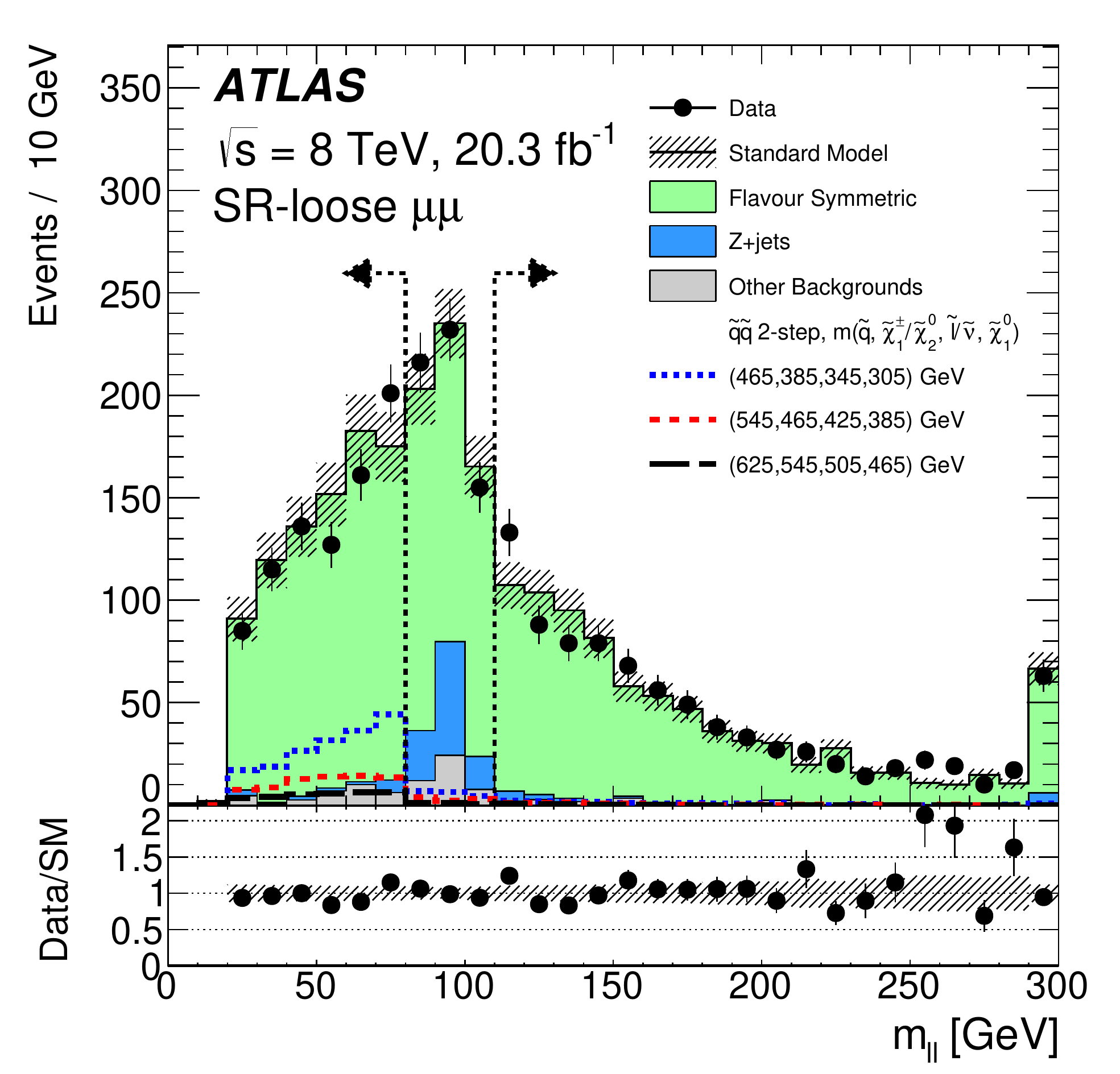}
\includegraphics[width=0.44\textwidth]{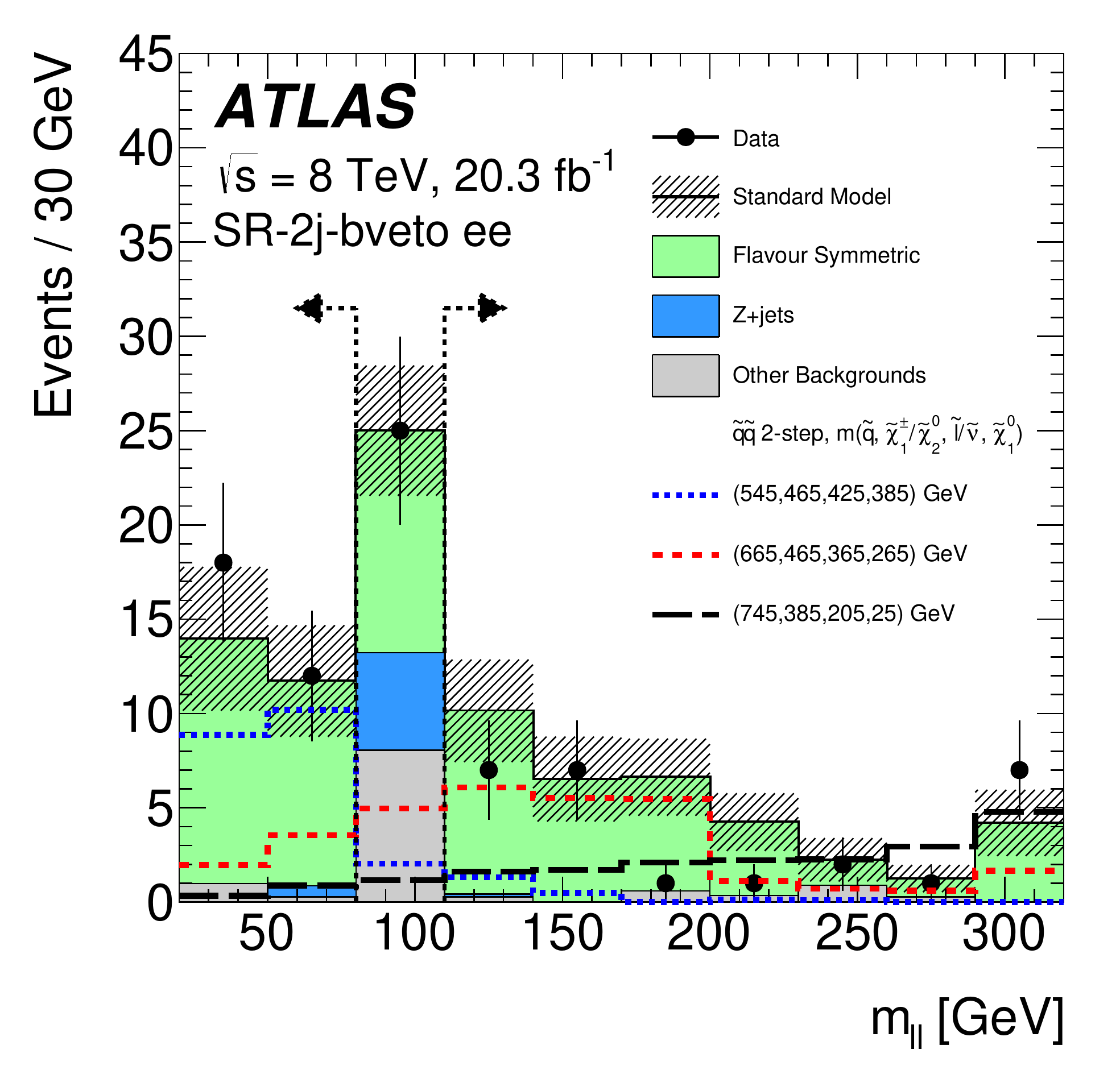}
\includegraphics[width=0.44\textwidth]{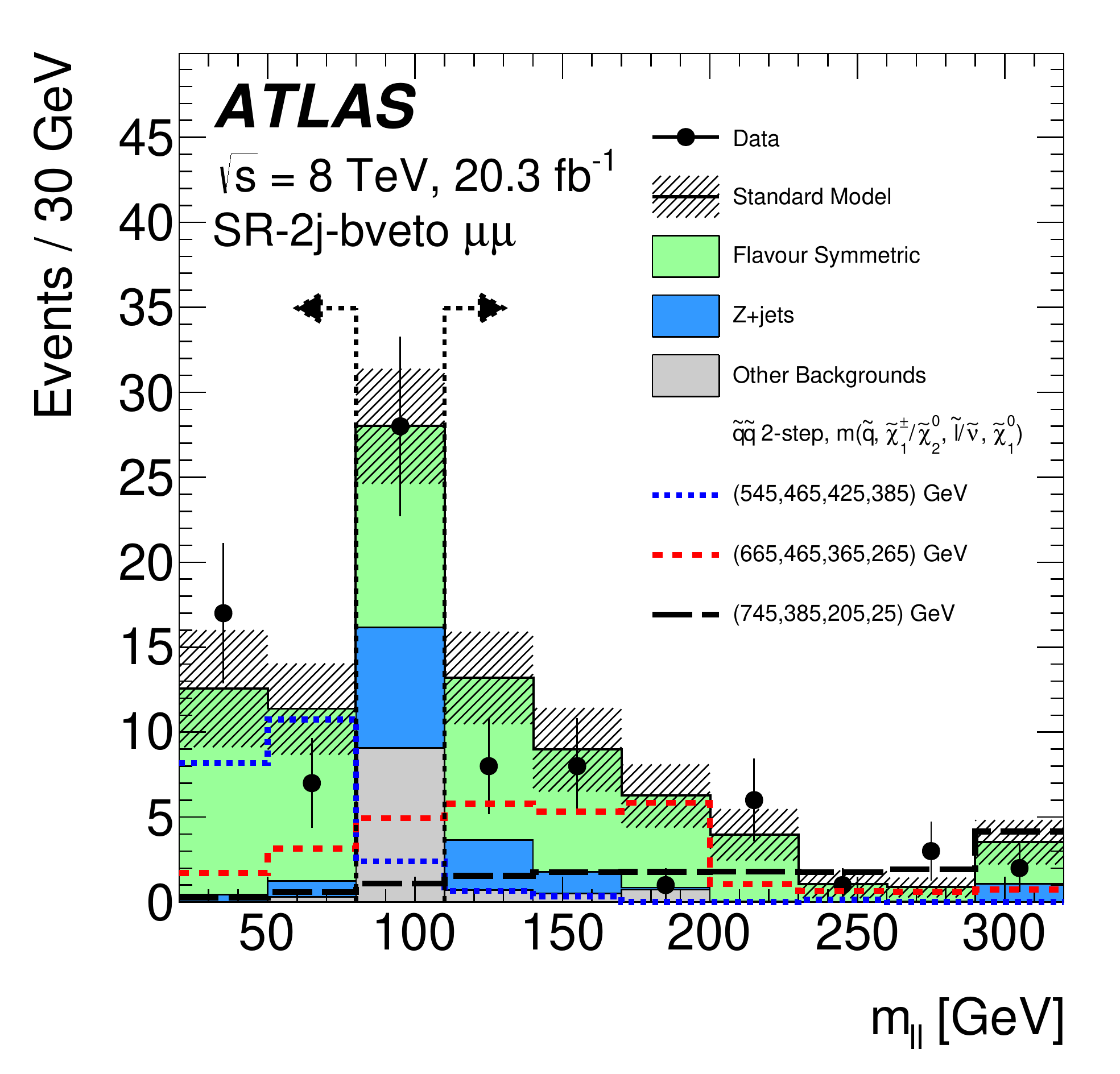}
\includegraphics[width=0.44\textwidth]{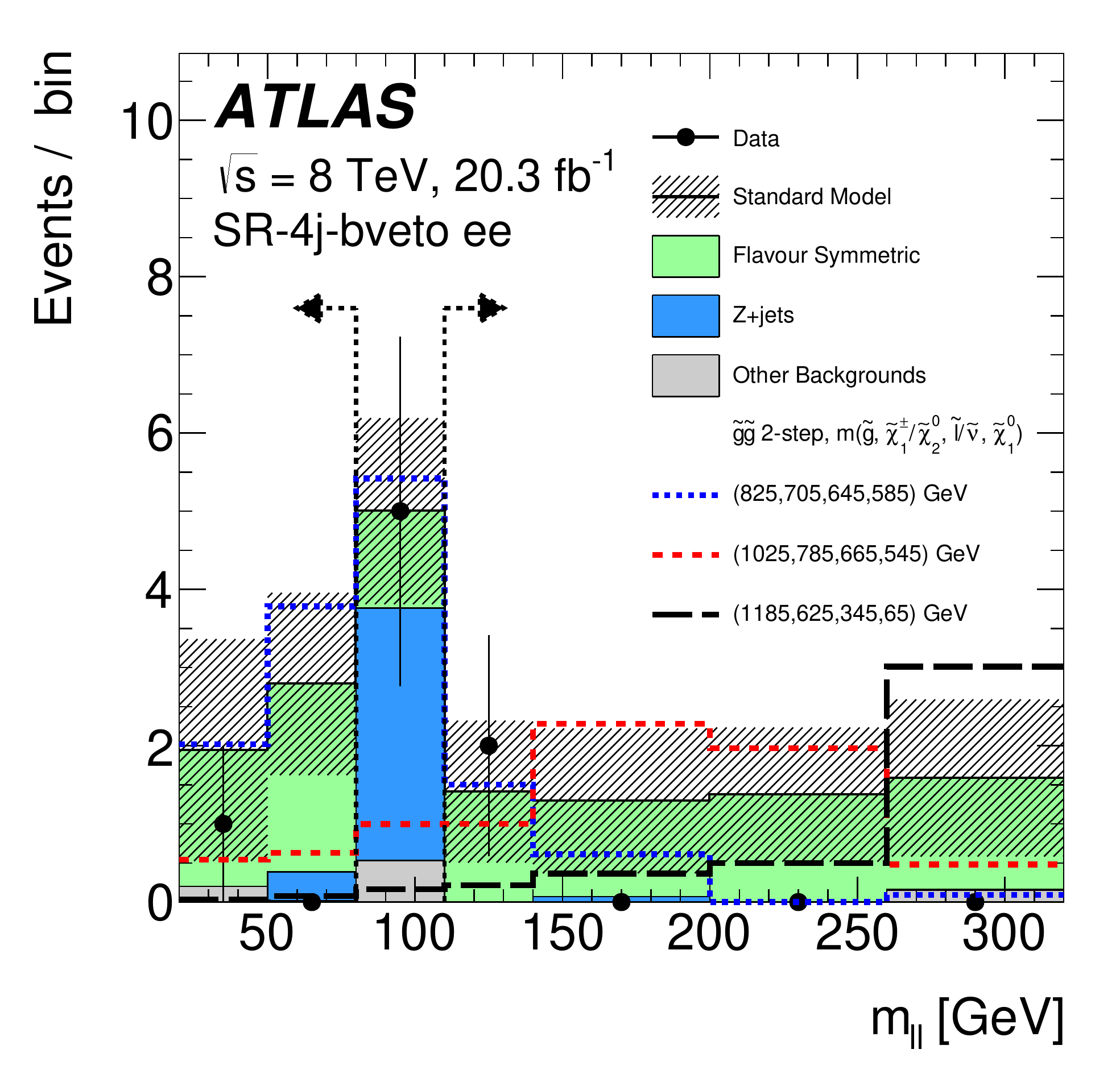}
\includegraphics[width=0.44\textwidth]{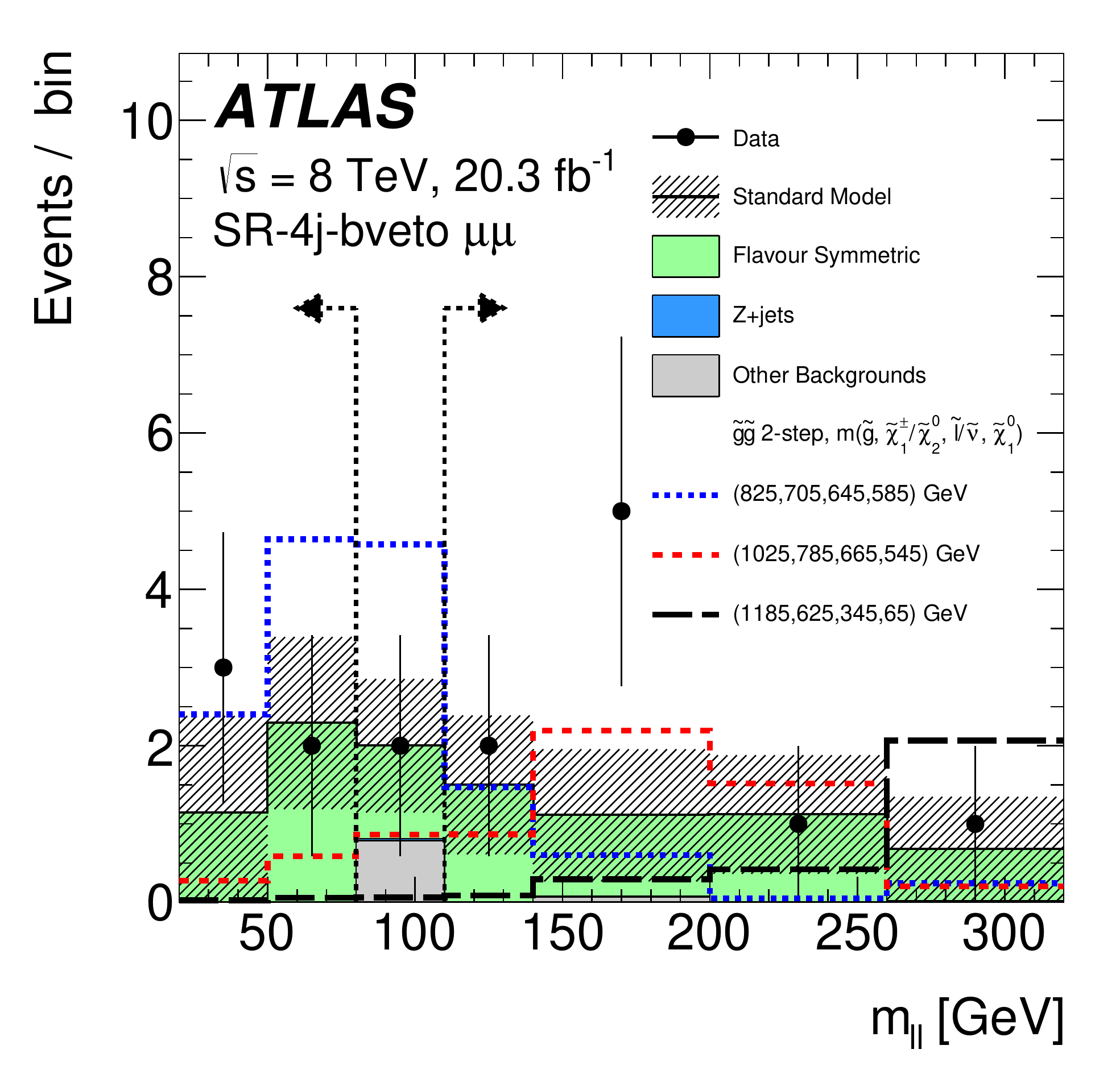}
\caption{ The observed and expected dilepton mass distributions in the off-$Z$ SR-loose (top), SR-2j-bveto (middle), and SR-4j-bveto (bottom). 
The vertical dashed lines indicate the $80<\mll<110$~\GeV\ region, which is used to normalise the \zjets\ background and is thus
not treated as a search region.
Example signal models (dashed lines) are overlaid, with $m(\tilde{q})/m(\tilde{g})$, $m(\tilde{\chi}^{0}_{2})/m(\tilde{\chi}^{\pm}_{1})$, $m(\tilde{\ell})/m(\tilde{\nu})$, and $m(\tilde{\chi}^{0}_{1})$ of each benchmark point being indicated in the figure legend. 
The last bin contains the overflow. 
All uncertainties are included in the hatched uncertainty band. 
\label{fig:edge-SR1}}
\end{figure*}

\clearpage

\begin{figure*}[!htbp]
\centering
\includegraphics[width=0.44\textwidth]{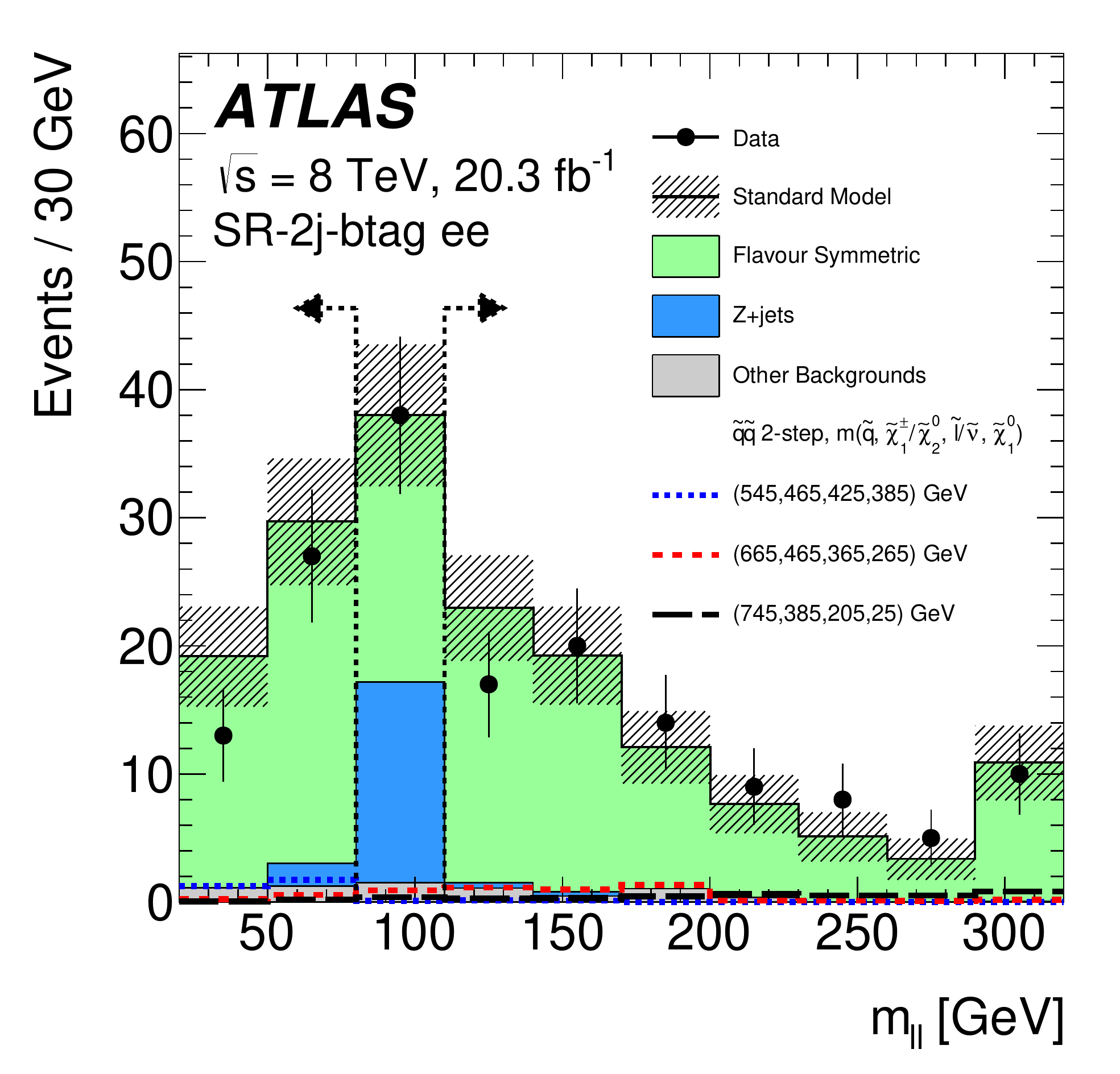}
\includegraphics[width=0.44\textwidth]{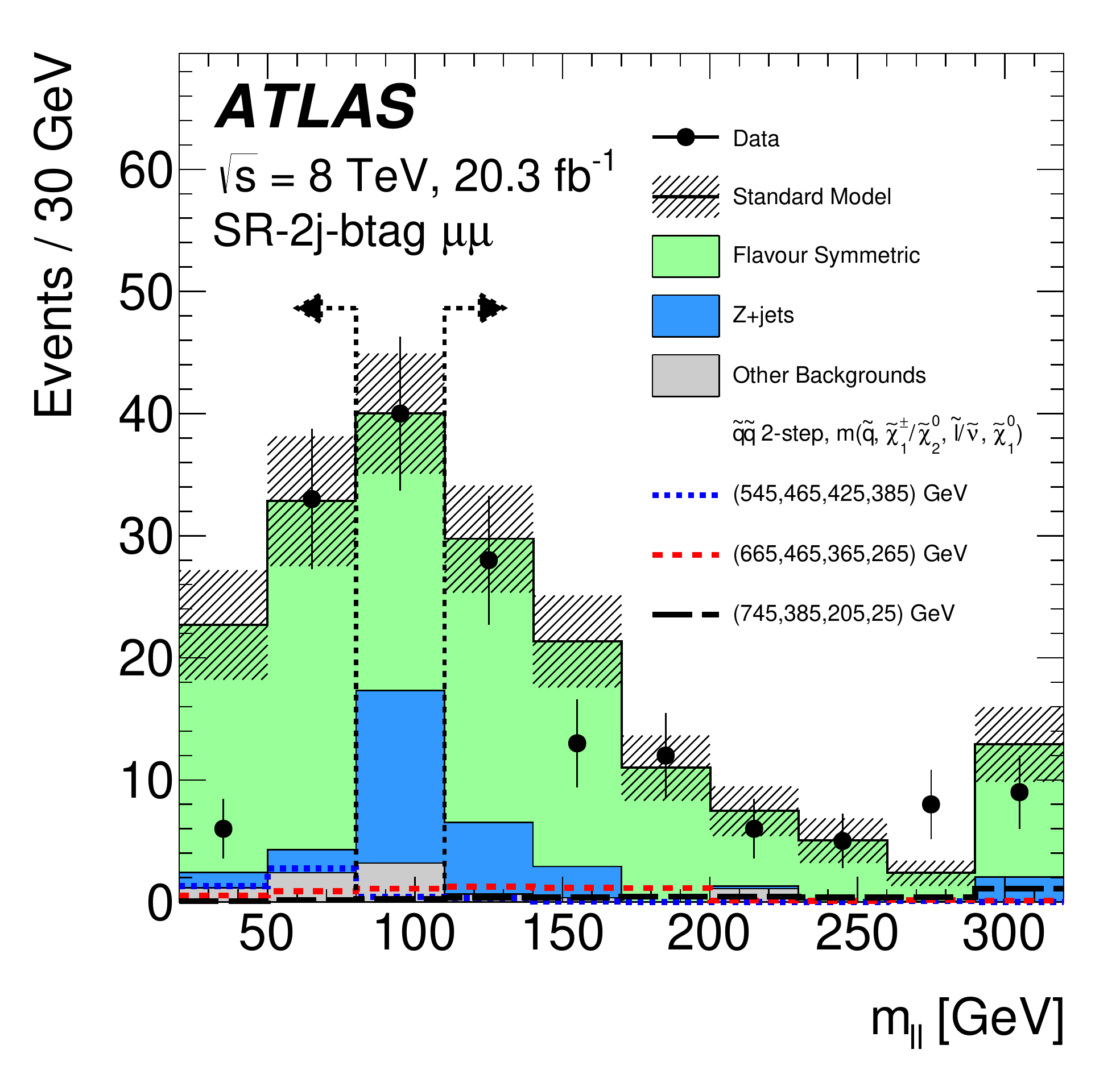}
\includegraphics[width=0.44\textwidth]{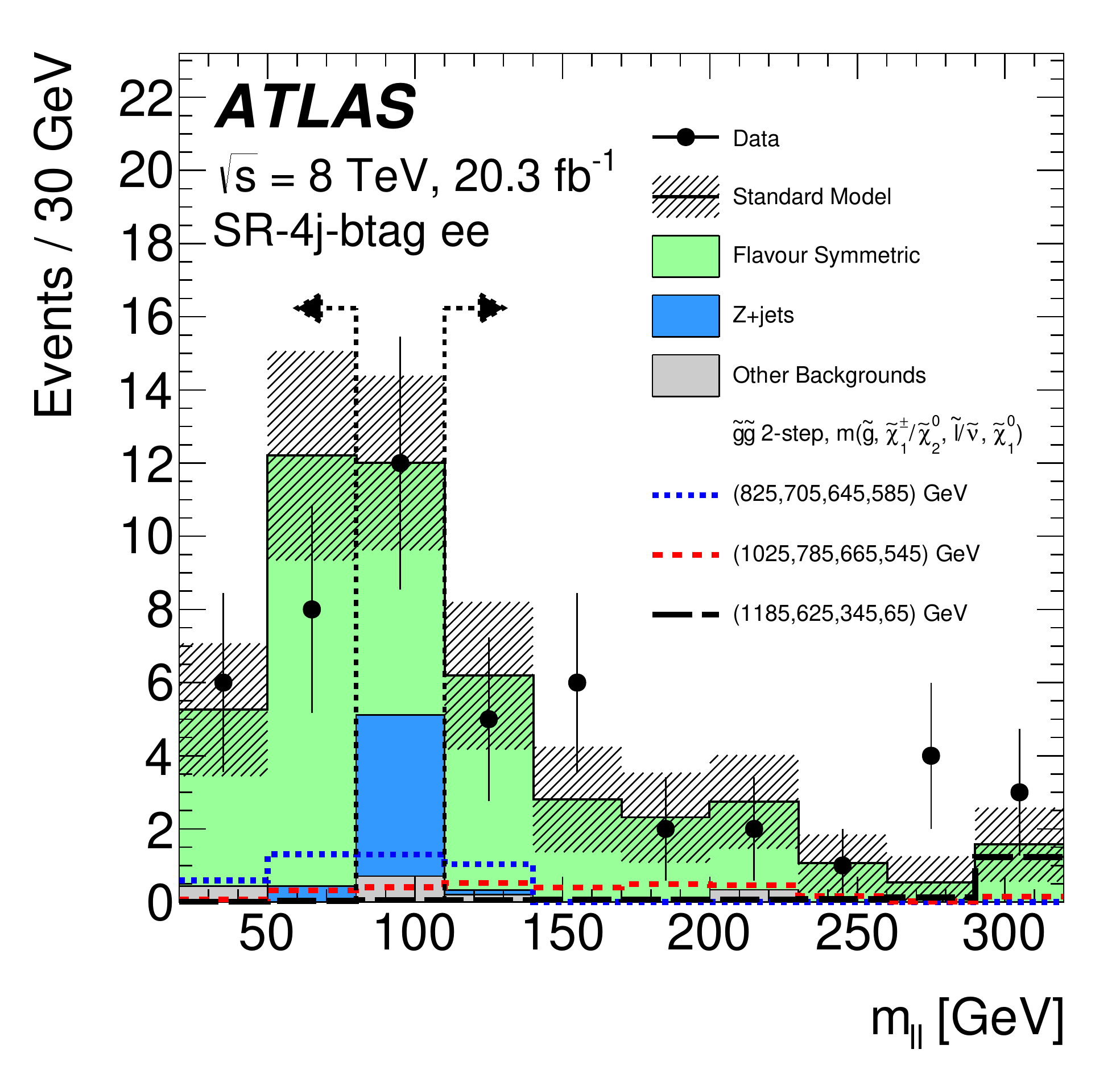}
\includegraphics[width=0.44\textwidth]{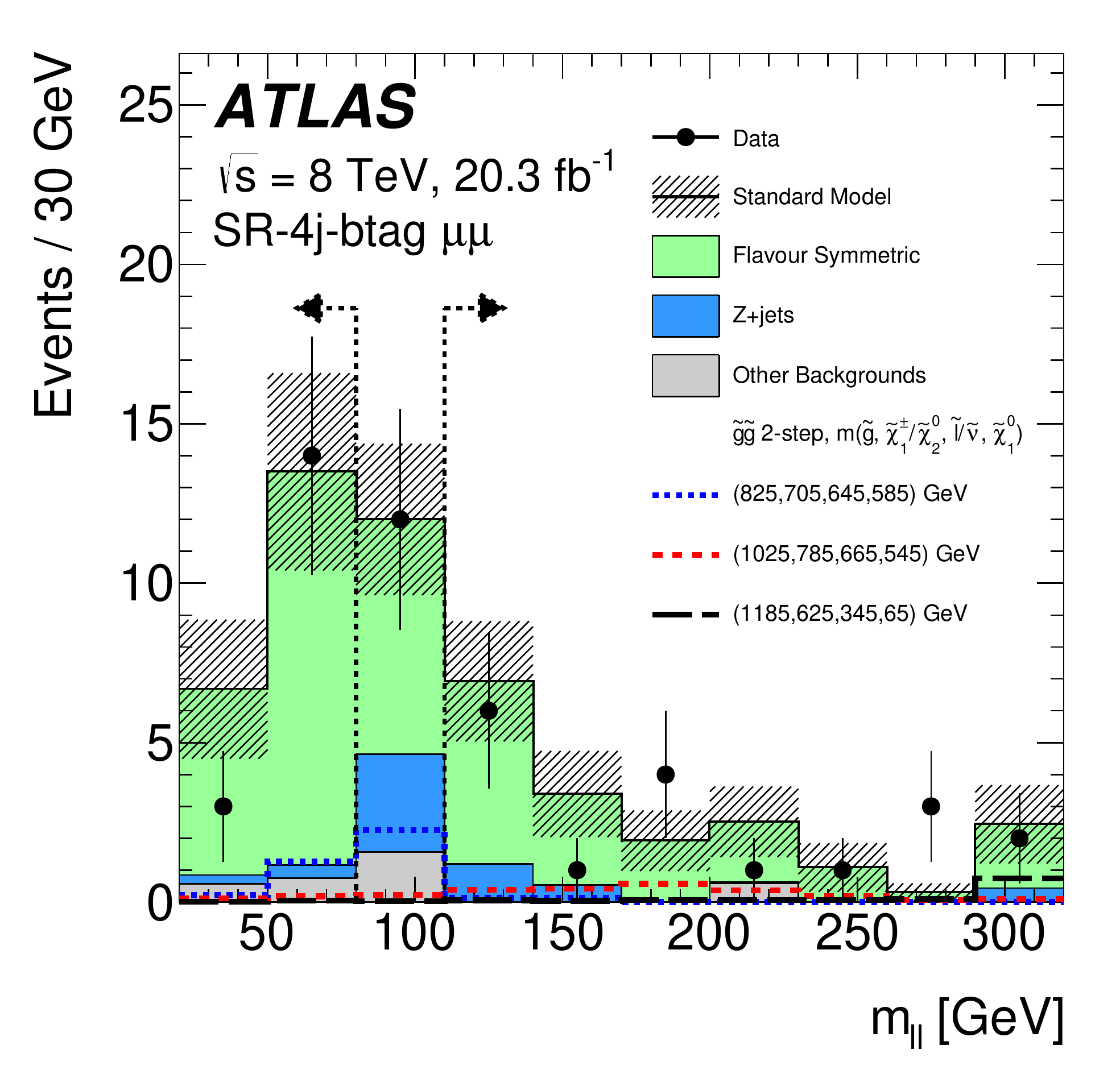}
\caption{ The observed and expected dilepton mass distributions in the SR-2j-btag (top) and SR-4j-btag (bottom) signal regions
of the off-$Z$ search. 
The vertical dashed lines indicate the $80<\mll<110$~\GeV\ region, which is used to normalise the \zjets\ background and is thus
not treated as a search region.
Example signal models of squark- or gluino-pair production (dashed lines) are overlaid, with $m(\tilde{g})$, $m(\tilde{\chi}^{0}_{2})/m(\tilde{\chi}^{\pm}_{1})$, $m(\tilde{\ell})/m(\tilde{\nu})$, and $m(\tilde{\chi}^{0}_{1})$ of each benchmark point being indicated in the figure legend. 
The last bin contains the overflow.
All uncertainties are included in the hatched uncertainty band. 
\label{fig:edge-SR2}}
\end{figure*}

\clearpage

\begin{figure*}[!htbp]
\centering
\includegraphics[width=01.0\textwidth]{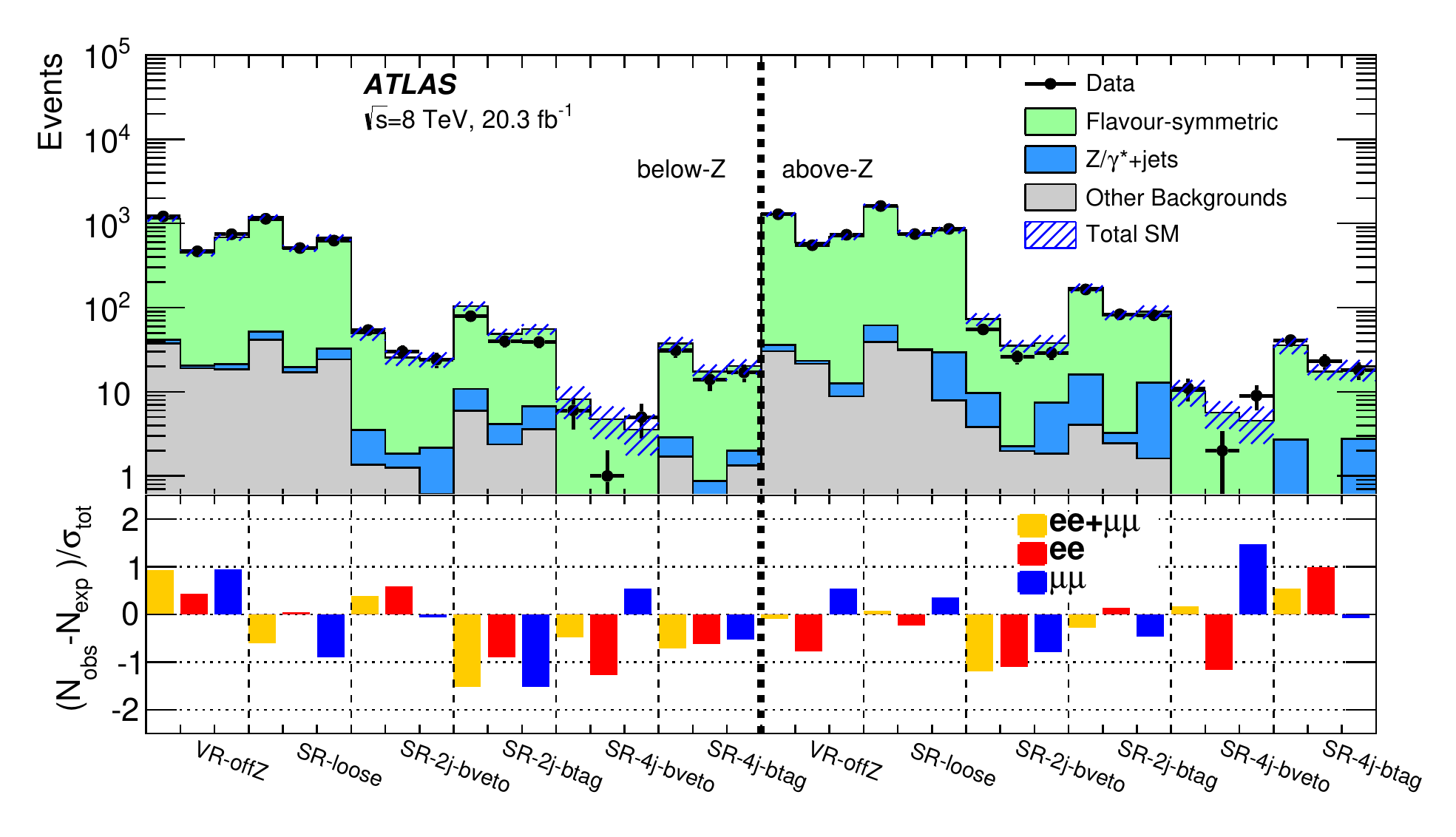}
\caption{ The observed and expected yields in the below-$Z$ (left) and above-$Z$ (right) dilepton mass regions, for the VR and five SRs of the off-$Z$ search.
Here below-$Z$ is $20<\mll<70$ \GeV\ for VR-offZ and SR-loose and otherwise $20<\mll<80$ \GeV; above-$Z$ is $\mll>110$ \GeV.
The bottom plot shows the difference in standard deviations between the observed and expected yields.
Results are shown for the $ee$ and $\mu\mu$ channels as well as for the sum.
\label{fig:edgeresults}}
\end{figure*}

\begin{table*}[h]
\begin{center}
\setlength{\tabcolsep}{0.0pc}
{\small
\begin{tabular*}{\textwidth}{@{\extracolsep{\fill}}lrrr}
\noalign{\smallskip}\hline\noalign{\smallskip}
Below-$Z$ ($20<\mll<80$ \GeV)      & SR-2j-bveto $ee$   & SR-2j-bveto $\mu\mu$ & SR-2j-bveto same-flavour             \\[-0.05cm]
               &        &          & combined             \\[-0.05cm]
\noalign{\smallskip}\hline\noalign{\smallskip}
          Observed events   &                       30   &                       24   &                       54  \\
\noalign{\smallskip}\hline\noalign{\smallskip}
Expected background events   &     $26 \pm 4 \pm 3$   &     $24 \pm 4 \pm 3$   &     $50 \pm 8 \pm 5$  \\
\noalign{\smallskip}\hline\noalign{\smallskip}
Flavour-symmetric backgrounds   &     $24 \pm 4 \pm 3$   &     $22 \pm 4 \pm 3$   &     $46 \pm 8 \pm 4$  \\
                  \dyjets   &      $0.6\pm 0.3\pm 0.7$   &      $1.6\pm 0.6\pm 1.4$   &      $2.2\pm 0.7\pm 1.7$  \\
                 Rare top   &      $<0.1$   &      $<0.1$   &      $<0.1$  \\
            $WZ/ZZ$ diboson   &      $0.6\pm 0.2\pm 0.1$   &      $0.6\pm 0.2\pm 0.2$   &      $1.2\pm 0.3\pm 0.2$  \\
             Fake leptons   &      $0.6\pm 0.9\pm 0.1$   &                    $<0.1$   &      $0.2\pm 0.9\pm 0.1$  \\
\noalign{\smallskip}\hline\noalign{\smallskip}
\noalign{\smallskip}\hline\noalign{\smallskip}
Above-$Z$ ($\mll>110$ \GeV)      & SR-2j-bveto $ee$   & SR-2j-bveto $\mu\mu$ & SR-2j-bveto same-flavour             \\[-0.05cm]
               &        &          & combined             \\[-0.05cm]
\noalign{\smallskip}\hline\noalign{\smallskip}
          Observed events   &                       26   &                       29   &                       55  \\
\noalign{\smallskip}\hline\noalign{\smallskip}
Expected background events   &     $35 \pm 5 \pm 4$   &     $38 \pm 4 \pm 8$   &     $73 \pm 9 \pm 9$  \\
\noalign{\smallskip}\hline\noalign{\smallskip}
Flavour-symmetric backgrounds   &     $33 \pm 4 \pm 4$   &     $30 \pm 4 \pm 3$   &     $63 \pm 8 \pm 5$  \\
                  \dyjets   &      $0.3\pm 0.2\pm 0.3$   &      $5.6\pm 0.6\pm 7.5$   &      $5.9\pm 0.7\pm 7.5$  \\
                 Rare top   &      $<0.1$   &      $<0.1$   &      $<0.1$  \\
            $WZ/ZZ$ diboson   &      $0.3\pm 0.1\pm 0.1$   &      $0.6\pm 0.2\pm 0.1$   &      $0.8\pm 0.2\pm 0.1$  \\
             Fake leptons   &      $1.7\pm 1.1\pm 0.2$   &      $1.3\pm 1.1\pm 0.5$   &      $3.0\pm 1.5\pm 0.4$  \\

\noalign{\smallskip}\hline\noalign{\smallskip}
\end{tabular*}
}
\end{center}
\caption{{\small Results in the off-$Z$ search region SR-2j-bveto, in the below-$Z$ range ($20<\mll<80$ \GeV, top) and above-$Z$ range ($\mll>110$ \GeV, bottom).
The flavour symmetric, \dyjets\ and fake lepton background components are all derived using data-driven estimates described in the text.
All other backgrounds are taken from MC simulation.
The first uncertainty is statistical and the second is systematic.}}

\label{tab:SR2jbveto}
\end{table*}

\begin{table*}[h]
\begin{center}
\setlength{\tabcolsep}{0.0pc}
{\small
\begin{tabular*}{\textwidth}{@{\extracolsep{\fill}}lrrr}
\noalign{\smallskip}\hline\noalign{\smallskip}
Below-$Z$ ($20<\mll<80$ \GeV)      & SR-4j-bveto $ee$   & SR-4j-bveto $\mu\mu$ & SR-4j-bveto same-flavour             \\[-0.05cm]
               &        &          & combined             \\[-0.05cm]
\noalign{\smallskip}\hline\noalign{\smallskip}
          Observed events   &                        1   &                        5   &                        6  \\
\noalign{\smallskip}\hline\noalign{\smallskip}
Expected background events   &      $4.7\pm 1.6\pm 1.1$   &      $3.6\pm 1.5\pm 1.0$   &      $8.2\pm 3.1\pm 1.4$  \\
\noalign{\smallskip}\hline\noalign{\smallskip}
Flavour-symmetric backgrounds   &      $4.1\pm 1.6\pm 1.1$   &      $3.5\pm 1.5\pm 1.0$   &      $7.7\pm 3.1\pm 1.3$  \\
                  \dyjets   &      $0.4\pm 0.2\pm 0.3$   &      $0.0\pm 0.0\pm 0.4$   &      $0.4\pm 0.2\pm 0.5$  \\
                 Rare top   &      $<0.1$   &      $<0.1$   &      $<0.1$  \\
            $WZ/ZZ$ diboson   &      $<0.1$   &      $<0.1$   &      $<0.1$  \\
             Fake leptons   &      $0.2\pm 0.3\pm 0.0$   &                    $<0.1$   &      $0.1\pm 0.3\pm 0.0$  \\
\noalign{\smallskip}\hline\noalign{\smallskip}
\noalign{\smallskip}\hline\noalign{\smallskip}
Above-$Z$ ($\mll>110$ \GeV)      & SR-4j-bveto $ee$   & SR-4j-bveto $\mu\mu$ & SR-4j-bveto same-flavour             \\[-0.05cm]
               &        &          & combined             \\[-0.05cm]
\noalign{\smallskip}\hline\noalign{\smallskip}
          Observed events   &                        2   &                        9   &                       11  \\
\noalign{\smallskip}\hline\noalign{\smallskip}
Expected background events   &      $5.7\pm 1.6\pm 1.2$   &      $4.5\pm 1.3\pm 1.7$   &     $10 \pm 3 \pm 2$  \\
\noalign{\smallskip}\hline\noalign{\smallskip}
Flavour-symmetric backgrounds   &      $5.5\pm 1.6\pm 1.2$   &      $4.3\pm 1.3\pm 1.0$   &      $9.8\pm 2.9\pm 1.4$  \\
                  \dyjets   &      $0.2\pm 0.1\pm 0.1$   &      $0.0\pm 0.0\pm 1.3$   &      $0.2\pm 0.1\pm 1.3$  \\
                 Rare top   &      $<0.1$   &      $<0.1$   &      $<0.1$  \\
            $WZ/ZZ$ diboson   &      $<0.1$   &      $0.2\pm 0.1\pm 0.0$   &      $0.2\pm 0.1\pm 0.0$  \\
             Fake leptons   &                    $<0.2$   &                    $<0.1$   &                    $<0.2$  \\
\noalign{\smallskip}\hline\noalign{\smallskip}
\end{tabular*}
}
\end{center}
\caption{{\small Results in the off-$Z$ search region SR-4j-bveto, in the below-$Z$ range ($20<\mll<80$ \GeV, top) and above-$Z$ range ($\mll>110$ \GeV, bottom).
Details are the same as in Table~\ref{tab:SR2jbveto}.}}
\label{tab:SR4jbveto}
\end{table*}

\begin{table*}[h]
\begin{center}
\setlength{\tabcolsep}{0.0pc}
{\small
\begin{tabular*}{\textwidth}{@{\extracolsep{\fill}}lrrr}
\noalign{\smallskip}\hline\noalign{\smallskip}
Below-$Z$ ($20<\mll<70$ \GeV)      & SR-loose $ee$   & SR-loose $\mu\mu$ & SR-loose same-flavour             \\[-0.05cm]
               &        &          & combined             \\[-0.05cm]
\noalign{\smallskip}\hline\noalign{\smallskip}
          Observed events   &                      509   &                      624   &                     1133  \\
\noalign{\smallskip}\hline\noalign{\smallskip}
Expected background events   &  $510 \pm 20 \pm 40$        &  $680 \pm 20 \pm 50$       & $1190 \pm 40 \pm 70$  \\
\noalign{\smallskip}\hline\noalign{\smallskip}
Flavour-symmetric backgrounds   &  $490 \pm 20 \pm 40$   &  $650 \pm 20\pm 50$   & $1140 \pm 40 \pm 70$  \\
                  \dyjets   &      $2.5\pm 0.8\pm 3.2$   &      $8\pm 2\pm 5$   &     $11\pm 2\pm 7$  \\
                 Rare top   &      $0.3\pm 0.0\pm 0.0$   &      $0.4\pm 0.0\pm 0.0$   &      $0.7\pm 0.0\pm 0.0$  \\
                    $WZ/ZZ$   &      $1.1\pm 0.3\pm 0.1$   &      $1.2\pm 0.2\pm 0.4$   &      $2.4\pm 0.4\pm 0.4$  \\
             Fake leptons   &     $16 \pm 4 \pm 2$       &     $23 \pm 5 \pm 1$       &     $38 \pm 6 \pm 4$  \\
\noalign{\smallskip}\hline\noalign{\smallskip}
\noalign{\smallskip}\hline\noalign{\smallskip}
Above-$Z$ ($\mll>110$ \GeV)      & SR-loose $ee$   & SR-loose $\mu\mu$ & SR-loose same-flavour             \\[-0.05cm]
               &        &          & combined             \\[-0.05cm]
\noalign{\smallskip}\hline\noalign{\smallskip}
          Observed events   &                      746   &                      859   &                     1605  \\
\noalign{\smallskip}\hline\noalign{\smallskip}
Expected background events   &  $760 \pm 20 \pm 60$   &  $830 \pm 20 \pm 70$   & $1600 \pm 40 \pm 100$  \\
\noalign{\smallskip}\hline\noalign{\smallskip}
Flavour-symmetric backgrounds   &  $730 \pm 20 \pm 60$   &  $800 \pm 20 \pm 60$   & $1500 \pm 40 \pm 100$  \\
                  \dyjets   &      $0.9\pm 0.2\pm 1.1$   &    $21\pm 3\pm 24$   &    $22\pm 3\pm 24$  \\
                 Rare top   &      $0.2\pm 0.0\pm 0.0$   &      $0.2\pm 0.0\pm 0.0$   &      $0.4\pm 0.0\pm 0.0$  \\
            $WZ/ZZ$ diboson   &      $0.6\pm 0.2\pm 0.2$   &      $1.0\pm 0.2\pm 0.1$   &      $1.6\pm 0.3\pm 0.2$  \\
             Fake leptons   &     $30 \pm 5 \pm 5$       &      $6.7\pm 3.7\pm 1.7$   &     $37 \pm 6 \pm 5$  \\
\noalign{\smallskip}\hline\noalign{\smallskip}
\end{tabular*}
}
\end{center}
\caption{{\small Results in the off-$Z$ search region SR-loose, in the below-$Z$ range ($20<\mll<70$ \GeV, top) and above-$Z$ range ($\mll>110$ \GeV, bottom).
Details are the same as in Table~\ref{tab:SR2jbveto}.
}}
\label{tab:SRCMS}
\end{table*}


\section{Interpretation of results}
\label{sec:interpretation}
In this section, exclusion limits are shown for the SUSY models described in Sect.~\ref{sec:data-selection}.
The asymptotic $CL_{\text{S}}$ prescription \cite{cls}, implemented in the HistFitter program \cite{Baak:2014wma}, is used
to determine upper limits at 95~\% confidence level (CL).
All signal and background uncertainties are
taken into account using a Gaussian model of nuisance parameter integration.
All uncertainties except that on the signal cross section are included in the limit-setting configuration. The impact
of varying the signal cross sections by their uncertainties is indicated separately. Numbers quoted in the text are
evaluated from the observed exclusion limit based on the nominal signal cross section minus its $1\sigma$ theoretical uncertainty.

For the on-$Z$ analysis, the data exceeds the background expectations in the $ee$ ($\mu\mu$) channel
with a significance of 3.0 (1.7) standard deviations.
Exclusion limits in specific models allow us to illustrate which regions of the model parameter space are 
affected by the observed excess, by comparing the expected and observed limits.
The results in SR-$Z$ $ee$ and SR-$Z$ $\mu\mu$ (Table~\ref{tab:zSR}) are considered simultaneously.
The signal contamination in CR$e\mu$ is found to be at the $\sim 1$~\% level, and is therefore neglected in this procedure. 
The expected and observed exclusion contours, in the plane of $\mu$ versus $m(\tilde{g})$ for the GGM model, are shown in Fig.~\ref{fig:GGMSF}. 
The $\pm1\sigma_{\mathrm{exp}}$ and  $\pm2\sigma_{\mathrm{exp}}$ experimental uncertainty bands indicate the impact on the expected limit of
all uncertainties considered on the background processes. 
The $\pm 1 \sigma_{\mathrm{theory}}^{\mathrm{SUSY}}$ uncertainty lines around the observed 
limit illustrate the change in the observed limit as the nominal signal cross section is scaled
up and down by the theoretical cross-section uncertainty. 
Given the observed excess of events with respect to the SM prediction, the observed limits are weaker than expected. 
In the case of the $\tan\beta=1.5$ exclusion contour, the on-$Z$ analysis is able to exclude gluino masses up to 850~\GeV\ for $\mu>450$~GeV,
whereas gluino masses of up to 820~\GeV\ are excluded for the $\tan\beta=30$ model for $\mu>600$~GeV. 
The lower exclusion reach for the $\tan\beta=30$ models is due to the fact that the branching fraction for $\tilde{\chi}^0_1 \rightarrow Z\tilde{G}$ is significantly smaller at $\tan\beta=30$ than at $\tan\beta=1.5$.

\begin{figure*}[ht]
\centering
\includegraphics[width=0.8\textwidth]{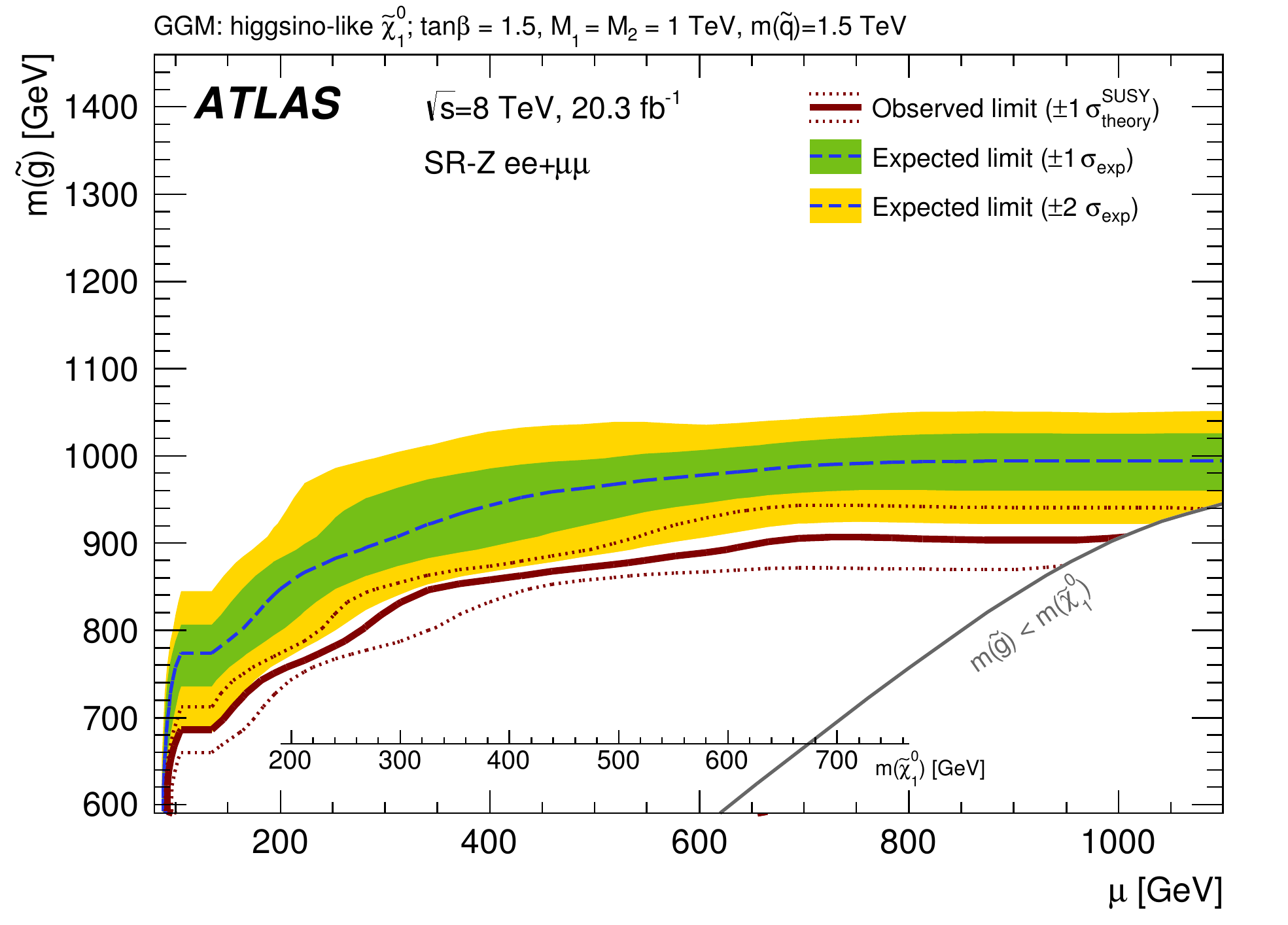}
\includegraphics[width=0.8\textwidth]{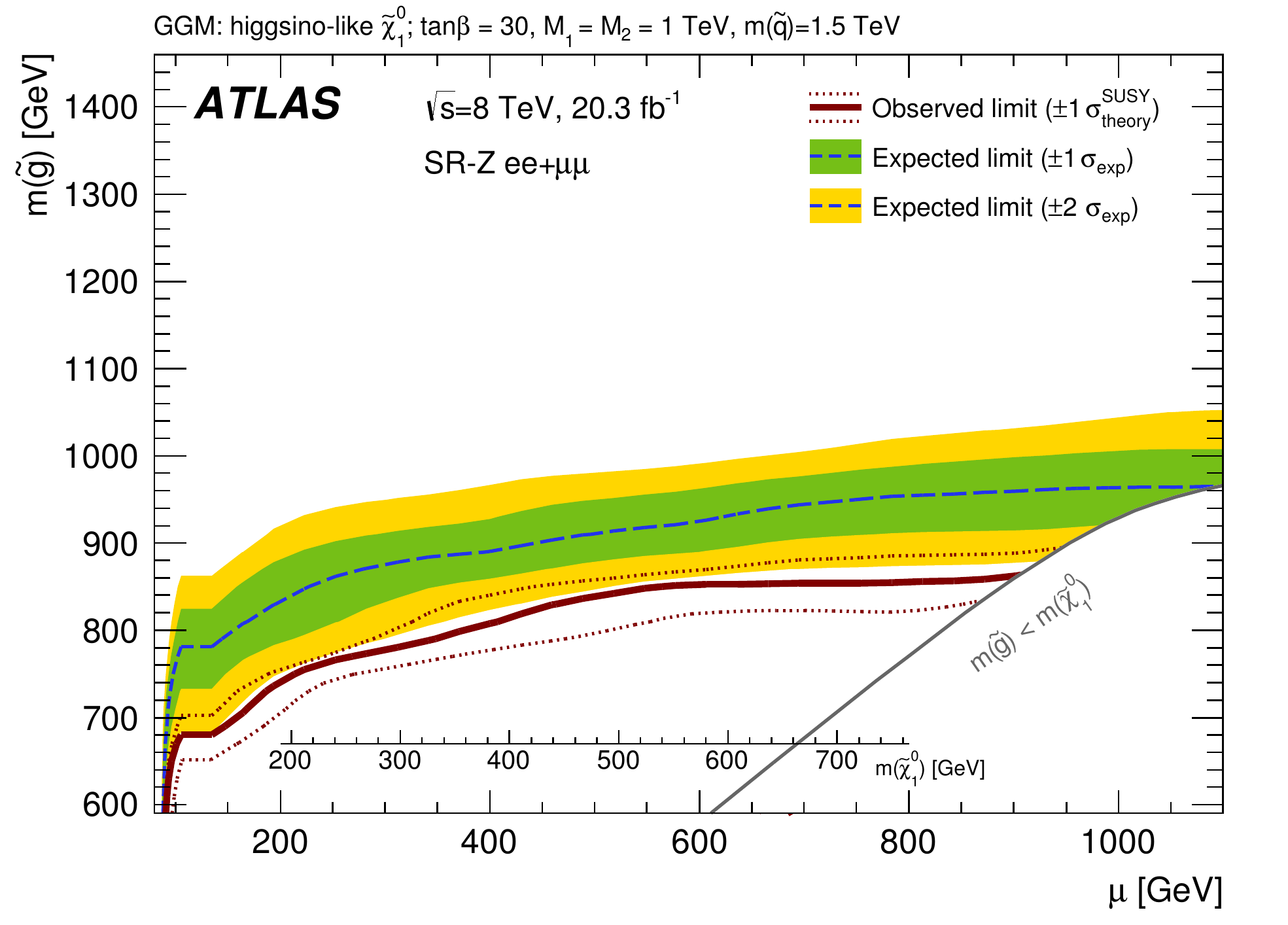}
\caption{{\small The 95~\% CL exclusion limit from the on-$Z$ combined same-flavour channels in the $\mu$ versus $m(\tilde{g})$ plane in the GGM model with $\tan\beta=1.5$ (top) and $\tan\beta=30$ (bottom). 
The dark blue dashed line indicates the expected limits at 95~\% CL and the green (yellow) bands show the $\pm 1 \sigma$ ($\pm 2 \sigma$) variation on the expected limit as a consequence of the experimental and theoretical uncertainties on the background prediction. 
The observed limits are shown by the solid red lines, with the dotted red lines indicating the limit obtained upon varying the signal cross section by $\pm 1 \sigma$.
The region below the grey line has the gluino mass less than the lightest neutralino mass and is hence not considered. The value of the lightest neutralino mass is indicated by the
$x$-axis inset. }}
\label{fig:GGMSF}
\end{figure*}

For the off-$Z$ search, the limits for the squark-pair (gluino-pair) model are based on the results of SR-2j-bveto (SR-4j-bveto).
The yields in the combined $ee$+$\mu\mu$ channels are used. Signal contamination in the $e\mu$ control region used for the flavour-symmetry method
is taken into account by subtracting the expected increase in the background prediction from the signal yields.
For each point in the signal model parameter space, limits on the signal strength are calculated using a ``sliding window'' approach.
The binning in SR-2j-bveto (SR-4j-bveto) defines 45 (21) possible dilepton mass windows to use for the squark-pair (gluino-pair) model interpretation,
of which the ten (nine) windows with the best expected sensitivity are selected.
For each point in the signal model parameter space, the dilepton mass window with the best expected
limit on the signal strength is selected. The excluded regions in the squark--LSP and gluino--LSP mass planes are shown in Fig.~\ref{fig:edgeint}.
The analysis probes squarks with masses up to 780 GeV, and gluinos with masses up to 1170 GeV.

\begin{figure*}[t]
  \begin{center}
     \includegraphics[width=0.8\textwidth]{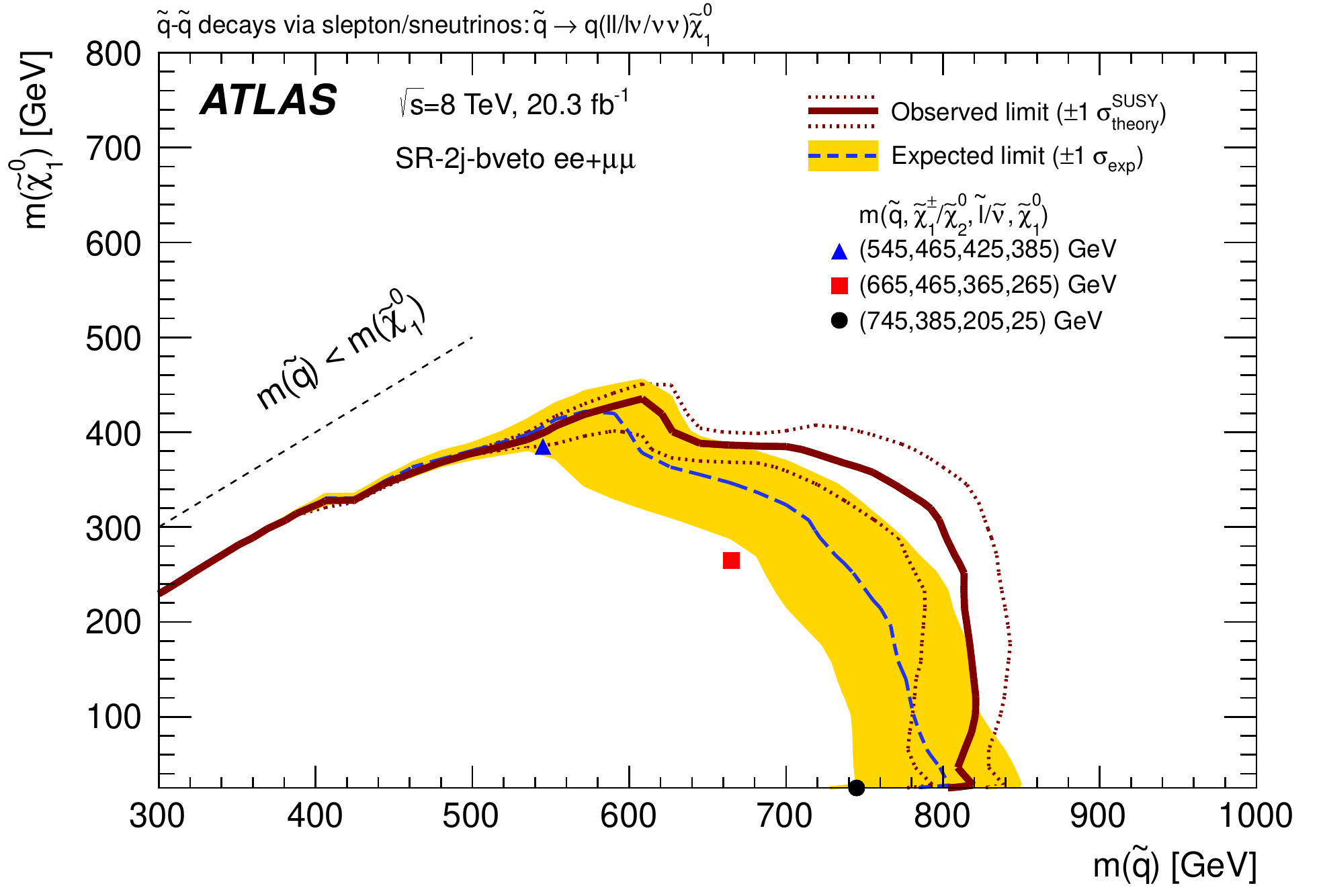}
    \includegraphics[width=0.8\textwidth]{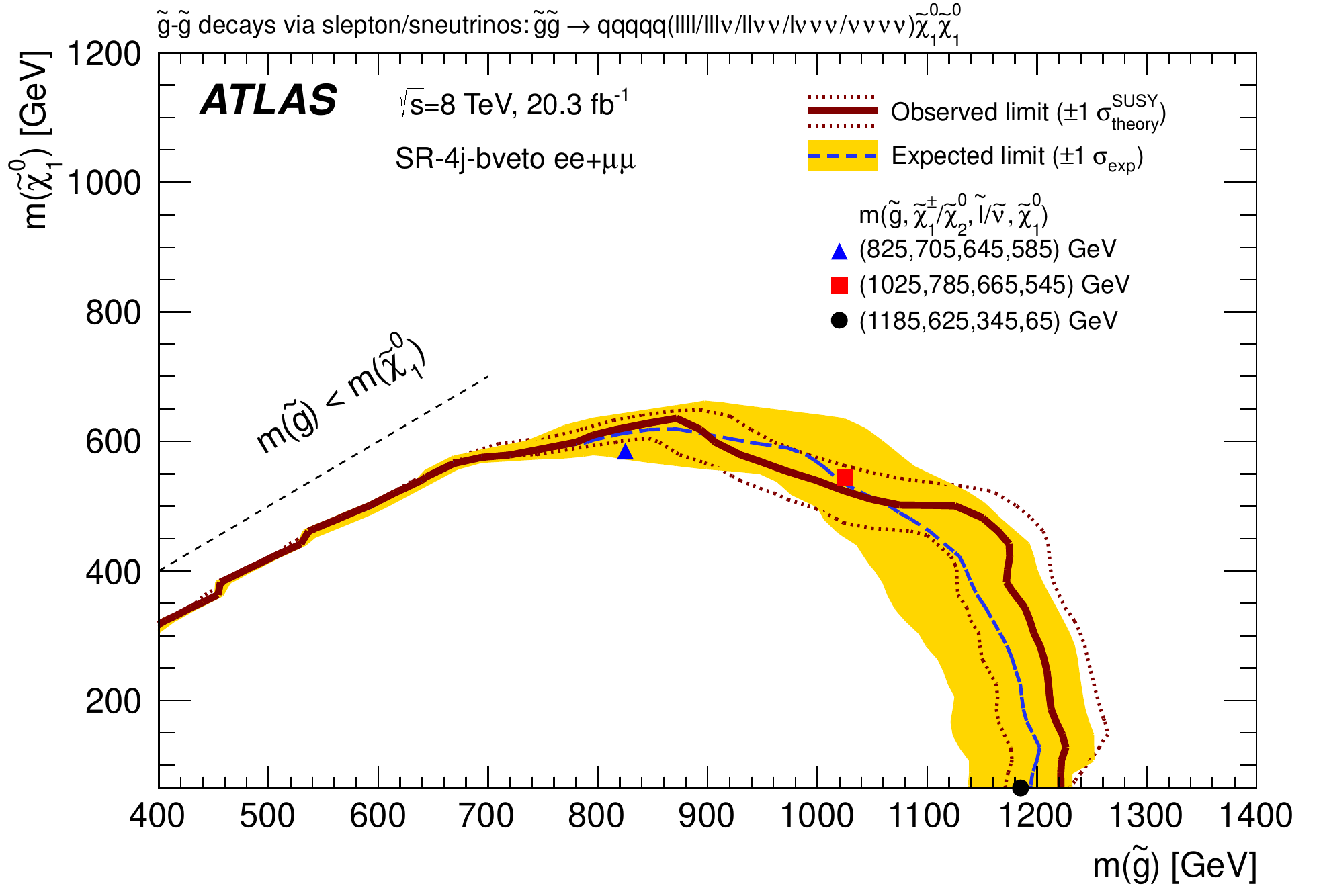}
  \end{center}
  \caption{\label{fig:edgeint} Excluded region in the (top) squark--LSP mass plane using the SR-2j-bveto results and (bottom) gluino--LSP mass plane
using the SR-4j-bveto results. The observed, expected, and $\pm1\sigma$ expected
exclusion contours are indicated. The observed limits obtained upon varying the signal cross section by $\pm 1 \sigma$ are also indicated.
The region to the left of the diagonal dashed line has the squark mass less than the LSP mass and is hence not considered. Three signal benchmark points are shown, with their SUSY particle masses indicated in parentheses. }
\end{figure*}

The signal regions in these analyses are also used to place upper limits on the allowed number of BSM events ($N_{\text{BSM}}$) in each region. 
The observed ($S_{\rm obs}^{95}$) and expected ($S_{\rm exp}^{95}$) 95~\% CL upper limits are also derived using the $CL_{\text{S}}$ procedure. 
These upper limits on $N_{\text{BSM}}$ can be interpreted as upper limits on the visible BSM cross section ($\langle\epsilon{\rm \sigma}\rangle_{\rm obs}^{95}$) by normalising $N_{\text{BSM}}$ by the total integrated luminosity. 
Here $\langle\epsilon{\rm \sigma}\rangle_{\rm obs}^{95}$ is defined as the product of the signal production cross section, acceptance and reconstruction efficiency. 
The results are obtained using asymptotic formulae~\cite{statforumlimits} in the case of the off-$Z$ numbers. 
For SR-Z, with a considerably smaller sample size, pseudo-experiments are used. 
These numbers are presented in Table~\ref{tab:upperlimits} for the on-$Z$ search. 
Model-independent upper limits on the visible BSM cross section in the below-$Z$
and above-$Z$ ranges of the five signal regions in the off-$Z$ search are presented in Tables~\ref{tab:upperLimitsBelow}
and \ref{tab:upperLimitsAbove}, respectively. Limits for the most sensitive dilepton mass windows of
SR-2j-bveto and SR-4j-bveto used for the squark- and gluino-pair model interpretations, respectively, are presented in Tables~\ref{tab:upperLimits-SR2j} and \ref{tab:upperLimits-SR4j}.
These tables also present the confidence level observed for the background-only hypothesis $CL_B$, and the one-sided discovery 
$p$-value, $p(s=0)$, which is 
the probability that the event yield obtained in a single hypothetical background-only experiment (signal, $s=0$) 
is greater than that observed in this dataset. The $p(s=0)$ value is truncated at 0.5.

\begin{table*}
\begin{center}
\small
\setlength{\tabcolsep}{0.0pc}
\begin{tabular*}{\textwidth}{@{\extracolsep{\fill}}llcccccc}
\noalign{\smallskip}\hline\noalign{\smallskip}
{\bf Signal region}    & Channel & $\langle\epsilon{\rm \sigma}\rangle_{\rm obs}^{95}$ [fb]  & $S_{\rm obs}^{95}$  & $S_{\rm exp}^{95}$ & $CL_{B}$ & $p(s=0)$ & Gaussian significance \\
\noalign{\smallskip}\hline\noalign{\smallskip}
SR-Z & $ee+\mu\mu$   & $1.46$ & $29.6$  & ${12}^{+5}_{-2}$ & $0.998$ & $0.0013$  & 3.0 \\
     & $ee$       & $1.00$ & $20.2$  &  ${8}^{+4}_{-2}$ & $0.998$ & $0.0013$     & 3.0 \\
     & $\mu\mu$   & $0.72$ & $14.7$  &  ${9}^{+4}_{-2}$ & $0.951$ & $0.0430$     & 1.7 \\
\noalign{\smallskip}\hline\noalign{\smallskip}
\end{tabular*}
\end{center}
\caption[Breakdown of upper limits for SR-Z.]{From left to right: 95~\% CL upper limits on the visible cross section
($\langle\epsilon{\rm \sigma}\rangle_{\rm obs}^{95}$) and on the number of signal events ($S_{\rm obs}^{95}$); 
the expected 95~\% CL upper limit on the number of signal events is denoted by $S_{\rm exp}^{95}$ 
and is derived from the expected number of background events (and the $\pm 1\sigma$ uncertainty on the expectation); 
two-sided $CL_B$ value, which is the confidence level observed for
the background-only hypothesis; the discovery $p$-value for 0 signal strength $s$ ($p(s = 0)$), and the Gaussian significance for the on-$Z$ search. 
\label{tab:upperlimits}}
\end{table*}

\begin{table*}
\begin{center}
\setlength{\tabcolsep}{0.0pc}
\begin{tabular*}{\textwidth}{@{\extracolsep{\fill}}llccccccc}
\noalign{\smallskip}\hline\noalign{\smallskip}
Signal region     & Channel & $N_\mathrm{data}$ & $N_\mathrm{bkg}$ &  $\langle\epsilon{\rm \sigma}\rangle_{\rm obs}^{95}$[fb]  &  $S_{\rm obs}^{95}$  & $S_{\rm exp}^{95}$ & $CL_{B}$ & $p(s=0)$  \\

\noalign{\smallskip}\hline\noalign{\smallskip}
SR-2j-bveto  & $ee+\mu\mu$ & 54 & $50\pm8\pm5$       & $1.38$ &  $28.0$  & ${24}^{+8}_{-5}$ & $0.66$ &  $0.35$  \\            
             & $ee$        & 30 & $26\pm4\pm3$       & $0.99$ &  $20.1$  & ${18}^{+3}_{-5}$ & $0.73$ &  $0.28$  \\            
             & $\mu\mu$    & 24 & $24\pm3\pm3$       & $0.88$ &  $17.8$  & ${18}^{+3}_{-6}$ & $0.50$ &  $0.50$  \\            
\noalign{\smallskip}\hline\noalign{\smallskip}                                                                                            
SR-2j-btag   & $ee+\mu\mu$ & 79 & $104\pm11\pm7$     & $0.98$ &  $19.8$  & ${30}^{+10}_{-9}$ & $0.06$ &  $0.50$  \\           
             & $ee$        & 40 & $49\pm6\pm4$       & $0.85$ &  $17.2$  & ${20}^{+8}_{-3}$ & $0.19$ &  $0.50$  \\            
             & $\mu\mu$    & 39 & $56\pm6\pm5$       & $0.63$ &  $12.8$  & ${20}^{+9}_{-3}$ & $0.06$ &  $0.50$  \\            
\noalign{\smallskip}\hline\noalign{\smallskip}                                                                                            

SR-4j-bveto  & $ee+\mu\mu$ & 6 & $8.2\pm3.1\pm1.4$         & $0.38$ &  $7.7$  & ${8.3}^{+3.2}_{-1.6}$ & $0.37$ &  $0.50$  \\
             & $ee$        & 1 & $4.7\pm1.6\pm1.1$         & $0.19$ &  $3.9$  & ${5.4}^{+2.0}_{-1.4}$ & $0.08$ &  $0.50$  \\
             & $\mu\mu$    & 5 & $3.6\pm1.5\pm1.0$         & $0.41$ &  $8.4$  & ${6.5}^{+2.9}_{-1.1}$ & $0.75$ &  $0.26$  \\
\noalign{\smallskip}\hline\noalign{\smallskip}                                                                                            
SR-4j-btag   & $ee+\mu\mu$ & 31 & $38\pm6\pm3$       & $0.85$ &  $17.3$  & ${19}^{+7}_{-4}$ & $0.25$ &  $0.50$  \\            
             & $ee$        & 14 & $18\pm3\pm2$       & $0.51$ &  $10.3$  & ${13}^{+6}_{-2}$ & $0.30$ &  $0.50$  \\            
             & $\mu\mu$    & 17 & $20\pm4\pm2$       & $0.54$ &  $10.9$  & ${15}^{+4}_{-5}$ & $0.33$ &  $0.50$  \\            
\noalign{\smallskip}\hline\noalign{\smallskip}
SR-loose       & $ee+\mu\mu$ & 1133 & $1190\pm40\pm70$ & $6.82$ &  $138.4$ & ${170}^{+50}_{-40}$ & $0.28$ &  $0.50$  \\         
             & $ee$        & 509 & $510\pm20\pm40$   & $4.88$ &  $99.0$  &  ${100}^{+40}_{-30}$ & $0.51$ &  $0.48$  \\         
             & $\mu\mu$    & 624 & $680\pm20\pm50$   & $4.10$ &  $83.3$  & ${110}^{+40}_{-30}$ & $0.18$ &  $0.50$  \\         
\noalign{\smallskip}\hline\noalign{\smallskip}                                                                                            
\end{tabular*}
\end{center}
\caption[Breakdown of upper limits.]{
Summary of model-independent upper limits for the five signal regions,
in the below-$Z$ region ($20<\mll<70$ \GeV\ for SR-loose, $20<\mll<80$ \GeV\ for all other signal regions), 
in the combined $ee+\mu\mu$ and individual $ee$ and $\mu\mu$ channels.
Left to right: the observed yield ($N_\mathrm{data}$), total expected background ($N_\mathrm{bkg}$), 95\ \% CL upper limits on the visible cross section
($\langle\epsilon\sigma\rangle_{\rm obs}^{95}$) and on the number of
signal events ($S_{\rm obs}^{95}$ ).  The fifth column
($S_{\rm exp}^{95}$) shows the 95\ \% CL upper limit on the number of
signal events, given the expected number (and $\pm 1\sigma$
excursions on the expectation) of background events.
The last two columns
indicate the $CL_B$ value, i.e. the confidence level observed for
the background-only hypothesis, and the discovery $p$-value ($p(s = 0)$).
For an observed number of events lower than expected, the discovery p-value has been truncated at 0.5.
\label{tab:upperLimitsBelow}}
\end{table*}

\begin{table*}
\begin{center}
\setlength{\tabcolsep}{0.0pc}
\begin{tabular*}{\textwidth}{@{\extracolsep{\fill}}llccccccc}
\noalign{\smallskip}\hline\noalign{\smallskip}
{\bf Signal region} & Channel & $N_\mathrm{data}$ & $N_\mathrm{bkg}$ &  $\langle\epsilon{\rm \sigma}\rangle_{\rm obs}^{95}$[fb]  &  $S_{\rm obs}^{95}$  & $S_{\rm exp}^{95}$ & $CL_{B}$ & $p(s=0)$  \\

\noalign{\smallskip}\hline\noalign{\smallskip}
SR-2j-bveto  & $ee+\mu\mu$ & 55 & $73\pm9\pm9$                  & $0.96$ &  $19.4$  & ${27}^{+8}_{-7}$ & $0.11$ &  $0.50$  \\           
             & $ee$        & 26 & $35\pm5\pm4$                  & $0.60$ &  $12.1$  & ${18}^{+3}_{-6}$ & $0.14$ &  $0.50$  \\           
             & $\mu\mu$    & 29 & $38\pm4\pm8$                  & $0.89$ &  $18.1$  & ${20}^{+8}_{-3}$ & $0.24$ &  $0.50$  \\           
\noalign{\smallskip}\hline\noalign{\smallskip}                                                                                                
SR-2j-btag   & $ee+\mu\mu$ & 164 & $171\pm14\pm16$              & $2.19$ &  $44.4$  & ${48}^{+15}_{-12}$ & $0.39$ &  $0.50$  \\         
             & $ee$        &  83 & $81\pm7\pm7$                 & $1.45$ &  $29.5$  & ${28.3}^{+10}_{-8}$ & $0.56$ &  $0.43$  \\           
             & $\mu\mu$    &  81 & $90\pm7\pm14$                & $1.49$ &  $30.2$  & ${36}^{+10}_{-9}$ & $0.33$ &  $0.50$  \\          
\noalign{\smallskip}\hline\noalign{\smallskip}                                                                                                
SR-4j-bveto  & $ee+\mu\mu$ &  11 & $10\pm3\pm2$                 & $0.56$ &  $11.4$  & ${10}^{+4}_{-3}$ & $0.61$ &  $0.42$ \\
             & $ee$        &   2 & $5.7\pm1.6\pm1.2$            & $0.20$ &  $4.1$  & ${6.0}^{+2.3}_{-1.8}$ & $0.13$ &  $0.50$  \\ 
             & $\mu\mu$    &   9 & $4.5\pm1.3\pm1.7$            & $0.61$ &  $12.3$  & ${7.7}^{+2.7}_{-1.6}$ & $0.91$ &  $0.08$  \\ 
\noalign{\smallskip}\hline\noalign{\smallskip}                                                                                                
SR-4j-btag   & $ee+\mu\mu$ &  41 & $36\pm6\pm5$                 & $1.27$ &  $25.7$  & ${20}^{+9}_{-3}$ & $0.72$ &  $0.29$  \\           
             & $ee$        &  23 & $18\pm3\pm2$                 & $0.96$ &  $19.5$  & ${15}^{+5}_{-4}$ & $0.83$ &  $0.17$  \\           
             & $\mu\mu$    &  18 & $19\pm3\pm4$                 & $0.85$ &  $17.2$  & ${17}^{+3}_{-6}$ & $0.50$ &  $0.50$  \\           
\noalign{\smallskip}\hline\noalign{\smallskip}
SR-loose     & $ee+\mu\mu$ & 1605 & $1600\pm40\pm100$          & $10.58$ &  $214.8$  & ${210}^{+30}_{-40}$ & $0.62$ &  $0.40$  \\      
             & $ee$        &  746 & $760\pm20\pm60$           & $6.63$ &  $134.6$  & ${140}^{+50}_{-40}$ & $0.42$ &  $0.50$  \\       
             & $\mu\mu$    &  859 & $830\pm20\pm70$           & $8.23$ &  $167.1$  & ${150}^{+50}_{-40}$ & $0.64$ &  $0.32$   \\      
\noalign{\smallskip}\hline\noalign{\smallskip}                                                                                              

\end{tabular*}
\end{center}
\caption[Breakdown of upper limits.]{
Summary of model-independent upper limits for the five signal regions,
in the above-$Z$ ($\mll>110$ \GeV) dilepton mass range, in the combined $ee+\mu\mu$ and individual $ee$ and $\mu\mu$ channels.
Details are the same as in Table~\ref{tab:upperLimitsBelow}.
\label{tab:upperLimitsAbove}}
\end{table*}

\begin{table*}
\begin{center}
\footnotesize
\setlength{\tabcolsep}{0.0pc}
\begin{tabular*}{\textwidth}{@{\extracolsep{\fill}}lccccccc|ccc}
\noalign{\smallskip}\hline\noalign{\smallskip}
$m_{\ell\ell}$ range [GeV] &  $N_\mathrm{data}$ & $N_\mathrm{bkg}$ &  $\langle\epsilon{\rm \sigma}\rangle_{\rm obs}^{95}$[fb]  &  $S_{\rm obs}^{95}$  & $S_{\rm exp}^{95}$ & $CL_{B}$ & $p(s=0)$ & $N_{\mathrm{sig}}^{545,385}$ &$N_{\mathrm{sig}}^{665,265}$ &$N_{\mathrm{sig}}^{745,25}$   \\
\noalign{\smallskip}\hline\noalign{\smallskip}
20--50   & 35  & $26 \pm 6 \pm 3$   & $1.32$ &  $26.9$  & ${20}^{+7}_{-4}$ & $0.85$ &  $0.15$   &      $17.1 \pm 1.6$   & $3.7 \pm 0.4$    & $0.6 \pm 0.1$   \\                
20--80   & 54  & $50 \pm 8 \pm 4$   & $1.38$ &  $28.0$  & ${24}^{+8}_{-5}$ & $0.66$ &  $0.35$   &    $[38.0 \pm 2.4]$   & $10.4 \pm 0.6$   & $2.1 \pm 0.2$   \\         
50--80   & 19  & $23 \pm 5 \pm 2$   & $0.63$ &  $12.8$  & ${17}^{+3}_{-7}$ & $0.30$ &  $0.50$   &      $20.9 \pm 1.8$   & $6.7 \pm 0.5$    & $1.5 \pm 0.2$   \\                
50--140  & 34  & $46 \pm 7 \pm 6$   & $0.83$ &  $16.9$  & ${20}^{+8}_{-3}$ & $0.14$ &  $0.50$   &      $27.3 \pm 2.0$   & $28.5 \pm 1.0$   & $6.9 \pm 0.3$   \\                
50--200  & 51  & $75 \pm 9 \pm 8$   & $0.89$ &  $18.1$  & ${26}^{+8}_{-7}$ & $0.05$ &  $0.50$   &      $28.2 \pm 2.1$   & $50.6 \pm 1.4$   & $14.2 \pm 0.5$  \\                
110--200 & 32  & $52 \pm 7 \pm 7$   & $0.69$ &  $14.1$  & ${20}^{+8}_{-3}$ & $0.05$ &  $0.50$  &       $2.8 \pm 0.6$    & $[34.0 \pm 1.1]$   & $10.5 \pm 0.4$  \\       
170--260 & 12  & $24 \pm 5 \pm 2$   & $0.40$ &  $8.2$  & ${12}^{+5}_{-4}$ & $0.03$ &  $0.50$  &       $0.4 \pm 0.2$    & $14.8 \pm 0.7$   & $11.9 \pm 0.4$  \\                
170--290 & 16  & $26 \pm 5 \pm 2$   & $0.43$ &  $8.7$  & ${13}^{+5}_{-4}$ & $0.08$ &  $0.50$  &       $0.4 \pm 0.2$    & $16.1 \pm 0.8$   & $16.8 \pm 0.5$  \\                
$>$170   & 25  & $34 \pm 6 \pm 3$   & $0.68$ &  $13.9$  & ${19}^{+3}_{-5}$ & $0.15$ &  $0.50$  &       $0.4 \pm 0.2$    & $18.5 \pm 0.8$   & $[25.7 \pm 0.6]$  \\       
$>$230   & 16  & $13.1 \pm 3.2 \pm 2.3$   & $0.88$ &  $17.9$  & ${14}^{+5}_{-4}$ & $0.72$ &  $0.29$  &       $0.3 \pm 0.2$    & $5.0 \pm 0.4$    & $17.8 \pm 0.5$  \\                
\noalign{\smallskip}\hline\noalign{\smallskip}
\end{tabular*}
\end{center}
\caption[Breakdown of upper limits.]{
Summary of model-independent upper limits for SR-2j-bveto, in the combined $ee+\mu\mu$ and individual $ee$ and $\mu\mu$ channels,
for the 10 dilepton mass windows used for the squark-pair interpretation.
Details are the same as in Table~\ref{tab:upperLimitsBelow}.
The last three columns indicate the expected signal yield for three squark-pair model benchmark points; the first (second) number indicates the squark (LSP) mass.
The signal yield in square brackets indicates the best selected dilepton mass window for the given benchmark point.
\label{tab:upperLimits-SR2j}}
\end{table*}

\begin{table*}
\begin{center}
\footnotesize
\setlength{\tabcolsep}{0.0pc}
\begin{tabular*}{\textwidth}{@{\extracolsep{\fill}}lccccccc|ccc}
\noalign{\smallskip}\hline\noalign{\smallskip}
$m_{\ell\ell}$ range [GeV] &  $N_\mathrm{data}$ & $N_\mathrm{bkg}$ &  $\langle\epsilon{\rm \sigma}\rangle_{\rm obs}^{95}$[fb]  &  $S_{\rm obs}^{95}$  & $S_{\rm exp}^{95}$ & $CL_{B}$ & $p(s=0)$ & $N_{\mathrm{sig}}^{825,585}$ &$N_{\mathrm{sig}}^{1025,545}$ &$N_{\mathrm{sig}}^{1185,65}$  \\
\noalign{\smallskip}\hline\noalign{\smallskip}

20--50   & 4  &  $3.1 \pm 2.3 \pm 0.9$    & $0.40$ &  $8.2$  & ${7.5}^{+2.0}_{-1.4}$ & $0.70$ &    $0.38$  &$4.4 \pm 0.7$   & $0.8 \pm 0.1$   & $0.1 \pm 0.0$               \\    
20--80   & 6  &  $8.2 \pm 3.1 \pm 1.4$    & $0.38$ &  $7.7$  & ${8.3}^{+3.2}_{-1.6}$ & $0.37$ &    $0.50$  &$[12.8 \pm 1.1]$  & $2.0 \pm 0.2$   & $0.2 \pm 0.0$      \\    
50--140  & 6  &  $8.2 \pm 2.7 \pm 1.4$    & $0.37$ &  $7.5$  & ${8.2}^{+2.9}_{-1.3}$ & $0.35$ &    $0.50$  &$21.4 \pm 1.4$  & $4.9 \pm 0.3$   & $0.7 \pm 0.1$               \\    
110--200 & 9  &  $5.6 \pm 2.3 \pm 1.4$    & $0.59$ &  $12.0$  & ${8.4}^{+3.5}_{-2.0}$ & $0.85$ &   $0.17$  &$4.2 \pm 0.6$   & $6.3 \pm 0.3$   & $1.0 \pm 0.1$               \\   
140--260 & 6  &  $5.0 \pm 2.1 \pm 0.8$    & $0.43$ &  $8.6$  & ${7.4}^{+3.0}_{-1.4}$ & $0.66$ &    $0.38$  &$1.3 \pm 0.4$   & $[8.0 \pm 0.4]$   & $1.6 \pm 0.1$      \\    
$>$20    & 17 & $18 \pm 4 \pm 3$          & $0.63$ &  $12.8$  & ${14}^{+4}_{-4}$ & $0.46$ &  $0.50$  &$27.4 \pm 1.6$  & $14.4 \pm 0.5$  & $7.4 \pm 0.2$               \\  
$>$140   & 7  &  $7.2 \pm 2.4 \pm 1.3$    & $0.41$ &  $8.3$  & ${8.2}^{+3.1}_{-1.3}$ & $0.52$ &    $0.50$  &$1.6 \pm 0.4$   & $8.6 \pm 0.4$   & $6.7 \pm 0.2$               \\    
$>$200   & 2  &  $4.8 \pm 1.8 \pm 1.1$    & $0.21$ &  $4.2$  & ${5.9}^{+2.2}_{-1.7}$ & $0.23$ &    $0.50$  &$0.4 \pm 0.2$   & $4.2 \pm 0.3$   & $6.0 \pm 0.2$               \\    
$>$260   & 1  &  $2.3 \pm 1.2 \pm 0.7$    & $0.19$ &  $3.9$  & ${4.2}^{+1.9}_{-0.3}$ & $0.34$ &    $0.50$  &$0.3 \pm 0.2$   & $0.7 \pm 0.1$   & $[5.1 \pm 0.1]$      \\    

\noalign{\smallskip}\hline\noalign{\smallskip}

\end{tabular*}
\end{center}
\caption[Breakdown of upper limits.]{
Summary of model-independent upper limits for SR-4j-bveto, in the combined $ee+\mu\mu$ and individual $ee$ and $\mu\mu$ channels,
for the nine dilepton mass windows used for the gluino-pair interpretation.
Details are the same as in Table~\ref{tab:upperLimitsBelow}.
The last three columns indicate the expected signal yield for three gluino-pair model benchmark points; the first (second) number indicates the gluino (LSP) mass.
The signal yield in square brackets indicates the best selected dilepton mass window for the given benchmark point.
\label{tab:upperLimits-SR4j}}
\end{table*}

\clearpage
\section{Summary}
\label{sec:summary}
This paper presents results of two searches for supersymmetric particles in events with two \SFOS\
leptons, jets, and \met, using 20.3~fb$^{-1}$ of 8~TeV $pp$ collisions recorded by the ATLAS detector at the LHC. The first search targets events with a lepton pair with invariant mass consistent with that of the $Z$ boson
and hence probes models in which the lepton pair is produced from the decay $Z\to\ell\ell$. 
In this search $6.4\pm2.2$ ($4.2\pm1.6$) events from SM processes are expected in the $\mu\mu$ ($ee$) SR-Z, as predicted using almost exclusively data-driven methods. 
The background estimates for the major and most difficult-to-model backgrounds are cross-checked using MC simulation normalised in data control regions, 
providing further confidence in the SR prediction.  
Following this assessment of the expected background contribution to the SR the number of events in data is higher than anticipated, 
with 13 observed in SR-Z $\mu\mu$ and 16 in SR-Z $ee$.
This corresponding significances are 1.7 standard deviations in the muon channel and 3.0 standard deviations in the electron channel. 
These results are interpreted in a supersymmetric model of general gauge mediation, and probe gluino masses up to 900 GeV.
The second search targets events with a lepton pair with invariant mass inconsistent with $Z$ boson decay,
and probes models with the decay chain $\tilde{\chi}_{2}^{0}\to\ell^+\ell^-\tilde{\chi}_{1}^{0}$.
In this case the data are found to be consistent with the expected SM backgrounds.
No evidence for an excess is observed in the region similar to that in which CMS reported a 2.6$\sigma$ excess~\cite{CMS-edge}.
The results are interpreted in simplified models with squark- and gluino-pair production, and probe
squark (gluino) masses up to about 780 (1170) GeV.

\section*{Acknowledgements}

We thank CERN for the very successful operation of the LHC, as well as the
support staff from our institutions without whom ATLAS could not be
operated efficiently.

We acknowledge the support of ANPCyT, Argentina; YerPhI, Armenia; ARC,
Australia; BMWFW and FWF, Austria; ANAS, Azerbaijan; SSTC, Belarus; CNPq and FAPESP,
Brazil; NSERC, NRC and CFI, Canada; CERN; CONICYT, Chile; CAS, MOST and NSFC,
China; COLCIENCIAS, Colombia; MSMT CR, MPO CR and VSC CR, Czech Republic;
DNRF, DNSRC and Lundbeck Foundation, Denmark; EPLANET, ERC and NSRF, European Union;
IN2P3-CNRS, CEA-DSM/IRFU, France; GNSF, Georgia; BMBF, DFG, HGF, MPG and AvH
Foundation, Germany; GSRT and NSRF, Greece; RGC, Hong Kong SAR, China; ISF, MINERVA, GIF, I-CORE and Benoziyo Center, Israel; INFN, Italy; MEXT and JSPS, Japan; CNRST, Morocco; FOM and NWO, Netherlands; BRF and RCN, Norway; MNiSW and NCN, Poland; GRICES and FCT, Portugal; MNE/IFA, Romania; MES of Russia and NRC KI, Russian Federation; JINR; MSTD,
Serbia; MSSR, Slovakia; ARRS and MIZ\v{S}, Slovenia; DST/NRF, South Africa;
MINECO, Spain; SRC and Wallenberg Foundation, Sweden; SER, SNSF and Cantons of
Bern and Geneva, Switzerland; NSC, Taiwan; TAEK, Turkey; STFC, the Royal
Society and Leverhulme Trust, United Kingdom; DOE and NSF, United States of
America.

The crucial computing support from all WLCG partners is acknowledged
gratefully, in particular from CERN and the ATLAS Tier-1 facilities at
TRIUMF (Canada), NDGF (Denmark, Norway, Sweden), CC-IN2P3 (France),
KIT/GridKA (Germany), INFN-CNAF (Italy), NL-T1 (Netherlands), PIC (Spain),
ASGC (Taiwan), RAL (UK) and BNL (USA) and in the Tier-2 facilities
worldwide.


\clearpage
\bibliographystyle{spphys}       
\bibliography{zedge-epjc.bib}

\begin{thebibliography}{10}
\providecommand{\url}[1]{{#1}}
\providecommand{\urlprefix}{URL }
\expandafter\ifx\csname urlstyle\endcsname\relax
  \providecommand{\doi}[1]{DOI \discretionary{}{}{}#1}\else
  \providecommand{\doi}{DOI \discretionary{}{}{}\begingroup
  \urlstyle{rm}\Url}\fi

\bibitem{Miyazawa:1966}
H.~Miyazawa, Prog. Theor. Phys. \textbf{36 (6)}, 1266 (1966).
\newblock \doi{10.1143/PTP.36.1266}

\bibitem{Ramond:1971gb}
P.~Ramond, Phys. Rev. D \textbf{3}, 2415 (1971).
\newblock \doi{10.1103/PhysRevD.3.2415}

\bibitem{Golfand:1971iw}
Y.A. Golfand, E.P. Likhtman, JETP Lett. \textbf{13}, 323 (1971).
\newblock [Pisma Zh.Eksp.Teor.Fiz.13:452-455,1971]

\bibitem{Neveu:1971rx}
A.~Neveu, J.H. Schwarz, Nucl. Phys. B \textbf{31}, 86 (1971).
\newblock \doi{10.1016/0550-3213(71)90448-2}

\bibitem{Neveu:1971iv}
A.~Neveu, J.H. Schwarz, Phys. Rev. D \textbf{4}, 1109 (1971).
\newblock \doi{10.1103/PhysRevD.4.1109}

\bibitem{Gervais:1971ji}
J.~Gervais, B.~Sakita, Nucl. Phys. B \textbf{34}, 632 (1971).
\newblock \doi{10.1016/0550-3213(71)90351-8}

\bibitem{Volkov:1973ix}
D.V. Volkov, V.P. Akulov, Phys. Lett. B \textbf{46}, 109 (1973).
\newblock \doi{10.1016/0370-2693(73)90490-5}

\bibitem{Wess:1973kz}
J.~Wess, B.~Zumino, Phys. Lett. B \textbf{49}, 52 (1974).
\newblock \doi{10.1016/0370-2693(74)90578-4}

\bibitem{Wess:1974tw}
J.~Wess, B.~Zumino, Nucl. Phys. B \textbf{70}, 39 (1974).
\newblock \doi{10.1016/0550-3213(74)90355-1}

\bibitem{Witten:1981nf}
E.~Witten, Nucl. Phys. B \textbf{188}, 513 (1981).
\newblock \doi{10.1016/0550-3213(81)90006-7}

\bibitem{Dine:1981za}
M.~Dine, W.~Fischler, M.~Srednicki, Nucl. Phys. B \textbf{189}, 575 (1981).
\newblock \doi{10.1016/0550-3213(81)90582-4}

\bibitem{Dimopoulos:1981au}
S.~Dimopoulos, S.~Raby, Nucl. Phys. B \textbf{192}, 353 (1981).
\newblock \doi{10.1016/0550-3213(81)90430-2}

\bibitem{Sakai:1981gr}
N.~Sakai, Zeit. Phys. C \textbf{11}, 153 (1981).
\newblock \doi{10.1007/BF01573998}

\bibitem{Kaul:1981hi}
R.~Kaul, P.~Majumdar, Nucl. Phys. B \textbf{199}, 36 (1982).
\newblock \doi{10.1016/0550-3213(82)90565-X}

\bibitem{Dimopoulos:1981zb}
S.~Dimopoulos, H.~Georgi, Nucl. Phys. B \textbf{193}, 150 (1981).
\newblock \doi{10.1016/0550-3213(81)90522-8}

\bibitem{Fayet:1976et}
P.~Fayet, Phys. Lett. B \textbf{64}, 159 (1976).
\newblock \doi{10.1016/0370-2693(76)90319-1}

\bibitem{Fayet:1977yc}
P.~Fayet, Phys. Lett. B \textbf{69}, 489 (1977).
\newblock \doi{10.1016/0370-2693(77)90852-8}

\bibitem{Farrar:1978xj}
G.R. Farrar, P.~Fayet, Phys. Lett. B \textbf{76}, 575 (1978).
\newblock \doi{10.1016/0370-2693(78)90858-4}

\bibitem{Fayet:1979sa}
P.~Fayet, Phys. Lett. B \textbf{84}, 416 (1979).
\newblock \doi{10.1016/0370-2693(79)91229-2}

\bibitem{Goldberg:1983nd}
H.~Goldberg, Phys. Rev. Lett. \textbf{50}, 1419 (1983).
\newblock \doi{10.1103/PhysRevLett.50.1419}

\bibitem{Ellis:1983ew}
J.~Ellis, J.~Hagelin, D.~Nanopoulos, K.~Olive, M.~Srednicki, Nucl. Phys. B
  \textbf{238}, 453 (1984).
\newblock \doi{10.1016/0550-3213(84)90461-9}

\bibitem{Hinchliffe:1996iu}
I.~Hinchliffe, F.~Paige, M.~Shapiro, J.~Soderqvist, W.~Yao, Phys. Rev. D
  \textbf{55}, 5520 (1997), \href{http://arxiv.org/abs/hep-ph/9610544}{{\tt
  arXiv:hep-ph/9610544 [hep-ph]}}.
\newblock \doi{10.1103/PhysRevD.55.5520}

\bibitem{Chatrchyan:2012qka}
{CMS} Collaboration, Phys. Lett. B \textbf{716}, 260 (2012),
  \href{http://arxiv.org/abs/1204.3774}{{\tt arXiv:1204.3774 [hep-ex]}}.
\newblock \doi{10.1016/j.physletb.2012.08.026}

\bibitem{CMS-edge}
{CMS} Collaboration, JHEP \textbf{04}, 124 (2015),
  \href{http://arxiv.org/abs/1502.06031}{{\tt arXiv:1502.06031 [hep-ex]}}

\bibitem{Chatrchyan:2012te}
{CMS} Collaboration, Phys. Lett. B \textbf{718}, 815 (2013),
  \href{http://arxiv.org/abs/1206.3949}{{\tt arXiv:1206.3949 [hep-ex]}}.
\newblock \doi{10.1016/j.physletb.2012.11.036}

\bibitem{Aad:2008zzm}
{ATLAS} Collaboration, JINST \textbf{3}, S08003 (2008).
\newblock \doi{10.1088/1748-0221/3/08/S08003}

\bibitem{atlastrigger}
{ATLAS} Collaboration, Eur. Phys. J. C \textbf{72}, 1849 (2012),
  \href{http://arxiv.org/abs/1110.1530}{{\tt arXiv:1110.1530 [hep-ex]}}.
\newblock \doi{10.1140/epjc/s10052-011-1849-1}

\bibitem{2011lumi}
{ATLAS} Collaboration, Eur. Phys. J. C \textbf{73}, 2518 (2013),
  \href{http://arxiv.org/abs/1302.4393}{{\tt arXiv:1302.4393 [hep-ex]}}.
\newblock \doi{10.1140/epjc/s10052-013-2518-3}

\bibitem{DYNNLO1}
S.~Catani, L.~Cieri, G.~Ferrera, D.~de~Florian, M.~Grazzini, Phys. Rev. Lett.
  \textbf{103}, 082001 (2009), \href{http://arxiv.org/abs/0903.2120}{{\tt
  arXiv:0903.2120 [hep-ph]}}

\bibitem{DYNNLO2}
S.~Catani, M.~Grazzini, Phys. Rev. Lett. \textbf{98}, 222002 (2007),
  \href{http://arxiv.org/abs/hep-ph/0703012}{{\tt arXiv:hep-ph/0703012
  [hep-ph]}}

\bibitem{Lai:2010vv}
H.L. Lai, et~al., Phys. Rev. D \textbf{82}, 074024 (2010),
  \href{http://arxiv.org/abs/1007.2241}{{\tt arXiv:1007.2241 [hep-ph]}}.
\newblock \doi{10.1103/PhysRevD.82.074024}

\bibitem{ttbarxsec1}
M.~Czakon, P.~Fiedler, A.~Mitov, Phys. Rev. Lett. \textbf{110}, 252004 (2013),
  \href{http://arxiv.org/abs/1303.6254}{{\tt arXiv:1303.6254 [hep-ph]}}

\bibitem{ttbarxsec2}
M.~Czakon, A.~Mitov, Comput. Phys. Commun. \textbf{185}, 2930,
  \href{http://arxiv.org/abs/1112.5675}{{\tt arXiv:1112.5675 [hep-ph]}}.
\newblock \doi{10.1016/j.cpc.2014.06.021}

\bibitem{Kidonakis:2010a}
N.~Kidonakis, Phys. Rev. D \textbf{81}, 054028 (2010),
  \href{http://arxiv.org/abs/1001.5034}{{\tt arXiv:1001.5034 [hep-ph]}}.
\newblock \doi{10.1103/PhysRevD.81.054028}

\bibitem{Kidonakis:2010b}
N.~Kidonakis, Phys. Rev. D \textbf{82}, 054018 (2010),
  \href{http://arxiv.org/abs/1005.4451}{{\tt arXiv:1005.4451 [hep-ph]}}.
\newblock \doi{10.1103/PhysRevD.82.054018}

\bibitem{Pumplin:2002vw}
J.~Pumplin, et~al., JHEP \textbf{0207}, 012 (2002),
  \href{http://arxiv.org/abs/hep-ph/0201195}{{\tt arXiv:hep-ph/0201195}}

\bibitem{Campbell:2012}
J.M. Campbell, R.K. Ellis, JHEP \textbf{1207}, 052 (2012),
  \href{http://arxiv.org/abs/1204.5678}{{\tt arXiv:1204.5678 [hep-ph]}}

\bibitem{Lazopoulos:2008}
A.~Lazopoulos, T.~McElmurry, K.~Melnikov, F.~Petriello, Phys. Lett. B
  \textbf{666}, 62 (2008), \href{http://arxiv.org/abs/0804.2220}{{\tt
  arXiv:0804.2220 [hep-ph]}}

\bibitem{diboson1}
J.M. Campbell, R.K. Ellis, Phys. Rev. D \textbf{60}, 113006 (1999),
  \href{http://arxiv.org/abs/hep-ph/9905386}{{\tt arXiv:hep-ph/9905386
  [hep-ph]}}

\bibitem{diboson2}
J.M. Campbell, R.K. Ellis, C.~Williams, JHEP \textbf{1107}, 018 (2011),
  \href{http://arxiv.org/abs/1105.0020}{{\tt arXiv:1105.0020 [hep-ph]}}

\bibitem{Alwall:2011uj}
J.~Alwall, M.~Herquet, F.~Maltoni, O.~Mattelaer, T.~Stelzer, JHEP
  \textbf{1106}, 128 (2011), \href{http://arxiv.org/abs/1106.0522}{{\tt
  arXiv:1106.0522 [hep-ph]}}.
\newblock \doi{10.1007/JHEP06(2011)128}

\bibitem{PYTHIA}
T.~Sjostrand, S.~Mrenna, P.~Skands, JHEP \textbf{0605}, 026 (2006),
  \href{http://arxiv.org/abs/hep-ph/0603175}{{\tt arXiv:hep-ph/0603175}}

\bibitem{PowhegBOX1}
P.~Nason, JHEP \textbf{0411}, 040 (2004),
  \href{http://arxiv.org/abs/hep-ph/0409146}{{\tt arXiv:hep-ph/0409146
  [hep-ph]}}

\bibitem{PowhegBOX2}
S.~Frixione, P.~Nason, C.~Oleari, JHEP \textbf{0711}, 070 (2007),
  \href{http://arxiv.org/abs/0709.2092}{{\tt arXiv:0709.2092 [hep-ph]}}

\bibitem{PowhegBOX3}
S.~Alioli, P.~Nason, C.~Oleari, E.~Re, JHEP \textbf{1006}, 043 (2010),
  \href{http://arxiv.org/abs/1002.2581}{{\tt arXiv:1002.2581 [hep-ph]}}

\bibitem{Pythia8}
T.~Sjostrand, S.~Mrenna, P.~Skands, Comput. Phys. Comm. \textbf{178}, 852
  (2008), \href{http://arxiv.org/abs/0710.3820}{{\tt arXiv:0710.3820 [hep-ph]}}

\bibitem{sherpa}
T.~Gleisberg, et~al., JHEP \textbf{0902}, 007 (2009),
  \href{http://arxiv.org/abs/0811.4622}{{\tt arXiv:0811.4622 [hep-ph]}}

\bibitem{AUET2}
{{ATLAS}} Collaboration, ATL-PHYS-PUB-2011-008  (2011).
\newblock \url{http://cdsweb.cern.ch/record/1345343}

\bibitem{pythiaperugia}
B.~Cooper, et~al., Eur. Phys. J. C \textbf{72}, 2078 (2011),
  \href{http://arxiv.org/abs/1109.5295}{{\tt arXiv:1109.5295 [hep-ph]}}.
\newblock \doi{10.1140/epjc/s10052-012-2078-y}

\bibitem{Meade:2009qv}
P.~Meade, M.~Reece, D.~Shih, JHEP \textbf{1005}, 105 (2010),
  \href{http://arxiv.org/abs/0911.4130}{{\tt arXiv:0911.4130 [hep-ph]}}.
\newblock \doi{10.1007/JHEP05(2010)105}

\bibitem{Mangano:2006rw}
M.L. Mangano, M.~Moretti, F.~Piccinini, M.~Treccani, JHEP \textbf{0701}, 013
  (2007), \href{http://arxiv.org/abs/hep-ph/0611129}{{\tt arXiv:hep-ph/0611129
  [hep-ph]}}.
\newblock \doi{10.1088/1126-6708/2007/01/013}

\bibitem{Djouadi:2002ze}
A.~Djouadi, J.L. Kneur, G.~Moultaka, Comput. Phys. Commun. \textbf{176}, 426
  (2007), \href{http://arxiv.org/abs/hep-ph/0211331}{{\tt arXiv:hep-ph/0211331
  [hep-ph]}}.
\newblock \doi{10.1016/j.cpc.2006.11.009}

\bibitem{Muhlleitner:2003vg}
M.~Muhlleitner, A.~Djouadi, Y.~Mambrini, Comput. Phys. Commun. \textbf{168}, 46
  (2005), \href{http://arxiv.org/abs/hep-ph/0311167}{{\tt arXiv:hep-ph/0311167
  [hep-ph]}}.
\newblock \doi{10.1016/j.cpc.2005.01.012}

\bibitem{Sherstnev:2007nd}
A.~Sherstnev, R.~Thorne, Eur. Phys. J. C \textbf{55}, 553 (2008),
  \href{http://arxiv.org/abs/0711.2473}{{\tt arXiv:0711.2473 [hep-ph]}}.
\newblock \doi{10.1140/epjc/s10052-008-0610-x}

\bibitem{Beenakker:1996ch}
W.~Beenakker, R.~H{\"o}pker, M.~Spira, P.~Zerwas, Nucl. Phys. B \textbf{492},
  51 (1997), \href{http://arxiv.org/abs/hep-ph/9610490}{{\tt
  arXiv:hep-ph/9610490 [hep-ph]}}.
\newblock \doi{10.1016/S0550-3213(97)00084-9}

\bibitem{Kulesza:2008jb}
A.~Kulesza, L.~Motyka, Phys. Rev. Lett. \textbf{102}, 111802 (2009),
  \href{http://arxiv.org/abs/0807.2405}{{\tt arXiv:0807.2405 [hep-ph]}}.
\newblock \doi{10.1103/PhysRevLett.102.111802}

\bibitem{Kulesza:2009kq}
A.~Kulesza, L.~Motyka, Phys. Rev. D \textbf{80}, 095004 (2009),
  \href{http://arxiv.org/abs/0905.4749}{{\tt arXiv:0905.4749 [hep-ph]}}.
\newblock \doi{10.1103/PhysRevD.80.095004}

\bibitem{Beenakker:2009ha}
W.~Beenakker, et~al., JHEP \textbf{0912}, 041 (2009),
  \href{http://arxiv.org/abs/0909.4418}{{\tt arXiv:0909.4418 [hep-ph]}}.
\newblock \doi{10.1088/1126-6708/2009/12/041}

\bibitem{Beenakker:2011fu}
W.~Beenakker, et~al., Int. J. Mod. Phys. A \textbf{26}, 2637 (2011),
  \href{http://arxiv.org/abs/1105.1110}{{\tt arXiv:1105.1110 [hep-ph]}}.
\newblock \doi{10.1142/S0217751X11053560}

\bibitem{:2010wqa}
{ATLAS} Collaboration, Eur. Phys. J. C \textbf{70}, 823 (2010),
  \href{http://arxiv.org/abs/1005.4568}{{\tt arXiv:1005.4568
  [physics.ins-det]}}.
\newblock \doi{10.1140/epjc/s10052-010-1429-9}

\bibitem{Agostinelli:2002hh}
{GEANT4} Collaboration, S.~Agostinelli, et~al., Nucl. Instrum. Meth. A
  \textbf{506}, 250 (2003).
\newblock \doi{10.1016/S0168-9002(03)01368-8}

\bibitem{atlfast}
{ATLAS} Collaboration, ATL-PHYS-PUB-2010-013  (2010).
\newblock \url{http://cds.cern.ch/record/1300517}

\bibitem{a2tune}
{{ATLAS}} Collaboration, ATL-PHYS-PUB-2011-014  (2011).
\newblock \url{http://cds.cern.ch/record/1400677}

\bibitem{electronref}
{ATLAS} Collaboration, Eur. Phys. J. C \textbf{74}, 2941 (2014),
  \href{http://arxiv.org/abs/1404.2240}{{\tt arXiv:1404.2240 [hep-ex]}}.
\newblock \doi{10.1140/epjc/s10052-014-2941-0}

\bibitem{muonref}
{ATLAS} Collaboration, Eur. Phys. J. C \textbf{74}, 3034 (2014),
  \href{http://arxiv.org/abs/1404.4562}{{\tt arXiv:1404.4562 [hep-ex]}}.
\newblock \doi{10.1140/epjc/s10052-014-3034-9}

\bibitem{ATLAS:2011ad}
{ATLAS} Collaboration, Phys. Rev. D \textbf{85}, 012006 (2012),
  \href{http://arxiv.org/abs/1109.6606}{{\tt arXiv:1109.6606 [hep-ex]}}.
\newblock \doi{10.1103/PhysRevD.85.012006}

\bibitem{Cacciari:2008gp}
M.~Cacciari, G.P. Salam, G.~Soyez, JHEP \textbf{0804}, 063 (2008),
  \href{http://arxiv.org/abs/0802.1189}{{\tt arXiv:0802.1189 [hep-ph]}}.
\newblock \doi{10.1088/1126-6708/2008/04/063}

\bibitem{CSCbook}
{ATLAS} Collaboration, CERN-OPEN-2008-020  (2009),
  \href{http://arxiv.org/abs/0901.0512}{{\tt arXiv:0901.0512 [hep-ex]}}

\bibitem{jetPU}
M.~Cacciari, G.~Salam, G.~Soyez, JHEP \textbf{0804}, 005 (2008),
  \href{http://arxiv.org/abs/0802.1188}{{\tt arXiv:0802.1188 [hep-ph]}}

\bibitem{JES}
{ATLAS} Collaboration, Eur. Phys. J. C \textbf{73}, 2304 (2013),
  \href{http://arxiv.org/abs/1112.6426}{{\tt arXiv:1112.6426 [hep-ex]}}

\bibitem{JES2}
{ATLAS} Collaboration, Eur. Phys. J. C \textbf{73}, 2305 (2013),
  \href{http://arxiv.org/abs/1203.1302}{{\tt arXiv:1203.1302 [hep-ex]}}

\bibitem{Aad:2013zwa}
{ATLAS} Collaboration, JINST \textbf{8}, P07004 (2013),
  \href{http://arxiv.org/abs/1303.0223}{{\tt arXiv:1303.0223 [hep-ex]}}.
\newblock \doi{10.1088/1748-0221/8/07/P07004}

\bibitem{ATLAS-CONF-2012-020}
{ATLAS} Collaboration, ATLAS-CONF-2012-020  (2012).
\newblock \url{http://cds.cern.ch/record/1430034}

\bibitem{ATLAS:2010069}
{ATLAS} Collaboration, ATLAS-CONF-2010-069  (2010).
\newblock \url{http://cds.cern.ch/record/1281344}

\bibitem{MV1}
{ATLAS} Collaboration, ATLAS-CONF-2014-046  (2014).
\newblock \url{http://cds.cern.ch/record/1741020}

\bibitem{met_conf}
{ATLAS} Collaboration, ATLAS-CONF-2013-083  (2013).
\newblock \url{http://cds.cern.ch/record/1570994}

\bibitem{ATLAS:2012ana}
{ATLAS} Collaboration, ATLAS-CONF-2012-123  (2012).
\newblock \url{http://cds.cern.ch/record/1473426}

\bibitem{Aad:2012fqa}
{ATLAS} Collaboration, Phys. Rev. D \textbf{87}, 012008 (2013),
  \href{http://arxiv.org/abs/1208.0949}{{\tt arXiv:1208.0949 [hep-ex]}}.
\newblock \doi{10.1103/PhysRevD.87.012008}

\bibitem{Aad:2014kra}
{ATLAS} Collaboration, JHEP \textbf{1411}, 118 (2014),
  \href{http://arxiv.org/abs/1407.0583}{{\tt arXiv:1407.0583 [hep-ex]}}.
\newblock \doi{10.1007/JHEP11(2014)118}

\bibitem{Aad:2014wea}
{ATLAS} Collaboration, JHEP \textbf{1409}, 176 (2014),
  \href{http://arxiv.org/abs/1405.7875}{{\tt arXiv:1405.7875 [hep-ex]}}.
\newblock \doi{10.1007/JHEP09(2014)176}

\bibitem{Aad:2015mia}
{ATLAS} Collaboration, JHEP \textbf{04}, 116 (2015),
  \href{http://arxiv.org/abs/1501.03555}{{\tt arXiv:1501.03555 [hep-ex]}}

\bibitem{matrixmethod}
{ATLAS} Collaboration, Eur. Phys. J. C \textbf{71}, 1577 (2011),
  \href{http://arxiv.org/abs/1012.1792}{{\tt arXiv:1012.1792 [hep-ex]}}

\bibitem{atlas-jer}
{ATLAS} Collaboration, Eur. Phys. J. C \textbf{73}, 2306 (2013),
  \href{http://arxiv.org/abs/1210.6210}{{\tt arXiv:1210.6210 [hep-ex]}}.
\newblock \doi{10.1140/epjc/s10052-013-2306-0}

\bibitem{ATLAS-CONF-2014-004}
{ATLAS} Collaboration, ATLAS-CONF-2014-004  (2014).
\newblock \url{http://cds.cern.ch/record/1664335}

\bibitem{ATLAS-CONF-2012-043}
{ATLAS} Collaboration, ATLAS-CONF-2012-043  (2012).
\newblock \url{http://cdsweb.cern.ch/record/1435197}

\bibitem{ATLAS-CONF-2012-040}
{ATLAS} Collaboration, ATLAS-CONF-2012-040  (2012).
\newblock \url{http://cdsweb.cern.ch/record/1435194}

\bibitem{Botje01}
M.~Botje, et~al.
\newblock {The PDF4LHC Working Group Interim Recommendations},
  \href{http://arxiv.org/abs/1101.0538}{{\tt arXiv:1101.0538 [hep-ph]}} (2011)

\bibitem{Corcella:2000bw}
G.~Corcella, et~al., JHEP \textbf{0101}, 010 (2001),
  \href{http://arxiv.org/abs/hep-ph/0011363}{{\tt arXiv:hep-ph/0011363
  [hep-ph]}}

\bibitem{Butterworth:1996zw}
J.~Butterworth, J.R. Forshaw, M.~Seymour, Z. Phys. C \textbf{72}, 637 (1996),
  \href{http://arxiv.org/abs/hep-ph/9601371}{{\tt arXiv:hep-ph/9601371
  [hep-ph]}}.
\newblock \doi{10.1007/s002880050286}

\bibitem{Kramer:2012bx}
M.~Kramer, et~al.
\newblock {Supersymmetry production cross sections in pp collisions at
  $\sqrt{s} = 7\TeV$}, \href{http://arxiv.org/abs/1206.2892}{{\tt
  arXiv:1206.2892 [hep-ph]}} (2012)

\bibitem{cls}
A.~Read, J. Phys. G: Nucl. Part. Phys. \textbf{28}(10), 2693 (2002).
\newblock \doi{10.1088/0954-3899/28/10/313}

\bibitem{Baak:2014wma}
M.~Baak, et~al., Eur. Phys. J. \textbf{C75}(4), 153 (2015),
  \href{http://arxiv.org/abs/1410.1280}{{\tt arXiv:1410.1280 [hep-ex]}}.
\newblock \doi{10.1140/epjc/s10052-015-3327-7}

\bibitem{statforumlimits}
G.~Cowan, K.~Cranmer, E.~Gross, O.~Vitells, Eur. Phys. J. C \textbf{71}, 1554
  (2011), \href{http://arxiv.org/abs/1007.1727}{{\tt arXiv:1007.1727
  [physics.data-an]}}.
\newblock \doi{10.1140/epjc/s10052-011-1554-0}

\end{thebibliography}


\clearpage
\appendix
\section{Additional results of off-$Z$ search}
\label{app:edge}

This section provides additional results of the off-$Z$ search.
The expected backgrounds and observed yields in the below-$Z$ and above-$Z$ regions for VR, SR-2j-btag, and SR-4j-btag,
are presented in Tables~\ref{tab:VR}, \ref{tab:SR2jbtag}, and \ref{tab:SR4jbtag}, respectively.
\begin{table*}[!htbp]
\begin{center}
\setlength{\tabcolsep}{0.0pc}
{\small
\begin{tabular*}{\textwidth}{@{\extracolsep{\fill}}lrrr}
\noalign{\smallskip}\hline\noalign{\smallskip}
Below-$Z$ ($20<\mll<70$ \GeV)      & VR-offZ $ee$   & VR-offZ $\mu\mu$ & VR-offZ same-flavour             \\[-0.05cm]
               &        &          & combined             \\[-0.05cm]
\noalign{\smallskip}\hline\noalign{\smallskip}
          Observed events   &                      465   &                      742   &                     1207  \\
\noalign{\smallskip}\hline\noalign{\smallskip}
Expected background events   &  $445 \pm 15 \pm 36$       &  $682 \pm 23 \pm 53$   & $1128 \pm 37 \pm 69$  \\
\noalign{\smallskip}\hline\noalign{\smallskip}
Flavour-symmetric backgrounds   &  $425\pm 15 \pm 36$   &  $661 \pm 22 \pm 53$   & $1086 \pm 37 \pm 68$  \\
                  \dyjets   &      $1.5\pm 0.5\pm 1.6$   &      $2.7\pm 0.6\pm 3.5$   &      $4.1\pm 0.8\pm 4.7$  \\
                 Rare top   &      $0.1\pm 0.0\pm 0.0$   &      $0.1\pm 0.0\pm 0.0$   &      $0.1\pm 0.0\pm 0.0$  \\
            $WZ/ZZ$ diboson   &      $0.6\pm 0.2\pm 0.1$   &      $0.8\pm 0.2\pm 0.1$   &      $1.4\pm 0.2\pm 0.2$  \\
             Fake leptons   &     $18 \pm 4 \pm 2$   &    $18 \pm 4 \pm 4$   &     $36 \pm 5 \pm 7$  \\
\noalign{\smallskip}\hline\noalign{\smallskip}
\noalign{\smallskip}\hline\noalign{\smallskip}
Above-$Z$ ($\mll>110$ \GeV)      & VR-offZ $ee$   & VR-offZ $\mu\mu$ & VR-offZ same-flavour             \\[-0.05cm]
               &        &          & combined             \\[-0.05cm]
\noalign{\smallskip}\hline\noalign{\smallskip}
          Observed events   &                      550   &                      732   &                     1282  \\
\noalign{\smallskip}\hline\noalign{\smallskip}
Expected background events   &  $594 \pm 18 \pm 48$   &  $696 \pm 21 \pm 55$   & $1290 \pm 38 \pm 79$  \\
\noalign{\smallskip}\hline\noalign{\smallskip}
Flavour-symmetric backgrounds   &  $571 \pm 17 \pm 48$   &  $684 \pm 21 \pm 55$   & $1254 \pm 38 \pm 79$  \\
                  \dyjets   &      $1.9\pm 0.7\pm 2.0$   &      $3.8\pm 0.4\pm 6.0$   &      $5.7\pm 0.8\pm 7.5$  \\
                 Rare top   &      $<0.1$   &      $<0.1$   &      $<0.1$  \\
            $WZ/ZZ$ diboson   &      $0.9\pm 0.2\pm 0.2$   &      $0.9\pm 0.2\pm 0.2$   &      $1.8\pm 0.3\pm 0.2$  \\
             Fake leptons   &     $21 \pm 4 \pm 2$   &      $7.9\pm 3.1\pm 2.9$   &     $29 \pm 5 \pm 4$  \\
\noalign{\smallskip}\hline\noalign{\smallskip}
\end{tabular*}
}
\end{center}
\caption{{\small Results in the off-$Z$ validation region (VR-offZ), in the below-$Z$ range ($20<\mll<70$ \GeV, top) and above-$Z$ range ($\mll>110$ \GeV, bottom).
The flavour symmetric, \dyjets\ and fake lepton background components are all derived using data-driven estimates described in the text.
All other backgrounds are taken from MC simulation. The first uncertainty is statistical and the second is systematic.}}
\label{tab:VR}
\end{table*}

\begin{table*}[h]
\begin{center}
\setlength{\tabcolsep}{0.0pc}
{\small
\begin{tabular*}{\textwidth}{@{\extracolsep{\fill}}lrrr}
\noalign{\smallskip}\hline\noalign{\smallskip}
Below-$Z$ ($20<\mll<80$ \GeV)      & SR-2j-btag $ee$   & SR-2j-btag $\mu\mu$ & SR-2j-btag same-flavour             \\[-0.05cm]
               &        &          & combined             \\[-0.05cm]
\noalign{\smallskip}\hline\noalign{\smallskip}
          Observed events   &                       40   &                       39   &                       79  \\
\noalign{\smallskip}\hline\noalign{\smallskip}
Expected background events   &     $49 \pm 6 \pm 4$   &     $56 \pm 6 \pm 5$   &   $104 \pm 11 \pm 7$  \\
\noalign{\smallskip}\hline\noalign{\smallskip}
Flavour-symmetric backgrounds   &     $45 \pm 5 \pm 4$   &     $49 \pm 6 \pm 5$   &    $94 \pm 11 \pm 7$  \\
                  \dyjets   &      $1.8\pm 1.0\pm 0.8$   &      $3.1\pm 1.3\pm 1.9$   &      $4.9\pm 1.6\pm 2.2$  \\
                 Rare top   &      $0.1\pm 0.0\pm 0.0$   &      $0.1\pm 0.0\pm 0.0$   &      $0.1\pm 0.0\pm 0.0$  \\
            $WZ/ZZ$ diboson   &      $<0.1$   &      $0.1\pm 0.1\pm 0.1$   &      $0.1\pm 0.1\pm 0.1$  \\
             Fake leptons   &      $2.3\pm 1.2\pm 0.3$   &      $3.4\pm 1.9\pm 0.2$   &      $5.7\pm 2.3\pm 0.6$  \\
\noalign{\smallskip}\hline\noalign{\smallskip}
\noalign{\smallskip}\hline\noalign{\smallskip}
Above-$Z$ ($\mll>110$ \GeV)      & SR-2j-btag $ee$   & SR-2j-btag $\mu\mu$ & SR-2j-btag same-flavour             \\[-0.05cm]
               &        &          & combined             \\[-0.05cm]
\noalign{\smallskip}\hline\noalign{\smallskip}
          Observed events   &                       83   &                       81   &                      164  \\
\noalign{\smallskip}\hline\noalign{\smallskip}
Expected background events   &     $81 \pm 7 \pm 7$   &    $90 \pm 7 \pm 14$   &  $171 \pm 14 \pm 16$  \\
\noalign{\smallskip}\hline\noalign{\smallskip}
Flavour-symmetric backgrounds   &     $78 \pm 7 \pm 7$   &     $77 \pm 7 \pm 7$   &  $155 \pm 13 \pm 10$  \\
                  \dyjets   &      $0.8\pm 0.5\pm 0.4$   &    $11 \pm 1 \pm 13$   &    $12 \pm 1 \pm 13$  \\
                 Rare top   &      $<0.1$   &      $<0.1$   &      $0.1\pm 0.0\pm 0.0$  \\
            $WZ/ZZ$ diboson   &      $<0.1$   &      $<0.1$   &      $<0.1$  \\
             Fake leptons   &      $2.4\pm 1.6\pm 0.8$   &      $1.6\pm 1.3\pm 0.2$   &      $4.0\pm 2.1\pm 0.7$  \\
\noalign{\smallskip}\hline\noalign{\smallskip}
\end{tabular*}
}
\end{center}
\caption{{\small Results in the off-$Z$ search region SR-2j-btag, in the below-$Z$ range ($20<\mll<80$ \GeV, top) and above-$Z$ range ($\mll>110$ \GeV, bottom).
Details are the same as in Table~\ref{tab:SR2jbveto}.}}
\label{tab:SR2jbtag}
\end{table*}

\begin{table*}[h]
\begin{center}
\setlength{\tabcolsep}{0.0pc}
{\small
\begin{tabular*}{\textwidth}{@{\extracolsep{\fill}}lrrr}
\noalign{\smallskip}\hline\noalign{\smallskip}
Below-$Z$ ($20<\mll<80$ \GeV)      & SR-4j-btag $ee$   & SR-4j-btag $\mu\mu$ & SR-4j-btag same-flavour             \\[-0.05cm]
               &        &          & combined             \\[-0.05cm]
\noalign{\smallskip}\hline\noalign{\smallskip}
          Observed events   &                       14   &                       17   &                       31  \\
\noalign{\smallskip}\hline\noalign{\smallskip}
Expected background events   &     $18 \pm 3 \pm 2$   &     $20 \pm 4 \pm 2$   &     $38 \pm 6 \pm 3$  \\
\noalign{\smallskip}\hline\noalign{\smallskip}
Flavour-symmetric backgrounds   &     $17 \pm 3 \pm 2$   &     $18 \pm 3 \pm 2$   &     $35 \pm 6 \pm 3$  \\
                  \dyjets   &      $0.5\pm 0.3\pm 0.5$   &      $0.7\pm 0.3\pm 0.9$   &      $1.2\pm 0.4\pm 1.1$  \\
                 Rare top   &      $<0.1$   &      $<0.1$   &      $0.1\pm 0.0\pm 0.0$  \\
            $WZ/ZZ$ diboson   &      $<0.1$   &      $<0.1$   &      $<0.1$  \\
             Fake leptons   &      $0.3\pm 0.6\pm 0.0$   &      $1.3\pm 1.2\pm 0.0$   &      $1.6\pm 1.4\pm 0.2$  \\
\noalign{\smallskip}\hline\noalign{\smallskip}
\noalign{\smallskip}\hline\noalign{\smallskip}
Above-$Z$ ($\mll>110$ \GeV)      & SR-4j-btag $ee$   & SR-4j-btag $\mu\mu$ & SR-4j-btag same-flavour             \\[-0.05cm]
               &        &          & combined             \\[-0.05cm]
\noalign{\smallskip}\hline\noalign{\smallskip}
          Observed events   &                       23   &                       18   &                       41  \\
\noalign{\smallskip}\hline\noalign{\smallskip}
Expected background events   &     $18 \pm 3 \pm 2$   &     $19 \pm 3 \pm 4$   &     $36 \pm 6 \pm 5$  \\
\noalign{\smallskip}\hline\noalign{\smallskip}
Flavour-symmetric backgrounds   &     $17 \pm 3 \pm 2$   &     $16 \pm 3 \pm 2$   &     $33 \pm 6 \pm 3$  \\
                  \dyjets   &      $0.2\pm 0.1\pm 0.2$   &      $2.4\pm 0.3\pm 4.0$   &      $2.7\pm 0.3\pm 4.1$  \\
                 Rare top   &      $<0.1$   &      $<0.1$   &      $<0.1$  \\
            $WZ/ZZ$ diboson   &      $<0.1$   &      $<0.1$   &      $<0.1$  \\
             Fake leptons   &                    $<0.6$   &      $0.3\pm 0.6\pm 0.1$   &      $0.0\pm 0.9\pm 0.2$  \\
\noalign{\smallskip}\hline\noalign{\smallskip}
\end{tabular*}
}
\end{center}
\caption{{\small Results in the off-$Z$ search region SR-4j-tag, in the below-$Z$ range ($20<\mll<80$ \GeV, top) and above-$Z$ range ($\mll>110$ \GeV, bottom).
Details are the same as in Table~\ref{tab:SR2jbveto}.}}
\label{tab:SR4jbtag}
\end{table*}

\clearpage
\begin{flushleft}
{\Large The ATLAS Collaboration}

\bigskip

G.~Aad$^{\rm 85}$,
B.~Abbott$^{\rm 113}$,
J.~Abdallah$^{\rm 152}$,
O.~Abdinov$^{\rm 11}$,
R.~Aben$^{\rm 107}$,
M.~Abolins$^{\rm 90}$,
O.S.~AbouZeid$^{\rm 159}$,
H.~Abramowicz$^{\rm 154}$,
H.~Abreu$^{\rm 153}$,
R.~Abreu$^{\rm 30}$
% ...
\bigskip
\\
% ...
$^{11}$ Institute of Physics, Azerbaijan Academy of Sciences, Baku, Azerbaijan\\
% ...
$^{30}$ CERN, Geneva, Switzerland\\
% ...
$^{85}$ CPPM, Aix-Marseille Universit{\'e} and CNRS/IN2P3, Marseille, France\\
% ...
$^{90}$ Department of Physics and Astronomy, Michigan State University, East Lansing MI, United States of America\\
% ...
$^{107}$ Nikhef National Institute for Subatomic Physics and University of Amsterdam, Amsterdam, Netherlands\\
% ...
$^{113}$ Homer L. Dodge Department of Physics and Astronomy, University of Oklahoma, Norman OK, United States of America\\
% ...
$^{152}$ Institute of Physics, Academia Sinica, Taipei, Taiwan\\
$^{153}$ Department of Physics, Technion: Israel Institute of Technology, Haifa, Israel\\
$^{154}$ Raymond and Beverly Sackler School of Physics and Astronomy, Tel Aviv University, Tel Aviv, Israel\\
% ...
$^{159}$ Department of Physics, University of Toronto, Toronto ON, Canada\\
% ...
$^{*}$ Deceased
\end{flushleft}


\end{document}